\documentclass[journal=jacsat,manuscript=article]{achemso}

\usepackage{chemformula} % Formula subscripts using \ch{}
\usepackage[T1]{fontenc} % Use modern font encodings
\usepackage{hyperref}
\usepackage{soul}
\usepackage[dvipsnames]{xcolor}

\author{Haotian Chen}
\affiliation{Michigan Institute for Data and AI in Society, University of Michigan, Ann Arbor, MI, 48109, United States}
\alsoaffiliation{Department of Aerospace Engineering, University of Michigan, Ann Arbor, MI, 48109, United States}
\author{Chenyang Huang}
\affiliation{Department of Mechanical Engineering, University of Michigan, Ann Arbor, MI, 48109, United States}

\author{Alexander Rodríguez}
\affiliation{Division of Computer Science and Engineering, University of Michigan, Ann Arbor, MI, 48109, United States}

\author{Aashutosh Mistry}
\affiliation{Department of Mechanical Engineering, Colorado School of Mines, Golden, CO, 80401, United States}
\alsoaffiliation{Payne Institute for Public Policy, Colorado School of Mines, Golden, CO, 80401, United States}

\author{Venkatasubramanian Viswanathan}
\affiliation{Department of Aerospace Engineering, University of Michigan, Ann Arbor, MI, 48109, United States}
\alsoaffiliation{Department of Mechanical Engineering, University of Michigan, Ann Arbor, MI, 48109, United States}
\email{venkvis@umich.edu}
\title{Differentiable Electrochemistry: A paradigm for uncovering hidden physical phenomena in electrochemical systems}

\abbreviations{BV,MH,MHC}

\begin{document}

\begin{tocentry}
\centering
\includegraphics[scale=0.22]{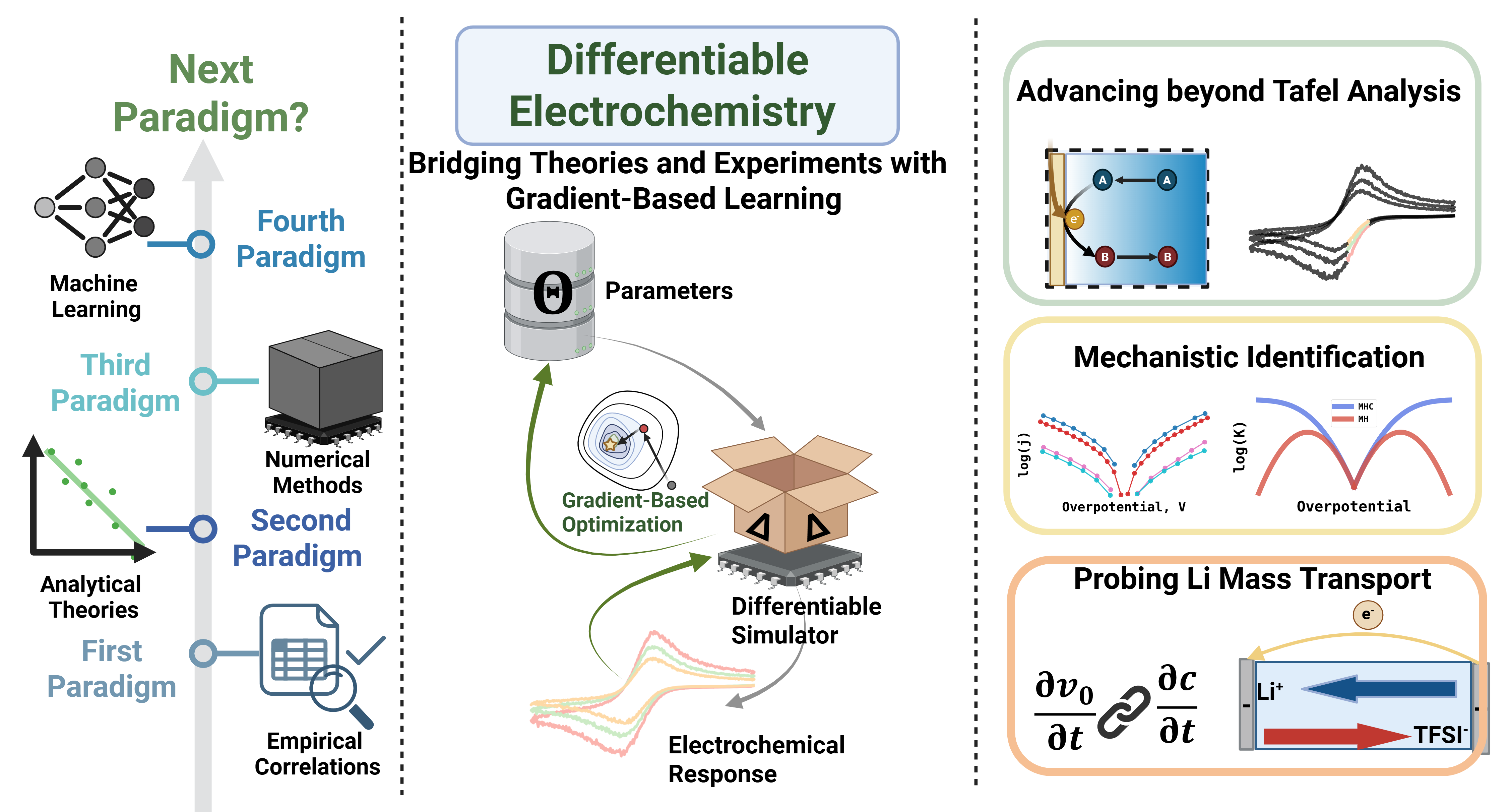}
\end{tocentry}

\begin{abstract}

Despite the long history of electrochemistry, there is a lack of quantitative algorithms that rigorously correlate experiment with theory. Electrochemical modeling has had advanced across empirical, analytical, numerical, and data-driven paradigms. Data-driven machine learning and physics based electrochemical modeling, however, have not been explicitly linked. Here we introduce Differentiable Electrochemistry, a mew paradigm in electrochemical modeling that integrates thermodynamics, kinetics and mass transport with differentiable programming enabled by automatic differentiation. By making the entire electrochemical simulation end-to-end differentiable, this framework enables gradient-based optimization for mechanistic discovery from experimental and simulation data, achieving approximately one to two orders of improvement over gradient-free methods. We develop a rich repository of differentiable simulators across diverse mechanisms, and apply Differentiable Electrochemistry to bottleneck problems in kinetic analysis. Specifically, Differentiable Electrochemistry advances beyond Tafel and Nicholson method by removing several limitations including Tafel region selection, and identifies the electron transfer mechanism in Li metal electrodeposition/stripping by parameterizing the full Marcus-Hush-Chidsey formalism. In addition, Differentiable Electrochemistry interprets \emph{Operando} X-ray measurements in concentrated electrolyte by coupling concentration and velocity theories. This framework resolves ambiguity when multiple electrochemical theories intertwine, and establishes a physics-consistent and data-efficient foundation for predictive electrochemical modeling.

\end{abstract}

\section{Main Text}

\begin{figure}
    \centering
    \includegraphics[width=1\linewidth]{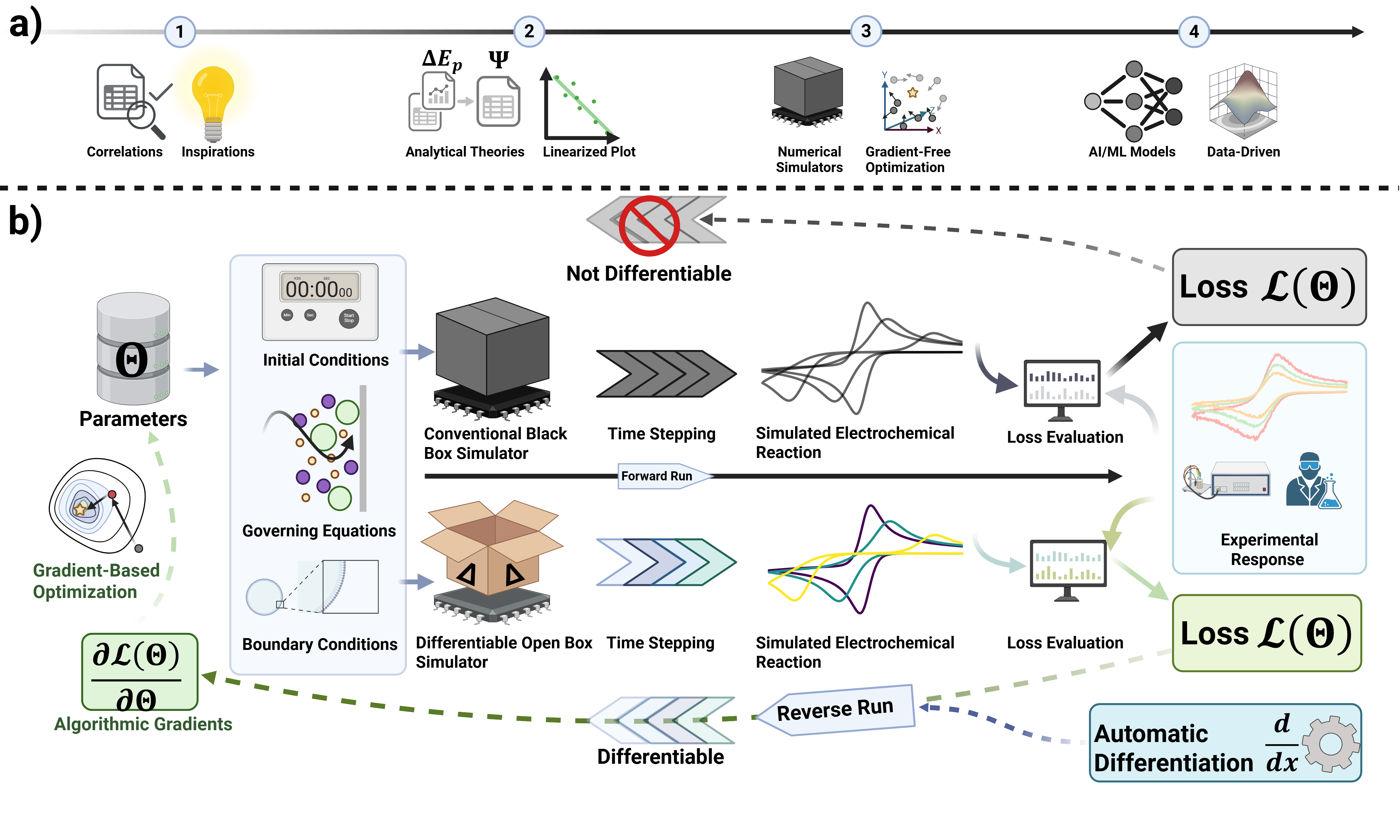}
    \caption{Differentiable Electrochemistry is proposed as the new paradigm of electrochemical modeling. (a) The four paradigms of scientific modeling: observations for correlation and inspiration, linearization of analytical theory (e.g. Nicholson method), numerical simulation with gradient-free optimization, and data-driven ML models. (b) The principle of Differentiable Electrochemistry simulations in comparison with conventional black-box simulators. The parameter set $\Theta$ populates initial conditions, boundary conditions, and governing equations of simulations for time stepping (forward run). The simulated responses are then compared with experimental responses to obtain loss values $\mathcal{L}\left(\Theta\right)$. In a conventional simulator, the losses are not differentiable with respect to $\Theta$, and thus requires parameter sweep or gradient-free optimization for parameter discovery. On the contrary, Differentiable Electrochemistry simulation utilizes differentiable programming and automatic differentiation to obtain the algorithmic gradients of loss with respect to parameters $\left(\frac{\partial \mathcal{L}\left(\Theta\right)}{\partial \Theta}\right)$ for gradient-based optimization.  }
    \label{fig:IntroducingDiffEC}
\end{figure}

Accurate modeling of electrochemical systems for reaction mechanism discovery, operation parameter identification, and material property evaluation underpins the success of modern electrochemical energy systems and their applications for electric mobility.\cite{RN1,RM1,RM2,RM3,RM4,Brucker2024Off-road}  As conceptualized by Kitchin, the evolution of scientific modeling can be traced through four major paradigms (Figure \ref{fig:IntroducingDiffEC}a). \cite{kitchin2025beyond}
The first paradigm relied on empirical correlations and tabulated data requiring little mechanistic understanding. Early electrochemical studies by Volta and Faraday exemplified this stage, where systematic experimentation revealed that current increases with applied potential and that charge transfer follows quantitative laws, forming the empirical foundation of electrochemistry. 
The second paradigm emerged with manual fitting of analytical expression to experimental data, enabling extraction of transport and kinetic parameters from linearized plots. Classical techniques like Tafel analysis and Nicholson method fall under this paradigm.\cite{tafel1905polarisation,Nicholson1965}
With the rise of digital computing, the third paradigm introduced nonlinear regression, multivariate analysis, and large-scale first-principle simulations. Continuum modeling combined with gradient-free optimization became a defining feature of this era.\cite{Coupling2023Bell,BondChemElectroChem}
More recently, the fourth paradigm, data-driven machine learning (ML), has transformed modeling through highly flexible, data-intensive methods capable of capturing complex correlations without explicitly embedding underlying physics. Neural networks, empowered by automatic differentiation (AD), exemplify this paradigm. However, purely data-driven ML model remains data-hungry and often functions as black box with limited physical interpretability.

Applications of ML in electrochemistry include Bayesian inference for electrode kinetic mechanisms,\cite{RN2} uncovering coupled (electro)chemical reaction mechanisms with computer vision,\cite{RN3} study of electro-descriptor for charge transfer mechanics of photocatalytic reactions,\cite{RN4} and quantification of ion transport properties.\cite{Mistry2024BO} Yet, despite these advances, AI/ML approaches are heavily data-dependent, often functioning as black-box models with limited physical interpretability and consistency. These limitations of purely data-driven approaches are particularly evident in electrochemical energy research, where multiscale and multiphysics interactions, incomplete mechanistic understanding, and fragmented experimental/theoretical insights hinder the development of predictive and interpretable models.\cite{RM5,AP20} There is a perceived paucity of quantitative studies in electrochemistry, due to limited bridging between experimental observation and mathematical modeling of electrochemical systems.\cite{RN5,MEMSIM1} As a result, they rarely provide causal insight and struggle to overcome fundamental bottlenecks in electrochemical systems.\cite{RM5,kitchin2025beyond} Thus, developing an accurate, fast, interpretable and data efficient mechanistic identification pipeline will be essential to bridge these gaps and to catalyze rigorous innovations in electrochemical systems and their technological applications. These challenges signals the emergence of a fifth paradigm, one that merges the physical interpretability of mechanistic modeling with the optimization efficiency of modern machine learning. As envisioned by Kitchin,\cite{kitchin2025beyond} differentiable programming enabled by AD forms the technical foundation of this paradigm, allowing gradients to propagate through every stage of an electrochemical simulation and thereby bridging physics-based model with data-driven learning.

Accordingly, electrochemical research is now entering the the fifth paradigm of scientific modeling, one that integrates rigorous physicochemical principles with the flexibility of differentiable programming. Here, we introduce Differentiable Electrochemistry simulations as a new paradigm designed to directly interpret experiments for system identification and to establish a new foundation for predictive and interpretable modeling of electrochemical systems.\cite{kitchin2025beyond,RN8} Differentiable simulation refers to the modeling of physical systems within a differentiable programming environment, where every computational step is constructed to be end-to-end differentiable. Differentiable simulations are compared with conventional black-box simulators in the context of electrochemistry and shown in Figure \ref{fig:IntroducingDiffEC}b. This enables the use of gradient-based optimization and learning algorithms directly on the simulation. In practice, electrochemical systems are governed by partial differential equations (PDEs) that describe coupled mass transport and interfacial reactions, for example:
\begin{equation}
    \frac{\partial u }{\partial t}=\mathcal{F}(u, \nabla u, \nabla^2 u, \dots; \Theta)
\end{equation}
where $u$ denotes the state variable (e.g., concentration $c(x,t)$, electric potential $\phi(x,t)$), $\mathcal{F}$ is an operator that denotes mass transport and electrochemical reaction processes, and $\Theta$ represents material or kinetic parameters including diffusion coefficients, electrochemical rate constants, transfer coefficients, reorganization energies, etc. Conventional simulations solve for $u(t,x)$ given $\Theta$, but they are not easily differentiable with respect to $\Theta$. As a result, parameter estimation or inverse design must rely on gradient-free optimization methods or surrogate models.  The former includes particle swarm optimization (PSO), Bayesian optimization (BO), covariance matrix adaptation evolution strategy (CMA-ES), Nelder–Mead (NM) method, etc. Neural networks, random forests, and gradient-boosting methods are the fourth paradigm surrogate models that builds a nonlinear relationship between parameters and data.\cite{RN22} These heuristic search methods and surrogate models are often computationally expensive, data-inefficient, non-interpretable, non-extrapolative, and vulnerable to the curse of dimensionality. By contrast, differentiable simulation makes the solution map $u(\Theta)$ differentiable, enabling efficient evaluation of gradient such as $\nabla_{\Theta}\mathcal{L}\left(u\left(\Theta\right),u_{exp}\right)$, 
where $\mathcal{L}$ is the loss function comparing simulation outputs to experimental measurements $u_{exp}$. Within this fifth paradigm, the dynamics of an electrochemical systems can be rigorously described by its governing PDEs, enabling direct gradient-based inference and optimization from experimental data, providing orders of magnitude improvement in efficiency and accuracy over conventional approaches.\cite{RN9, martins2021engineering}
This approach has already gained traction in  robotics,\cite{RN9} computational fluid dynamics,\cite{RN10,RM9} thermodynamic modeling,\cite{RM8,AP19} quantum chemistry,\cite{RN11} and molecular dynamics,\cite{RN12} but has not yet been systematically applied to electrochemical systems.

Major electrochemistry simulation software including commercial ones like DigiSim,\cite{DigiSim} DigiElch,\cite{DigiElch} and COMSOL Electroanalysis module,\cite{COMSOL} or open-source software like MEMSim,\cite{MEMSIM1} and FreeSim,\cite{FreeSimPaper} are not differentiable and rely on either parameter sweep or gradient-free optimization for system identification Table S 2. This limitation reflects the intrinsic challenges of electrochemistry, where interfacial reactions, coupled mass transport, and time-dependent dynamics introduce significant barriers to differentiable formulations. As a result, the lack of rigorous quantitative methods that directly solve multiphysics electrochemical problems to correlate experimental observations with underlying parameters remains a central bottleneck, hindering the translation of experiments into fundamental mechanistic understanding.

 This work contributes along two complementary perspectives. First, we developed five differentiable electrochemistry simulators (see Table S 1) that span fundamental and advanced electrochemical mechanisms, mass transport regimes (diffusion, migration, and convection), electrokinetic model (Nernst, Butler-Volmer, and Marcus-Hush-Chidsey), nonlinear kinetics (second-order chemical kinetics and Langmuir adsorption/desorption), and electrode geometries (macroelectrode, microsphere, microcylinder, and rotating disk). These differentiable simulators are provided as general-purpose, easily adaptable modules, accompanied by open-source data and code, and detailed documentation in the Supporting Information. Notably, the Differentiable Electrochemistry for Voltammetry of Adsorbed Species section in the Supporting Information demonstrates simultaneous estimation of 10 parameters, a tremendous challenge for stochastic methods due to the large parameter space and the associated curse of dimensionality.\cite{BondChemElectroChem} The Proof-of-Concept Case Study section in the Supporting Information provides three theoretical and experimental cases, with a notable example solving Poisson-Nernst-Planck (PNP) equation for differentiable parameter estimation of voltammetry in weakly supported media like battery electrolyte. This paper thus equips readers with the necessary resources and practical guidance to apply Differentiable Electrochemistry simulations to their own research problems, while laying the groundwork for broader adoption of differentiable simulation as a transformative paradigm across electrochemical energy system and beyond.

\begin{figure}
    \centering
    \includegraphics[width=1\linewidth]{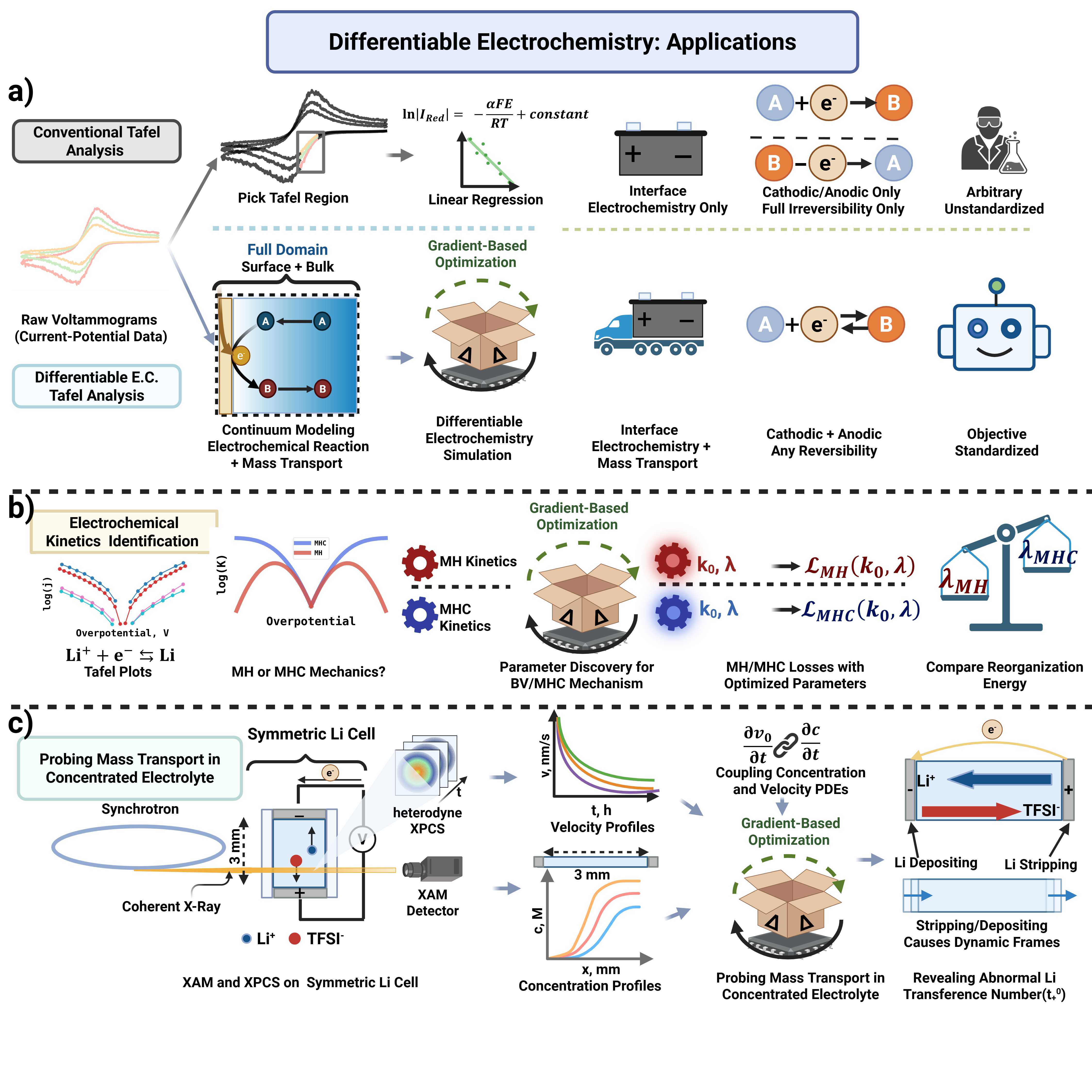}
    \caption{Three applications of Differentiable Electrochemistry. (a) Differentiable Electrochemistry offers a better alternative to traditional Tafel analysis and Nicholson method (not shown in this figure). Compared with conventional Tafel analysis that selects an arbitrary Tafel region, considers only cathodic or anodic reaction, and neglects mass transport, Differentiable Electrochemistry simulation enables a differentiable continuum modeling of the entire reaction domain to consider the entire current-potential data, with full electrokinetics and mass transport. Thus, Differentiable Electrochemistry removes the limitations of conventional Tafel analysis, offering a more accurate, objective, and standardized alternative. (b) Differentiable Electrochemistry enables mechanistic identification of Li metal anode electrodeposition and stripping. Specifically, it allows direct parameterization of Marcus-Hush-Chidsey (MHC) kinetics without closed-form approximations. (c) \emph{Operando} Heterodyne XPCS and XAM on symmetric Li cell generates dynamic  velocity and concentration profiles. Differentiable Electrochemistry couples concentrated transport theory with a hydrodynamic continuity constraint to interpret the two fields to identify and explain abnormal Li transference number $t_+^0$. }
    \label{fig:ApplyingDiffEC}
\end{figure}

In the second perspective, we leverage Differentiable Electrochemistry to overcome long-standing bottlenecks in system identification, and thereby advance mechanistic understanding of electrochemical phenomena. The three applications (see Figure \ref{fig:ApplyingDiffEC}) are: (i) Advancing from conventional Tafel analysis and Nicholson method; (ii) Mechanistic identification of Li metal anode electrodeposition/stripping; (iii) Interpreting \emph{Operando} X-ray data for $\ch{Li+}$ transport in concentrated electrolyte.

The first opportunity lies in advancing beyond  conventional Tafel analysis and Nicholson method, two classical techniques that have been the workhorses of electrochemical kinetics but suffer from fundamental limitations (see Figure \ref{fig:ApplyingDiffEC}a). Conventional Tafel analysis represents a clear example electrochemical research bottlenecks, as it cannot fully account for coupled multiphysics effects such as mass transport and time-dependent interfacial kinetics. Tafel analysis extracts transfer coefficients to elucidate electrocatalytic mechanisms and evaluate kinetics, and the International Union of Pure and Applied Chemistry (IUPAC) definitions of cathodic and anodic Tafel analysis are given in eqs. \ref{eq:TafelCathodic}-\ref{eq:TafelAnodic} :\cite{IUPACTafel1,IUPACTafel2}
\begin{equation}
    \alpha = -\frac{RT}{F}\frac{\mathrm{d} \ln(I)}{\mathrm{d}E}
    \label{eq:TafelCathodic}
\end{equation}
\begin{equation}
    \beta = \frac{RT}{F} \frac{\mathrm{d} \ln(I)}{\mathrm{d}E}
    \label{eq:TafelAnodic}
\end{equation}
where $\alpha$ and $\beta$ are the cathodic and  transfer coefficients, respectively; $E$ and $I$ are overpotential and current. Despite its widespread use, Yang and colleagues recently noted that Tafel analysis is often misapplied and that its limitations are underrecognized.\cite{YangTafel} There are three major limitations of the conventional Tafel analysis:
(i) Tafel analysis is an approximation to Butler-Volmer kinetics, which isolates only cathodic or anodic processes and is applicable only at sufficiently high overpotentials;
(ii) it neglects mass transport effects, and thus fails when transport interferes or dominates at high overpotentials; and
(iii) there is no universal guideline for selecting the portion of current-potential data suitable for analysis, making the procedure highly arbitrary and prone to human error. 
While numerous methods have been proposed to address these issues,\cite{AP1,KristinaTafel,Compton2020Mass,Compton2018Tafel,Kristof2023Tafel} no method to date has fully resolved all three (see Table S 3). The third paradigm simulation approaches such as ref.\cite{Coupling2023Bell}, couples boundary layer simulation with CMA-ES for Tafel analysis, but it is limited to cathodic reaction and relies on approximated mass transport with steady-state boundary layer transport and fitted boundary layer thickness. Furthermore, CMA-ES (a gradient-free optimization method) is applied in boundary layer simulation for parameterization because its COMSOL-MATLAB workflow does not provide analytical gradients, making gradient-based optimization infeasible. Electrochemistry-Informed Neural Network (ECINN) is an example of the fourth paradigm AI/ML models, and is a variation of Physics-Informed Neural Network that embeds electrochemical theories into neural network, offering an alternative to traditional Tafel analysis.\cite{chen2024discovering} However, ECINN is not a simulator, and is not immune to the inherent challenges of neural networks, which are highly vulnerable to spectra bias and highly sensitive to hyperparameters, initial guesses, and training protocols.\cite{chen2022critical} Collectively, the limitations of conventional Tafel analysis, the constraints of third-paradigm simulations, and the inherent drawbacks of AI/ML models underscore the need for a unified, physics-consistent, and gradient-based approach.

Differentiable Electrochemistry simulation is thus introduced to overcome these limitations by directly solving the coupled time-dependent mass transport equation with complete Butler-Volmer kinetics. This enables efficient gradient-based parameter identification without arbitrary data selection, steady-state assumption, or sole reliance on gradient-free methods for systematic prediction of cathodic/anodic transfer coefficients, electrochemical rate constants, and/or diffusion coefficients. The $\ch{Fe^{3+}}/\ch{Fe^{2+}}$ redox couple is the first experimental validation of Differentiable Electrochemistry, where both cathodic and anodic transfer coefficients, electrochemical rate constants, and diffusion coefficients are simultaneously discovered from cyclic voltammograms at different scan rates. Hydrogen evolution reaction (HER) on the rotating Pt electrode is the second experimental validation, where transfer coefficient of HER and the electrochemical rate constants are discovered by differentiating the convection-diffusion equation.

Beyond addressing the limitations of Tafel analysis, Differentiable Electrochemistry also resolves the long-standing challenge of reliably estimating electrochemical rate constants in the quasi-reversible regime. Nicholson method is the predominant technique for determining standard electrochemical rate constants ($k_0$)  of quasi-reversible redox couples for over 50 years.\cite{Nicholson1965,Radhul2025Nicholson} Nicholson method only requires peak-to-peak separation($\Delta E_p$) to derive a dimensionless kinetic parameter $\Psi$, and $k_0$ is extracted as the coefficient of regressing $\Psi$ against $\left(\frac{n\pi DF\nu}{RT}\right)^{-\frac{1}{2}}$, where $n$, $D$, and $\nu$ are the number of electron transferred, the diffusion coefficient, and scan rate. Obtaining the relationship between $\Delta E_p$ and $\Psi$, however, is nontrivial. As curated and evaluated by Agarwal,\cite{Radhul2025Nicholson} there are at least three datasets and twelve empirical equations that try to map $\Delta E_p$ to $\Psi$, and none of them are fully accurate across the entire $\Delta E_p$ range. The assumption that $\Delta E_p$ is independent from transfer coefficient and switching potential, the narrow but different ranges of $\Delta E_p$ validity, the ambiguity among multiple $\Psi-\Delta E_p$ datasets and equations, and the inapplicability to very slow electron-transfer kinetics are some of the major restrictions of Nicholson methods. As pointed by Agarwal, Bond, and numerous other studies,\cite{Radhul2025Nicholson,Bond2014Investigation,Tamas2023Determination} traditional Nicholson method should be discarded as a reliable quantitative kinetic analysis in favor of modern simulation-based methods. Differentiable Electrochemistry simulation is proposed as an ideal solution that relieves all restrictions on Nicholson method and  systematically discovers $k_0$ and $\alpha$ with gradient-based optimization. The $\ch{Fe^{3+}}/\ch{Fe^{2+}}$ redox couple on Pt electrode with quasi-reversible kinetics serves as the experimental example.

The second opportunity lies in the mechanistic identification of Li metal anode electrodeposition/stripping ($\ch{Li+ + e- <=> Li}$). While the Butler-Volmer (BV), Marcus-Hush (MH), and Marcus-Hush-Chidsey (MHC) models are central theories of electrochemical kinetics, determining which framework best captures Li electrodeposition and stripping remains an open challenge (see Figure \ref{fig:ApplyingDiffEC}b). Boyle and Cui collected transient voltammetry data using ultramicroelectrode for $\ch{LiPF6}$ in four solvents, and preferred MH kinetics over BV kinetics to explain the current-potential data.\cite{boyle2020transient} Sripad et al. continued this debate, argued that MH formalism was developed for homogeneous electron transfer and  MH has an inverted region at high overpotential, where the rate of electrochemical reaction decrease with overpotential. Thus, the MH formalism tends to overestimate the reorganization energy and is valid primarily at low overpotentials.\cite{sripad2020kinetics,bai2014charge} Instead, Sripad et al. argued in favor of MHC formalism (eqs. \ref{eq:MHCRates}-\ref{eq:MHCActivation} ) for heterogeneous electron transfer, as it explicitly accounts for Fermi-Dirac distribution of electrons on the metal electrode. The MHC kinetics is:
\begin{equation}
    k_{red/ox}^{MHC} = k_0\frac{\boldsymbol{I}\left(\theta,\Lambda\right)}{\boldsymbol{I}\left(0,\Lambda\right)}
    \label{eq:MHCRates}
\end{equation}
where $\theta$ and $\Lambda$ are the dimensionless overpotential and reorganization energy defined as $\Lambda=\frac{F}{RT}\lambda$ and $\theta=\frac{F}{RT}\left(E-E^0_f\right)$. $\boldsymbol{I}$ is the integration over electronic energy levels ($\epsilon$):
\begin{equation}
    \boldsymbol{I}\left(\theta,\Lambda\right) = \int^\infty_{-\infty} \frac{\exp{\left(-\frac{G^{\ddagger}\left(\theta, \Lambda, \epsilon\right)}{RT}\right)}}{1+\exp{\left(\mp\epsilon\right)}}d\epsilon
    \label{eq:MHCIntegration}
\end{equation}
where $G^{\ddagger}$ is the activation energy of the reduction/oxidation process for each electronic level. The upper sign in $\mp$ represents reduction and the lower sign represents oxidation. $G^{\ddagger}$s is calculated from:
\begin{equation}
    \frac{\Delta G^{\ddagger}}{RT} = \frac{\Lambda}{4}\left(1\pm \frac{\theta+\epsilon}{\Lambda} \right)^2
    \label{eq:MHCActivation}
\end{equation}
Although MHC formalism is widely recognized as the more accurate framework for describing heterogeneous electron transfer kinetics,\cite{Chidsey1991MHC,laborda2013asymmetric,UV4} its practical parameterization remains a challenging due to the need to evaluate the integral (eq. \ref{eq:MHCIntegration}). Consequently, closed-form approximations are frequently employed to circumvent this computational barrier, albeit at the cost of reduced accuracy and introducing additional approximation. For example, the expression proposed by Bazant and colleagues is frequently adopted, \cite{RN13,sripad2020kinetics} though it exhibits noticeable deviation ($5\%-10\%$ depending on the scale of reorganization energy) at low overpotentials. With the emerging fifth paradigm of modeling, Differentiable Electrochemistry simulation offers a transformative alternative. By directly differentiating the full MHC formalism using Gauss-Hermite quadrature (see Simulation Theory section in the Supporting Information), it enables efficient and accurate gradient-based parameterization without approximations. By doing so, Differentiable Electrochemistry transcends traditional trade-offs between physical fidelity and computational efficiency, marking a decisive step towards quantitatively predictive and mechanistically grounded electrochemical science.

The third opportunity arises from the interpretation of \emph{Operando} X-ray measurements of ion transport in concentrated electrolyte, where complex coupling between ion motion, solvent flow, and concentration gradients challenges classical transport theories (see Figure \ref{fig:ApplyingDiffEC}c). In polymer electrolytes such as LiTFSI/PEO, \emph{Operando} X-ray photon correlation spectroscopy (XPCS) and X-ray adsorption microscopy (XAM) have simultaneously revealed solvent velocity and ion concentration fields during electrochemical polarization, offering a rare window into the multiscale mechanisms governing ionic transport under realistic electrochemical conditions.\cite{AP8} A key feature of this system is that the electrolyte itself moves under polarization, causing electrodeposition in the the cathode and stripping in the anode, forming a moving electrolyte frame in which the polymer matrix and solvated ion drifts collectively in response to charge transport. This motion arises from  electro-osmotic and volume conservation effects that coupled the migration of charged species with solvent displacement. Accounting for this moving frame adds a critical layer of complexity, as the governing transport equations must capture not only diffusive and migrative fluxes, but also convective contributions tied to the evolving solvent velocity field.

To interpret such dynamic \emph{Operando} fields, conventional modeling approaches rely on fitting experimental data to solutions of the concentrated-solution theory using heuristic or gradient-free optimization.\cite{AP8} While these analysis have demonstrated that the concentrated-solution framework can predict solvent velocity and concentration gradient with a single set of transport parameters, they remain computationally intensive and non-differentiable. Consequently, a gradient-free optimization framework treats the forward simulation and inverse parameter estimation as separate process, limiting their efficiency and interpretability.

Differentiable Electrochemistry reformulates this problem by embedding the governing mass transport equations into an end-to-end differentiable pipeline. The spatiotemporal evolution of the salt concentration $c(x,t)$ and solvent velocity $v_0(x,t)$ in the moving electrode reference frame is governed by eqs.\ref{eq:concField}-\ref{eq:velField} \cite{AP2}
\begin{equation}
    \frac{\partial c}{\partial t} = \frac{\partial}{\partial x} \left( 
    D \left( 1 - \frac{\mathrm{d} \ln c_0}{\mathrm{d} \ln c} \right) 
    \frac{\partial c}{\partial x} 
    - t_+^0 \frac{i}{F} - c v_0 \right)
    \label{eq:concField}
\end{equation}
\begin{equation}
    \frac{\partial v_0}{\partial x} = \bar{V} \frac{\partial}{\partial x} \left(
    D \left( 1 - \frac{\mathrm{d} \ln c_0}{\mathrm{d} \ln c} \right) 
    \frac{\partial c}{\partial x} 
    - t_+^0 \frac{i}{F} \right)
    \label{eq:velField}
\end{equation}
where $D$ is the salt diffusivity, $t_+^0$ is the cation transference number relative to solvent motion, $c_0$ is the solvent concentration, $i$ is the current density, $\bar{V}$ is the salt partial molar volume. The gradients of the loss function comparing simulated and \emph{Operando} X-ray fields enables direct, efficient optimization without heuristic search. Differentiable Electrochemistry quantitatively connects \emph{Operando} observations with multiphysics transport mechanism, further reveals a negative transference number at high salt concentrations as a possible indicator of strong ion-solvent and ion-ion interactions that conventional continuum models fail to produce.\cite{AP8}

Together, these three applications demonstrate how Differentiable Electrochemistry unifies mechanistic modeling, data interpretation, and gradient-based learning within a single framework as a new paradigm of electrochemical modeling.

\begin{figure}
    \centering
    \includegraphics[width=1\linewidth]{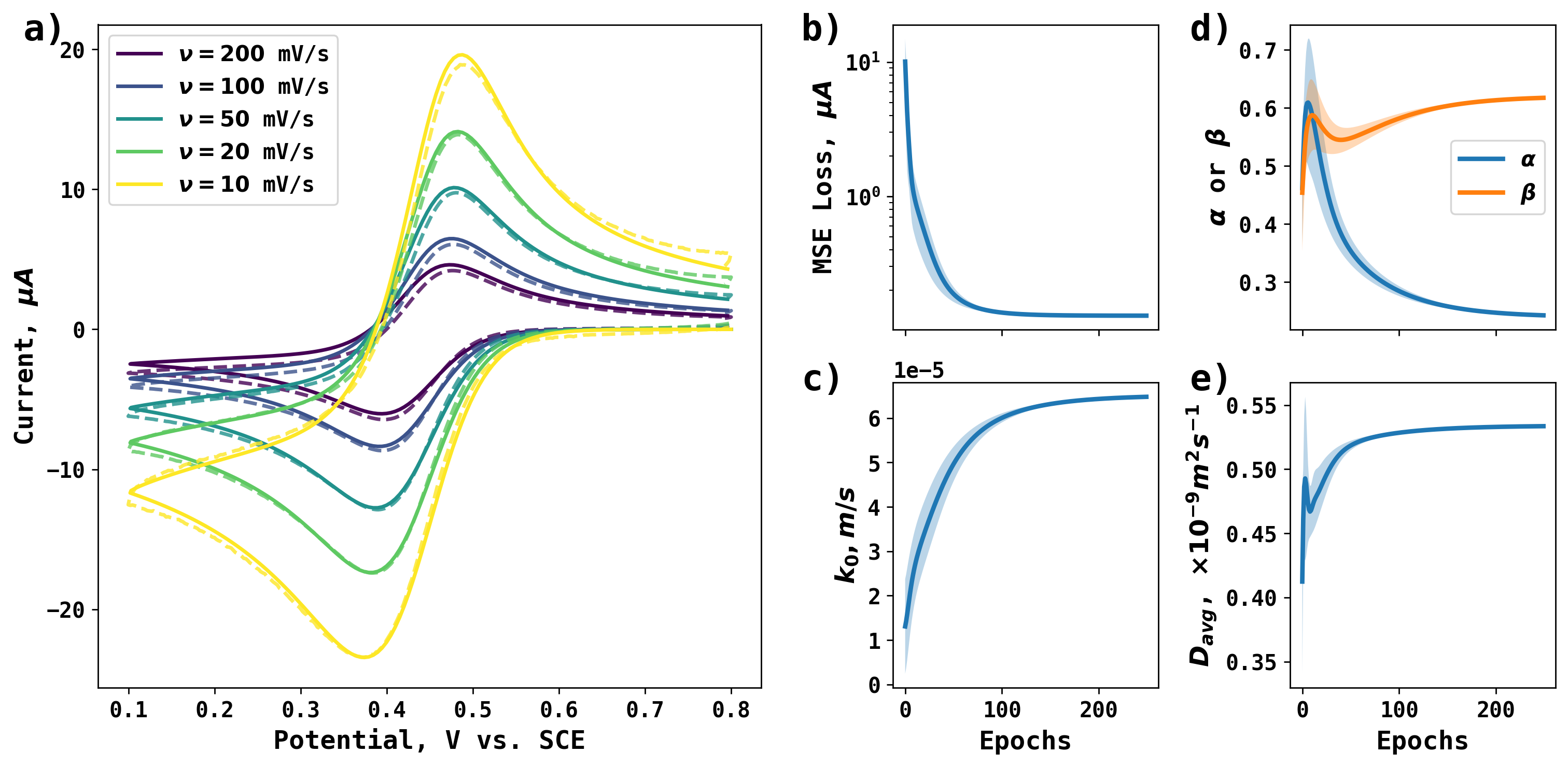}
    \caption{Differentiable Electrochemistry simulation and optimization of $\ch {Fe^{3+}}$/$\ch{Fe^{2+}}$ redox couple as an example to advance Tafel analysis and Nicholson method. a) The experimental voltammograms (dashed lines) are overlaid with simulated voltammograms (solid lines) using parameters estimated by Differentiable Electrochemistry. Subplots b-e are the optimization trajectories of Differentiable Electrochemistry. The mean and standard deviation of 30 runs are shown in solid lines and shadowed regions, respectively. The trajectories of (b) mean squared error of currents, (c) electrochemical rate constant, (d) transfer coefficients, and (e) average diffusion coefficient are shown.}
    \label{fig:DiffECFe}
\end{figure}

\paragraph{Advancing from conventional Tafel analysis and Nicholson method}
Tafel analysis and the Nicholson method are two cornerstones techniques for extracting transfer coefficients and electrochemical rate constants from voltammetric data. Despite their historical importance, both rely on simplifying assumptions and thus face fundamental limitations that restrict their applicability to complex electrochemical systems. Awareness of these shortcomings has grown in recent years, motivating sustained efforts to refine or replace these traditional approaches. \cite{YangTafel,Radhul2025Nicholson} A central challenge underlying both methods is their inability to consistently account for the coupled multiphysics nature of electrochemical systems, where charge transfer, mass transport, and interfacial kinetics are inherently intertwined. To illustrates this approach, we start with the $\ch{Fe^{3+} + e- <=> Fe^{2+}}$ redox couple data adapted from ref.\cite{chen2024discovering} , when both cathodic and anodic transfer coefficients ($\alpha$, $\beta$), electrochemical rate constant ($k_0$) and diffusion coefficient ($D_{avg}$) are simultaneously extracted from cyclic voltammograms at different scan rates.  Differentiable Electrochemistry simulation fully accounts for coupled mass transport and interfacial kinetics, and performs gradient-based optimization by differentiating the error of simulated currents with respect to parameters. Figure \ref{fig:DiffECFe}a shows the experimental voltammograms (dashed lines) of $\ch{Fe^{3+}}$ reduction on a Pt macroelectrode in $1.0\ \mathrm{M}\ \ch{H2SO4}$ at five scan rates from $10$ to $200\ \mathrm{mV/s}$. As shown in Figure \ref{fig:DiffECFe}a, the peak-to-peak separations ($\Delta E_p$) are $\sim0.1\ \mathrm{V}$, which is larger than the $\Delta E_P=57\ \mathrm{mV}$ for a fully reversible couple, suggesting that the full Butler-Volmer kinetics is necessary to account for the quasi-reversible kinetics. Thus, the diffusion governing equations (eq. \ref{eq:FeMassTransport}) for Differentiable Electrochemistry simulations are:
\begin{equation}
\begin{aligned}
    \frac{\partial c_{\ch{Fe^{3+}}}}{\partial t} &= \Theta_{D_{avg}}\frac{\partial^2c_{\ch{Fe^{3+}}}}{\partial x^2} \\
    \frac{\partial c_{\ch{Fe^{2+}}}}{\partial t} &= \Theta_{D_{avg}}\frac{\partial^2c_{\ch{Fe^{2+}}}}{\partial x^2} \\
\end{aligned}
\label{eq:FeMassTransport}
\end{equation}
where $x$ is the distance to the electrode surface. The electrode boundary condition is characterized by Butler-Volmer kinetics (eq. \ref{eq:FeKinetics}):
\begin{equation}
\begin{aligned}
    \Theta_{D_{avg}}\left(\frac{\partial c_{\ch{Fe^{3+}}}}{\partial x}\right)_{x=0} &=  \Theta_{k_0}\exp{\left(-\frac{\Theta_{\alpha}F\left(E-E_{ref}\right)}{RT}\right)} c_{\ch{Fe^{3+}}} - \Theta_{k_0}\exp{\left(\frac{\Theta_\beta F\left(E-E_{ref}\right)}{RT}\right)}c_{\ch{Fe^{2+}}} \\
    \Theta_{D_{avg}}\left(\frac{\partial c_{\ch{Fe^{2+}}}}{\partial x}\right)_{x=0} &= -\Theta_{D_{avg}}\left(\frac{\partial c_{\ch{Fe^{3+}}}}{\partial x}\right)_{x=0}
\end{aligned}
\label{eq:FeKinetics}
\end{equation}
where $\Theta_{\alpha}$, $\Theta_{\beta}$, $\Theta_{k_0}$ and $\Theta_{D_{avg}}$ are the elements of the parameter set $\Theta$ to be discovered. Solving eqs. \ref{eq:FeMassTransport}-\ref{eq:FeKinetics} gives the concentration profile $\left(c_{\ch{Fe3+}}\left(x,t\right)\right)$, the concentration gradient at electrode surface $\left(\frac{\partial c_{\ch{Fe3+}}}{\partial x}\right)_{x=0}$, and ultimately the simulated voltammetric current ($\hat{I}$) as a function of $\Theta$. Differentiable Electrochemistry simulation enables calculation of the algorithmic gradients (eq. \ref{eq:FeGradient}) of the mean squared error (MSE) between experimental currents ($I_{exp}$) and simulated currents with respect to $\Theta$ for gradient-based optimization:
\begin{equation}
    \nabla_{\Theta}\frac{1}{n}\sum_{\nu\in\mathcal{V}}\sum_{i=0}^n\left(I_{exp,i}\left(\nu\right)-\hat{I}_i\left(\Theta,\nu\right)\right)^2
    \label{eq:FeGradient}
\end{equation}
where $\mathcal{V}$ is the set of scan rates and $n$ is the total number of discrete current-potential data points in a single voltammogram. Note that summation over voltammetric data at different scan rates enables simultaneous fitting across multiple experimental conditions, thereby enforcing the robustness and identifiability of the extracted parameters. The gradients of the losses with respect to parameters are the central quantities that guide the optimization process, linking experimental observables to the underlying physicochemical parameters through differentiable simulation.

During Differentiable Electrochemistry simulations, 30 initial guesses of $\Theta$ are randomly sampled within physically reasonable bounds to initialize the optimization and to mitigate sensitivity to initial conditions. In addition, predictions with different initial guesses enable epistemic uncertainty evaluation. Each optimization step involves simulations, loss and gradient evaluation, and gradient-based parameter optimization, and a total of 250 optimization steps are performed with stochastic gradient descent (SGD) optimizer. The Differentiable Electrochemistry optimization trajectories are shown in Figure \ref{fig:DiffECFe}b-e, where the mean trajectories are shown in solid line and the shaded areas represent the range of one standard deviation around the mean. As shown in Figure \ref{fig:DiffECFe}b, the MSE in reproducing voltammetric current is $\sim 0.1\ \mathrm{\mu A}$ after 250 epochs of optimization and satisfactorily converged. Figure \ref{fig:DiffECFe}c-e shows the trajectories, and the converged parameters are: $k_0=(6.54\pm0.02)\times10^{-5}\ \mathrm{m/s}$, $\alpha=0.248\pm0.002$, $\beta=0.612\pm0.002$, and $D_{avg}=(5.33\pm0.01)\times10^{-10}\ \mathrm{m^2s^{-1}}$. Using the estimated parameters, the simulated voltammograms (solid lines) are overlaid with experimental ones and shown in \ref{fig:DiffECFe}a, showing good agreements especially in the cathodic scan region. The estimated $k_0$ in $1\ \mathrm{M}\ \ch{H2SO4}$ on Pt agrees in principle with literature $k_0$ of $3\times10^{-5}\ \mathrm{m/s}$ in $0.5\ \mathrm{M}\ \ch{H2SO4}$.\cite{weber1978effect}  The estimated diffusion coefficient agrees well with literature value of $5.5\times10^{-10}\ \mathrm{m^2s^{-1}}$ for $\ch{Fe^{3+}}$ in $0.5\ \mathrm{M} \ch{H2SO4}$.\cite{angell1972kinetics} The cathodic transfer coefficient obtained via conventional Tafel analysis or its derivatives, however, ranged widely in literature from 0.33,\cite{gil1996diffusion} 0.5,\cite{angell1972kinetics} to 0.62.\cite{wijnen1960square} This broad variability underscores the inherent limitations of traditional extraction methods, which often neglect the coupled influence of mass transport and interfacial kinetics. In contrast, Differentiable Electrochemistry framework self-consistently accounts for these effects by directly optimizing these parameters through physics-informed gradients, ensuring the extracted $\alpha$, $\beta$, and $k_0$ are consistent with both the governing equations and experimental observables. The asymmetry between $\alpha$ and $\beta$ reflects the complex interfacial energetic and possible influence of reorganization effects. Apart from the small epistemic uncertainties estimated via the ensemble method, the robustness of Differentiable Electrochemistry is systematically evaluated in the Differentiable Electrochemistry with Noisy, Sparse, or Partial Data section in the Supporting Information. In particular, Differentiable Electrochemistry has the flexibility to differentiate arbitrarily partial current-potential data to the parameters as shown in Figures S 16-17. The capability to simultaneously resolve voltammograms for kinetic and transport characteristics via differentiable simulation marks a key advancement from traditional Tafel analysis or Nicholson method toward quantitatively interpretable and data-consistent electrochemical modeling.

\begin{figure}
    \centering
    \includegraphics[width=1\linewidth]{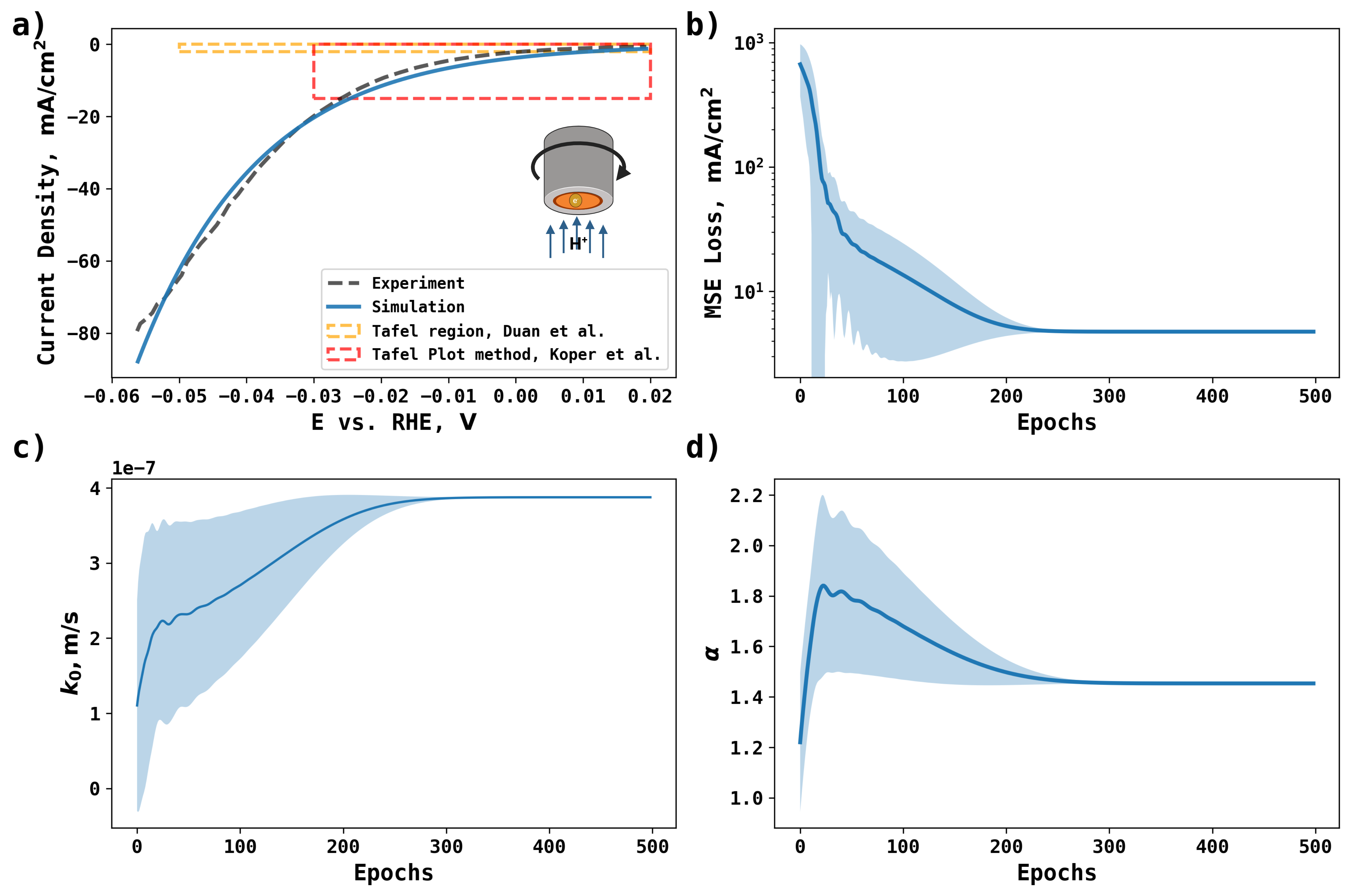}
    \caption{Differentiable Electrochemistry simulation for acidic HER on a rotating disk electrode. (a) The experimental linear sweep voltammogram (black dashed line) of acidic HER on a rotating Pt electrode in $1\ \mathrm{M}\ \ch{HClO4}$ at $2\ \mathrm{mV/s}$ and 2500 rpm reported by Koper et al.\cite{AP1}, and the simulated voltammogram (blue solid line) using parameters provided by Differentiable Electrochemistry. The Tafel region suggested by Duan et al. and the Tafel region used by Koper et al. are shown in orange and red dashed boxes.\cite{wan2024unraveling,AP1} The Differentiable Electrochemistry learning trajectories for (b) MSE of current density, (c) electrochemical rate constant, and (d) cathodic transfer coefficient. The mean trajectories and standard deviations are shown in solid lines and dashed trajectories, respectively.}
    \label{fig:DiffECHER}
\end{figure}

Having established the reliability of Differentiable Electrochemistry in resolving coupled kinetic-transport parameters for the $\ch{Fe^{3+}}$/$\ch{Fe^{2+}}$ redox systems, we next extend the framework to model the hydrogen evolution reaction (HER) on a rotating disk electrode (RDE), where convective mass transport challenges conventional parameter extraction. HER is a fundamental electrochemical process of the cathodic reaction of water electrolysis to produce green hydrogen.\cite{AP14} Understanding the reaction mechanisms, kinetics, and material properties involved in HER enables the design of more efficient and durable systems. The analysis of the transfer coefficient of HER in acidic media remains one of the most actively debated topics in contemporary  electrocatalysis. For example, Duan et al. have noted the Tafel region dependence of Tafel analysis and that limited mass transport suppresses the continued increase of HER current at higher overpotential, leading conventional Tafel analysis to overestimate the Tafel slope and, equivalently, underestimate the transfer coefficient. Therefore, they suggested limiting Tafel analysis to low overpotential ($<50\ \mathrm{mV}$) and current density ($<2\ \mathrm{mA/cm^2}$).\cite{wan2024unraveling} More recently, Koper et al. introduced the Tafel slope plot method, suggesting that a constant Tafel slope region at low current density indicated kinetic meaningfulness.\cite{AP1}  However, these approaches seek to avoid the influence of convective mass transport, whereas a comprehensive understanding requires explicit modeling its coupling with electrode kinetics. To overcome these limitations, we extend the Differentiable Electrochemistry framework to explicitly couple electrokinetics with convection-diffusion mass transport and apply it to analyze the linear sweep voltammogram (LSV) data of acidic HER on a rotating Pt disk electrode ($r=2.5\ \mathrm{mm}$) reported by Koper et al. as shown in Figure \ref{fig:DiffECHER}a. The Tafel region suggested by Duan and Koper are also shown in Figure \ref{fig:DiffECHER}A as orange and red dashed boxes, respectively.\cite{wan2024unraveling,AP1} The electrokinetics is shown in eq. \ref{eq:HERKinetics}
\begin{equation}
    \left(\frac{\partial c_{\ch{H+}}}{\partial y}\right)_{y=0}= \Theta_{k_0}\exp{\left(-\frac{\Theta_{\alpha}F\left(E-E_{ref}\right)}{RT}\right)}c_{\ch{H+}}
    \label{eq:HERKinetics}
\end{equation}
where $y$ is the distance to the electrode surface and the reference potential is $E_{ref}= 0\ \mathrm{V\ vs.\ RHE}$. $\Theta_{k_0}$ and $\Theta_{\alpha}$ are the electrochemical rate constant and transfer coefficient for parameter discovery.  In this analysis, only the cathodic branch of the HER is considered, consistent with the experimental data reported by Koper et al., which exclusively captures the cathodic responses. This focus does not reflect a limitation of the Differentiable Electrochemistry framework which can accommodate both anodic and cathodic reaction steps. The mass transport equation is described by one-dimensional convection-diffusion equation (eq. \ref{eq:HERConvectionDiffusion}), as the electrode dimension is on the millimeter scale:
\begin{equation}
    \begin{aligned}
        \frac{\partial c_{\ch{H+}}}{\partial t} &= D_{\ch{H+}}\frac{\partial^2c_{\ch{H^+}}}{\partial y^2} - v_y\frac{\partial c_{\ch{H^+}}}{\partial y}  \\ 
        v_y &=  -Ly^2, L=0.51023(2\pi f)^{3/2} \nu^{-1/2}
    \end{aligned}
\label{eq:HERConvectionDiffusion}
\end{equation}
where $L$ is the hydrodynamic constant derived by Levich.\cite{Levich_1963}. $f$ is the rotational frequency in Hertz, and $\nu$ is the kinematic viscosity. Details of simulation, including dimensionless equations, discretization and simulation parameters are shown in Differentiable Electrochemistry for Hydrogen Evolution Reaction section in the Supporting Information. Differentiable Electrochemistry simulations are initialized with 30 randomly sampled $\Theta$ values from reasonable bounds and Adam optimizers with a learning rate of $5\times10^{-2}$ for 500 epochs. Note that the entire LSV as shown in Figure \ref{fig:DiffECHER}a is solved and no Tafel region is required. The optimization trajectories are shown in Figure \ref{fig:DiffECHER}b-d with the mean and one standard deviation. The MSE of current densities decreases from $10^{-1}\ \mathrm{mA/cm^2}$ to $10^{-3}\ \mathrm{mA/cm^2}$ at the end of optimization. The discovered parameters are $\alpha=1.45\pm0.01$ (or $40.7\ \mathrm{mV/dec}$) and $k_0=\left(3.87\pm 0.01\right)\times10^{-7}\ \mathrm{m/s}$. The extracted $\alpha$ agrees well with the derivation and literature value of $\alpha=1.5$ (or $40\ \mathrm{mV/dec}$) for acidic HER on Pt electrode, which follows a Volmer-Heyrovsky mechanism.\cite{UV4} The simulated LSV is overlaid and agrees well with the entire experimental LSV as shown in Figure \ref{fig:DiffECHER}a and the MSE loss $4.75\ \mathrm{mA/cm^2}$. The slightly larger deviation at higher overpotentials likely arises from increased noise and non-ideal mass transport behavior of the experimental ground truth LSV in this regime.

Compared with conventional Tafel analysis or Tafel-Slope analysis methods, Differentiable Electrochemistry resolves two key bottlenecks in electrokinetic characterization: (1) it explicitly incorporates convective mass transport into the modeling framework and steady-state assumption is unnecessary; and (2), it eliminates the ambiguity associated with selecting a Tafel region by leveraging the full LSV to objectively extract intrinsic kinetic parameters. This generalized approach provides a rigorous and extensible paradigm for mechanistic identification in other electrochemical systems governed by coupled kinetic and transport processes.

\begin{figure}
    \centering
    \includegraphics[width=1\linewidth]{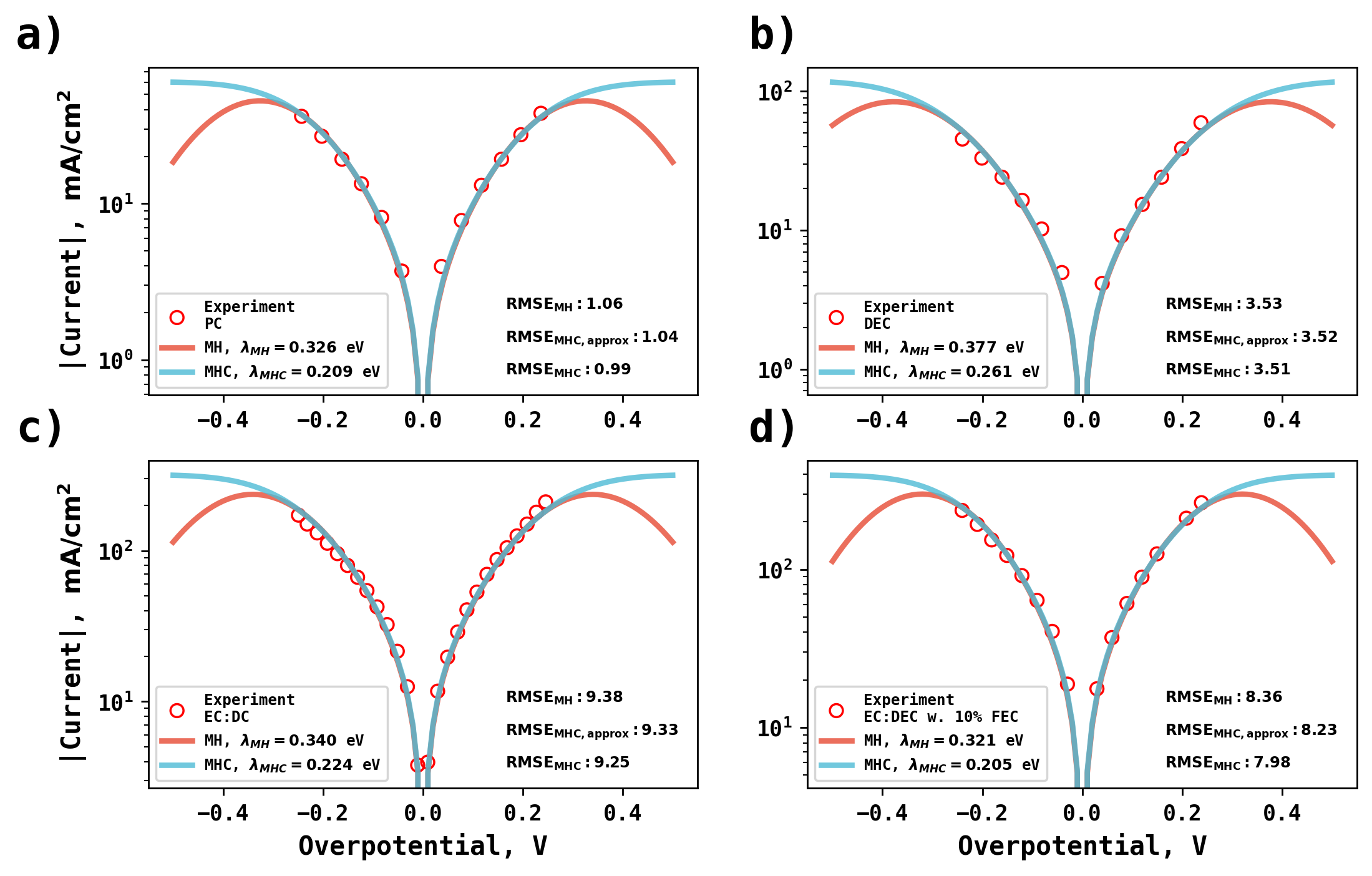}
    \caption{The predictions of experimental current densities (red scatters) reported by Boyel et al.\cite{boyle2020transient} using Differentiable Electrochemistry with MH (red lines) and MHC formalisms (blue lines) for four different solvents (a) PC, (b) DEC, (c) EC:DEC, (d) EC:DC w. 10\% FEC. The reorganization energies of MH ($\lambda_{MH}$) and MHC ($\lambda_{MHC}$) formalisms are reported. The root mean square errors of MH ($\mathrm{RMSE_{MH}}$), closed form approximation of MHC ($\mathrm{RMSE_{MHC,approx}}$), and MHC ($\mathrm{RMSE_{MHC}}$) are also shown in each subplot.}
    \label{fig:DiffECLiMech}
\end{figure}

\paragraph{Mechanistic identification of Li metal anode electrodeposition/stripping}Lithium metal electrode is a promising strategy to improve the energy density of rechargeable battery to meet specific energy requirement of electrifying ground transportation and aviation. \cite{sripad2017performance, bills2020performance} Understanding and modeling the kinetics of lithium metal electrode during electrodeposition and stripping are essential to predict battery behavior, and the kinetic model that best describe these processes is actively debated between Marcus-Hush (MH) and Marcus-Hush-Chidsey (MHC) theories.\cite{boyle2020transient,sripad2020kinetics} While Boyel et al. preferred MH formalism over Butler-Volmer kinetics, Sripad et al. analyzed the same data and argued for the preference of MHC formalism over MH formalism, as the latter was not developed for heterogeneous electron transfer, and inclined to overestimate reorganization energy.\cite{sripad2020kinetics} However, parameterizing MHC formalism was very challenging due to its integral (eq. \ref{eq:MHCIntegration}) so that Sripad et al. relied on an closed form approximation. \cite{RN13} Differentiable Electrochemistry, on the contrary, enables gradient-based parameterization of the full MHC formalism and is applied to the transient voltammetry data collected by Boyel et al.\cite{boyle2020transient} for $\ch{LiPF6}$ in four solvents: (i) propylene carbonate (PC), (ii) diethyl carbonate (DEC), (iii), 1:1 by volume ethylene carbonate: diethyl carbonate (EC:DEC), and (iv) EC:DEC with 10\% fluoroethylene carbonate (EC:DEC w. 10\% FEC), as shown in Figure \ref{fig:DiffECLiMech} as scatter points. Differentiable Electrochemistry optimizes reorganization energy ($\lambda_{MHC}$) and exchange current density ($j_{0,MHC}$) by optimizing the mean square error of fitted current density ($\mathrm{MSE_{MHC}}$). In practice, Differentiable Electrochemistry simulations are initialized with 30 different sets of reorganization energies and exchange current densities with MH, MHC, and the closed form MHC approximations (denoted as $\mathrm{MHC, approx}$), and trained for 5000 epochs with Adam optimizer ($lr=10^{-2}$). The optimization trajectories and equations are shown in Differentiable Li Kinetics section in the Supporting Information. The fitted current densities using MH and MHC formalism are shown in Figure \ref{fig:DiffECLiMech}, and agree qualitatively well with experiments at low overpotentials. The fitted current densities at high overpotentials ($\eta>0.4\ \mathrm{V}$) and the reorganization energies, however, are different between MH and MHC kinetics. In specific, MH formalism has an inverted region at high overpotentials, limiting its applicability at high overpotential and leading to $\lambda_{MH}$ exceeding $\lambda_{MHC}$ by $\sim 0.12\ \mathrm{eV}$. The root mean square error (RMSE) of MHC formalism  $\mathrm{RMSE_{MHC}}$ are benchmarked against $\mathrm{RMSE_{MH}}$ and $\mathrm{RMSE_{MHC,approx}}$ (see Figure \ref{fig:DiffECLiMech}), and MHC formalism outperforms the MH and approximated MHC models in all four solvents by having the lowest RMSE. The outperformance suggests that the full MHC formalism captures the interplay between solvent reorganization and charge-transfer energetics more faithfully than MH or approximate MHC models. By directly optimizing the full MHC integrals, Differentiable Electrochemistry circumvents the need for approximations, enabling rigorous, data-driven extraction of mechanistic parameters. This study establishes a generalizable workflow for mechanistic identification of heterogeneous electrochemical reactions, starting from experimental transient data, followed by differentiable simulation, and concluding with model discrimination and parameter inferences, showing great promise of extending the approach to more complex electrochemical systems.

\begin{figure}
    \centering
    \includegraphics[width=1.0\textwidth]{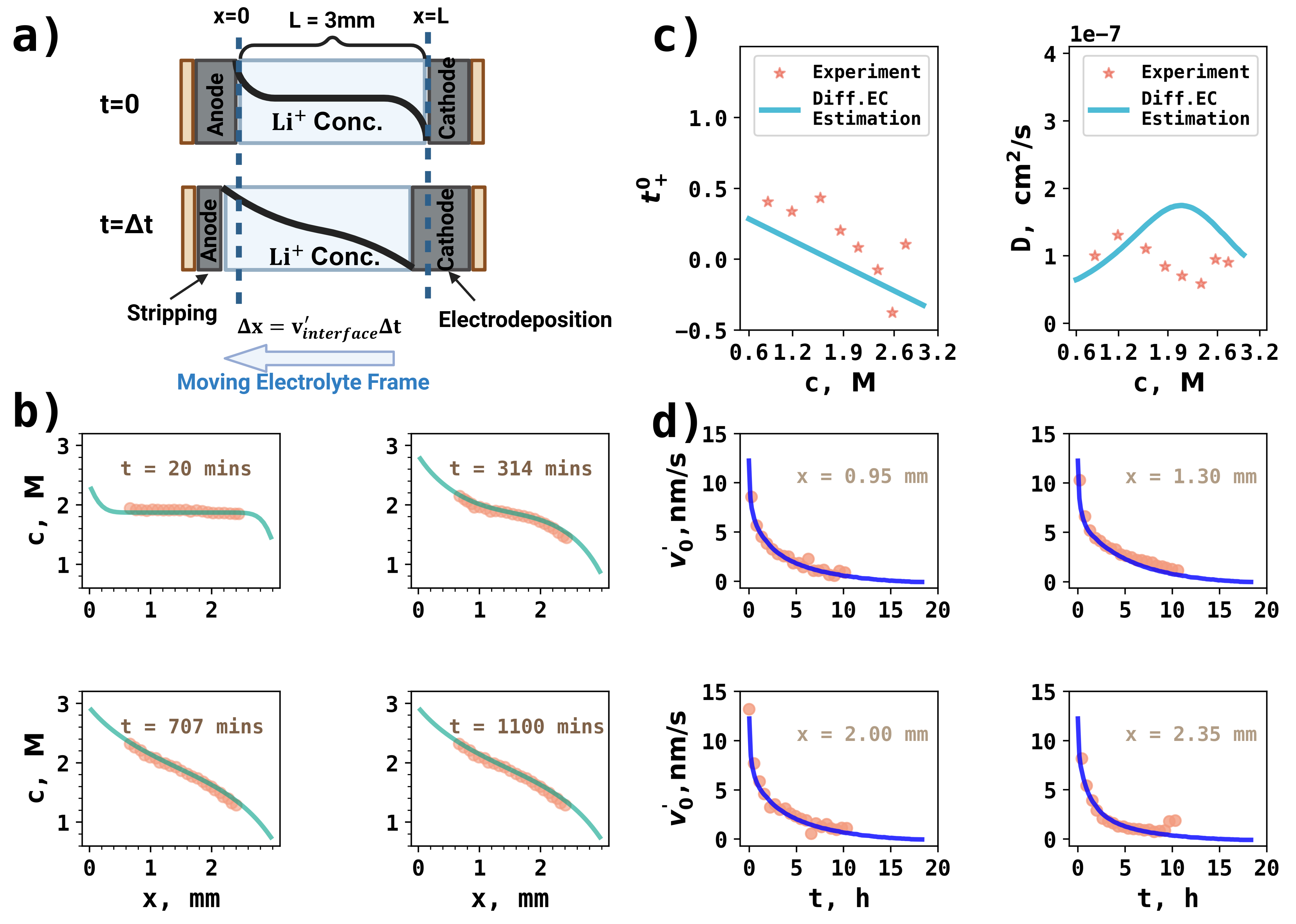}
    \caption{Energy applications of differentiable electrochemistry to mass transport in Li-cell e-h. (a) A scheme of Li symmetric cell during potentiostatic polarization. $v_{\mathrm{interface}}^{\prime}$ is the velocity of Li/electrolyte interface in the stationary laboratory frame, due to the Li electrodeposition/stripping.  (b) XAM-measured concentration (orange scatters) at four time points and simulated concentration profile using (c) transference number and (d) diffusivity estimated using differentiable simulation (blue). The orange scatters are transference number and diffusivity from literature.\cite{AP5}  (d) Comparing predicted solvent velocities (blue lines) against XPCS measurements  (red scatters) at different locations.}
    \label{fig:LiSymmetric}
\end{figure}

\paragraph{Interpreting \emph{Operando} X-ray data for $\ch{Li+}$ transport in concentrated electrolyte}
Lithium batteries are known for their high energy density, high coulombic efficiency and wide applications in energy systems,\cite{AP15} and understanding $\mathrm{Li}^+$ mass transport is highly relevant for extreme fast charging/discharging,\cite{AP16,AP25,Mistry2020XFC} low temperature operations,\cite{AP17} and mitigating dendrite formation.\cite{AP18,Mistry2018ConfinedElectrolyte} Taking a step further, Differentiable Electrochemistry simulation seamless integrates in battery research to accurately understand electrolyte transport and to facilitate rational exploration of electrochemical system.\cite{AP6} Figure \ref{fig:LiSymmetric}a shows a schematic of a Li symmetric cell with LiTFSI in poly (ethylene oxide) electrolyte under potentiostatic polarization. Lithium deposits and strips on the cathode and anode, respectively, causing a concentration gradient of $\ch{Li+}$ in the electrolyte. With recent advances in \emph{Operando} techniques,\cite{AP7,AP8} the \emph{Operando} salt concentrations and solvent velocities are measured during the same experiment using X-ray absorption microscopy (XAM) and X-ray photo correlation spectroscopy (XPCS) and the salt concentrations at four time points are shown as orange scatters in Figure \ref{fig:LiSymmetric}b. Under the constraint Newman's concentrated solution theory (eq. \ref{eq:concField}), differentiable simulation determines the best diffusivity and transference number by comparing the \emph{Operando} concentration fields with the simulated ones to estimate the underlying mass transport properties. Subsequently, the solvent velocity field (eq. \ref{eq:velField}) is predicted and compared against the XPCS measurements to justify the predictiveness of the ion transport theory. As shown in Figure S 21 in the section Estimating Diffusivity and Transference Number from \emph{Operando} Fields of Supporting Information, the loss to predict concentration fields converged in 12 iterations to give a simulated concentration profile that agreed well with the XAM measured concentration profiles as shown in \ref{fig:LiSymmetric}c and the XPCS measured velocity profile as shown in \ref{fig:LiSymmetric}d. 

The estimated concentration-dependent cation transference number and salt diffusivity shown in Figure \ref{fig:LiSymmetric}c faithfully capture the spatiotemporal evolution of the salt concentration field. A noteworthy aspect of this analysis is that unlike the traditional electrochemical property measurements\cite{AP5} require at least one electrolyte sample for every concentration of interest, the differentiable electrochemistry approach requires \emph{Operando} data for only one electrolyte and simultaneously estimates the concentration-dependent mass transport properties. As shown in \ref{fig:LiSymmetric}c, the predicted transference number ranges from $\sim$ 0.3 at a low salt concentration (0.6 M) to $\sim -$0.3 at a high salt concentration (3.0 M). While historically the negative transference number has been rationalized by arguing the presence of negatively charged clusters,\cite{AP26,AP27} recent studies have put forth an alternate explanation. Macroscopically, the negatively charged $\mathrm{Li}^+$ were invoked to explain the cation motion opposite to the current flow (in the absence of concentration gradients). However, this argument overlooked the contributions from the solvent motion (the Nernst Planck view predicts no solvent for closed systems like the lithium lithium symmetric cells). Once the physics of solvent motion is accounted for in electrolyte transport \cite{Mistry2022v0effect}, the negative cation transference number does not necessarily lead to cation motion toward the positive electrode. This continuum-scale finding prompts us to revise the molecular picture of ion transport, and a subsequent study\cite{Mistry2023MolecularMotions} shows that thermodynamically-consistent negative cation transference number can even originate from uncorrelated motions, and not just due to correlated molecular motions like ion clusters.

Obtaining deeper insights does not necessarily come at higher computational costs. In the context of interpreting \emph{Operando} X-ray data, Differentiable Electrochemistry simulations are benchmarked with four advanced gradient-free optimization methods including particle swarm optimization (PSO), Bayesian optimization (BO), covariance matrix adaptation evolution strategy (CMA-ES), and Nelder-Mead method (NM) in terms of wall time and the number of function evaluations. As shown in Figure S 25, Differentiable Electrochemistry outperform PSO and BO by approximately two orders of magnitude, and CMA-ES and NM by about one order of magnitude.

Looking back, scientific and electrochemical modeling has evolved through four major paradigms from empirical correlation, analytical fitting, numerical simulation, to data-driven machine learning. Each stage advanced the precision and scope of modeling, yet none fully reconciled the interpretability of coupled physical theories with the experimental observations. Differentiable Electrochemistry emerges as the fifth paradigm, bridging the experimental observations by embedding automatic differentiation within the governing transport-kinetic equations, enabling end-to-end gradient-based optimization that preserves physical consistency while learning from data. This work contributes along two complementary perspectives. First, we developed and open-sourced five differentiable simulators that span diverse electrochemical mechanisms, transport regimes, and kinetic models, and provided case studies concerning highly nonlinear problems like electrochemical migrations and adsorption/desorption. This open framework provides an accessible foundation for researchers to readily implement and extend Differentiable Electrochemistry in their own investigations.

More importantly, making simulations differentiable addresses long-standing bottlenecks in electrochemical energy research, particularly when multiple physical processes are intertwined.  First, Differentiable Electrochemistry simulation is proposed to advance from the traditional Tafel analysis and Nicholson method by simultaneously extract transfer coefficients and electrochemical rate constants by integrating electrokinetics and mass transport for gradient-based optimization. Using the $\ch{Fe^{3+}}$/$\ch{Fe^{2+}}$ redox couple as an example, the framework embeds Butler-Volmer kinetics with diffusion and accurately resolves both cathodic and anodic transfer coefficients, electrochemical rate constant, and diffusion coefficient across multiple scan rates. Extending this approach to HER on a rotating Pt disk electrode, Differentiable Electrochemistry explicitly incorporates convective mass transport and extracts kinetic parameters from the entire voltammogram and identifies Volmer-Heyrovsky mechanism for HER from the extracted $\alpha=1.46$. By obviating the need to select a Tafel region for Tafel analysis or $\Psi$ for Nicholson method and removing their approximations, Differentiable Electrochemistry rigorously and consistently identifies mechanistic parameters in electrochemical systems and addresses the long-standing gaps in electrochemical characterization. In the context of lithium metal electrodeposition and stripping, Differentiable Electrochemistry directly optimizes the full MHC formalism, accurately capturing heterogeneous electron kinetics in a wider potential window and outperforming MH and approximate MHC models. Lastly, differentiable simulation of \emph{Operando} X-ray data enables direct, quantitative inference of concentration-dependent $\ch{Li+}$ transport properties in concentrated electrolyte, capturing both diffusivity and cation transference number from a single experiment. Differentiable Electrochemistry simulation reproduces the spatiotemporal evolution of salt concentrations and velocity fields measured by XAM and XPCS, respectively, demonstrating excellent agreement with experimental observations. Importantly, this approach solves the ambiguity of the negative transference numbers by explicitly accounting for solvent motion and continuum-scale transport instead of attributing to ion clusters. These results illustrate how integrating both concentration and velocity field with \emph{Operando} measurements enhances the understanding and prediction of ion transport in battery electrolytes, facilitating rational electrolyte design for high-rate, low temperature, and dendrite-resistant operations.

Looking ahead, Differentiable Electrochemistry simulation represents a paradigm shift in how electrochemical systems are modeled and interpreted. By unifying physical laws with differentiable computation, it overshadows the empirical fitting,surrogate modeling and gradient-free optimization that dominate current practice, and provides a viable alternative for parameterizing and interpreting the more complex modern electrochemical systems. We believe that Differentiable Electrochemistry builds a new information highway between experiments and electrochemical theories, constituting a regime switch for uncovering ambiguous physical mechanisms and accelerating the discovery of next-generation electrochemical systems.

\section{Computational Methods}

The differentiable simulation program is implemented using JAX in Python.\cite{RN25} Just-in-time (JIT) complication and vectorization are adopted to speed up computation. All PDEs and variables are converted into dimensionless forms and solved in dimensionless systems. The non-linear mass transport equations for voltammetry are solved using an implicit finite difference method by discretizing the equations using an expanding spatial grid and uniform temporal grid. For chronoamperometry simulations, the temporal grid is also expanding. The highly non-linear PDEs including the PNP equations, are solved using Newton-Raphson method for at most 10 iterations with absolute tolerance of $10^{-12}$. The mass transport in the lithium symmetric cell is solved using finite volume method.  Dimensionless parameters, equations, and discretization schemes are reported in the Discretization Scheme section of Supporting Information. The convergence of the simulations was verified and reported in the Test and Validation section in the Supporting Information. A verifiability checklist is provided in the Supporting Information.\cite{Mistry2021Verifiability}  Figures are plotted with Matplotlib and BioRender(https://BioRender.com). All simulation programs and data are available at \url{https://github.com/BattModels/DiffEC}.

\begin{acknowledgement}
H.C. thanks the support of Schmidt AI in Science Fellowship Program from Schmidt Sciences, LLC. This work used Bridges-2 at Pittsburgh Supercomputing Center through allocation CTS180061 from the Advanced Cyberinfrastructure Coordination Ecosystem: Services \& Support (ACCESS) program. 
\end{acknowledgement}

\begin{suppinfo}
The Supporting Information is available. 

\begin{itemize}
  \item Introduction to Differentiable Simulation for Electrochemistry, A survey of improved Tafel analysis method, Proof-of-Concept Case Studies, Simulation Theories, Discretization Schemes, Differentiable Electrochemistry for Voltammetry of Adsorbed Species, Differentiable Electrochemistry for $\ch{Fe^{3+}}/\ch{Fe^{2+}}$ Redox Couple, Differentiable Electrochemistry with Noisy, Sparse, or Partial Data, Differentiable Electrochemistry for $\ch{Ru(NH3)_6^{3+}}$/$\ch{Ru(NH3)_6^{2+}}$ redox couple, Differentiable Li Kinetics, Differentiable Electrochemistry for Hydrogen Evolution Reaction, Estimating Diffusivity and Transference Number from Operando Fields, Test and Validation, Computational Methods, Verifiability Checklist. 
\end{itemize}

\end{suppinfo}

\bibliography{Library}

\end{document}

\tableofcontents


\clearpage
\section{Introduction to Differentiable Simulation for Electrochemistry}
A differentiable simulator is a computational model that represents a physical system within a differentiable programming framework, such that every operation in the simulation pipeline, from solving governing equations to evaluating observables, is continuously differentiable with respect to its parameters and initial conditions. Given an electrochemical system governed by: 
\begin{equation} 
\frac{\partial u }{\partial t}=\mathcal{F}(u, \nabla u, \nabla^2 u, \dots; \Theta)
\end{equation}
the simulator numerically solves for the system state $u(x,t)$ under specific boundary and initial conditions. A differentiable simulator ensures that the entire solution map:
\begin{equation}
    u=\mathcal{S}\left(\Theta\right)
\end{equation}
is differentiable with respect to model parameters $\Theta$, allowing efficient computation of parameter gradients:
\begin{equation}
\nabla_{\Theta}\mathcal{L}\left(u\left(\Theta\right),u_{exp}\right)
\end{equation}
where $\mathcal{L}$ quantifies the discrepancy between simulated and experimental observables. 

\begin{figure}
    \centering
    \includegraphics[width=0.8\linewidth]{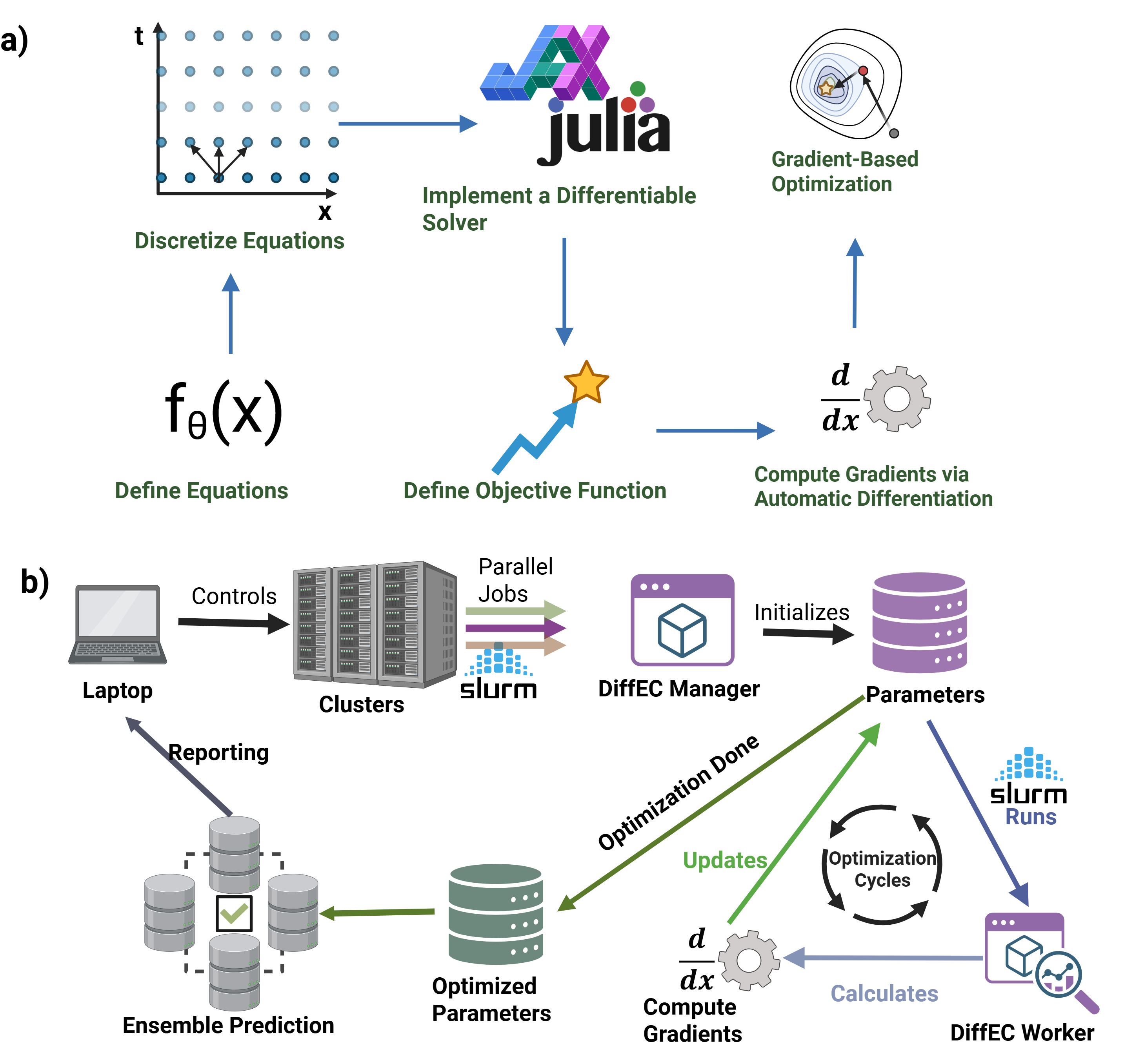}
    \caption{A schematic workflow of (a) formulating a Differentiable Electrochemistry simulator and (b) running Differentiable Electrochemistry simulations on a supercomputer with manager-worker topology. }
    \label{fig:DiffECWorkFlow}
\end{figure}

Differentiable simulation leverages Automatic Differentiation (AD) to compute exact derivatives of model outputs with respect to input parameters through the computational graph of a numerical solver. AD applies the chain rule systematically across all operations during simulation. This yields machine-precision derivatives with computational cost compared to a few forward evaluations of the original model. JAX provides exact derivatives of any function expressed as a composition of differentiable primitives. If the simulator is implemented purely with JAX's functional primitives,
\begin{equation}
    \mathcal{S}\left(\Theta\right) = s_k\circ s_{k-1}\circ \cdots \circ s_1\left(\Theta\right)
\end{equation}
where $s_1,s_2,\cdots,f_k$ are the individual computational/timesteps of the simulator. JAX's jax.grad or jax.jacobian function traverses this computation graph to evaluate the derivative efficiently and without numerical differencing. A differentiable simulator requires that all components of the numerical model are implemented in a way that supports gradient computation with respect to parameters $\Theta$. This includes spatial discretization (finite differences and finite elements) and time integration (forward and backward Euler) must use differentiable linear or nonlinear solver. Boundary and initial conditions must also depend on $\Theta$ when parameters influence them. Kinetic laws  including Butler-Volmer, Marcus-Hush-Chidsey, must be coded in a differentiable way. Lastly, sufficient memory must be allocated to store intermediate states and computational graphs required by AD, to ensure accurate and efficient gradient computation. The Differentiable Electrochemistry simulation workflow consists primarily six steps: definition of model, discretization of model, implementation of a differentiable solver, defining the loss function (possibly by comparing with experimental ground truth or other desired values), gradient calculation, and gradient-based optimization, as shown in Figure S \ref{fig:DiffECWorkFlow}a. The parallel computation workflow is shown in Figure S \ref{fig:DiffECWorkFlow}b, featuring a manager-worker computational workflow for efficient utilization of supercomputer facility. The manager starts the workflow with different initial guesses, and runs a sequence of workers for every step of optimization. The manager uses a small amount of computational resources but runs for hours or days. The worker uses a large amount of computational resources but runs only for minutes.

Given that differentiable simulation is a new regime for electrochemical modeling as suggested by Kitchin et al.,\cite{kitchin2025beyond} the authors have prepared five general purpose simulators (Table S \ref{tab:DiffECSimulators})for the readers to show efficient implementation of Differentiable Electrochemistry simulations. Readers are encouraged to inspect the JAX implementation of spatial grid construction, time stepping with implicit Euler method, and construction of coefficient matrices. 

Table S \ref{tab:DiffECSimulators} lists Differentiable Electrochemistry simulators that are available to the users in the project GitHub repository \url{https://github.com/BattModels/DiffEC}. These differentiable simulators are modular and general purpose, and can be easily modified for more applications. 
 
Table S \ref{tab:ConventionalSimulators} also reviews common electrochemical simulators (commercial or free), to see if they are differentiable and how they perform parameter estimation. It is evident that major electrochemical simulators are not differentiable, and they use either gradient-free optimization or parameter sweep for parameter discovery.

\begin{table}
    \fontsize{8pt}{8pt}\selectfont
    \centering
    \begin{tabular}{| p{0.15\linewidth} | p{0.15\linewidth} | p{0.15\linewidth} | p{0.15\linewidth} | p{0.15\linewidth} | p{0.15\linewidth}|}
    \toprule
         Name& Mechanism & Mass Transport & Electrokinetics, Electro-thermodynamics & Electrode Geometry & Differentiable w.r.t.\\ \midrule
         Fundamental Cyclic Voltammetry& $\ch{A + e^- <=> B}$ & Linear/radial Diffusion & Nernst, BV & Macroelectrode, (hemi-)sphere, and (hemi-)cylinder&$k_0$, $\alpha$, $\beta$, $D_A$, $D_B$, $E^0_f$  \\ \midrule
         Voltammetry in weakly-supported media& $
         \ch{A + e^- <=> B}$, Supporting electrolyte $M$ and $X$ are considered. Electric field $\phi$ is considered. & Migration-Diffusion (Nernst-Planck-Poisson equation)& Nernst, BV, MHC & Macroelectrode & $k_0$, $\alpha$, $\beta$, $D_A$, $D_B$, $D_M$, $D_X$, $E^0_f$, $\lambda$\\ \midrule
         Dissociative EC& $\ch{A <=> [k_f][k_b]B + C}$ $\ch{B + e^- <=> D}$ & Radial diffusion & BV  & (Hemi-)sphere, and (hemi-)cylinder & $k_0$, $k_f$, $k_b$, $\alpha$, $\beta$, $D_A$, $D_B$, $D_C$, $D_D$, $E^0_f$  \\  \midrule
         Hydrodynamic Voltammetry & $\ch{A + e^- <=> B}$ & Convention-Diffusion & Nernst, BV & Rotating disk & $k_0$, $\alpha$, $\beta$, kinematic viscosity, $D_A$, $D_B$, $E^0_f$\\ \midrule
         Electrochemical Adsorption/Desorption & See Figure S \ref{fig:AdsorptionMechanism}, Langmuir isotherm for adsorption modeling & Diffusion & BV & Macroelectrode & $k_0$, $\alpha$, $\beta$, $k_{0,ads}$, $\alpha_{ads}$, $\beta_{ads}$, $K_{A, ads}$, $K_{B,ads}$, $K_{A,des}$, $K_{B,des}$, $D_A$, $D_B$, $E^0_f$, $\lambda$ \\ \midrule

        \multicolumn{6}{l}{BV: Butler-Volmer; MHC: Marcus-Hush-Chidsey} \\ 
        \multicolumn{6}{l}{$k_0$: Standard electrochemical rate constant; $\alpha$ and $\beta$: Cathodic and anodic transfer coefficient;  } \\
        \multicolumn{6}{l} {$D_{{A,B,C,D}}$: Diffusion coefficient of species $A,B,C,D$;} \\
        \multicolumn{6}{l}{$K_{0,ads}$, $\alpha_{ads}$, $\beta_{ads}$: Electrochemical rate constant, cathodic, and anodic transfer coefficients of adsorbed species;} \\
        \multicolumn{6}{l}{$K_{A, ads}$, $K_{B,ads}$: Adsorption rate constant of A and B, respectively;} \\
        \multicolumn{6}{l}{$K_{A, des}$, $K_{B,des}$: Desorption rate constant of A and B, respectively;} \\
        \multicolumn{6}{l}{$E^0_f$: Formal potential; $\lambda$: Reorganization energy (MHC mechanism) }  \\ 
    \end{tabular}
    \caption{General-purpose Differentiable Electrochemistry simulators provided with this paper to maximize readers' utility.}
    \label{tab:DiffECSimulators}
\end{table}

\begin{table}
    \fontsize{9pt}{9pt}\selectfont
    \centering
    \begin{adjustbox}{width=\textwidth}
    
    \begin{tabular}{|p{0.15\linewidth} | p{0.12\linewidth} | p{0.15\linewidth} | p{0.20\linewidth} | p{0.2\linewidth} |}
        \toprule
        Simulator Name &Differentiable &Availability& Expertise & Parameter Estimation \\ \midrule
        Monash Electrochemistry Simulator (MECSim) & N & Free, open-source\cite{MEMSIM1} & AC voltammetry & Random parameter sweep and Bayesian regression* \\  \midrule
        DigiElch & N & Proprietary, distributed by Gamry Instrument\cite{DigiElch} (Parameter estimation is only available in the professional distribution.) & AC/DC Voltammetry, chronoamperometry, and Electrochemical Impedance Spectroscopy (EIS) & Nonlinear regression (Gauss-Newton)/Brute force parameter sweep \\ \midrule
        KISSA & N & Distributed by ProSense and BASi or by contacting the team\cite{KISSA}& Voltammetry and chronoamperometry &Not supported\\ \midrule
        ElectroKitty& N & Free, open-source\cite{ElectroKitty} & Electrochemical Process with Non-Langmuir Adsorption & CMA-ES (Covariance Matrix Adaptation Evolution Strategy) from pycma package\\ \midrule
        FreeSim & N & Free, open-source\cite{FreeSim} & Voltammetry and Chronoamperometry &Not supported\\ \midrule
        DigiSim& N & Distributed by BASi\cite{DigiSim} &Voltammetry& Not supported \\ \midrule
        DiffEC (Differentiable Electrochemistry), this work & Y & Free, open-source \url{https://github.com/BattModels/DiffEC} & Voltammetry, supporting diffusion, migration and convection mass transport; Nernst, Butler-Volmer, Marcus-Hush-Chidsey kinetics & Native support from differentiable simulation, allowing any gradient-based optimization. \\ \midrule
        \multicolumn{5}{l}{*From MECSim Tutorial 3: Multiple Fits To Data.}\\
        \multicolumn{5}{l}{Accessed via \url{https://www.garethkennedy.net/MECSimDocs.html} on 22-Aug-2025}\\
        \bottomrule
    \end{tabular}
    \end{adjustbox}
    \caption{Comparing electrochemical simulators and their parameter estimation capabilities.}
    \label{tab:ConventionalSimulators}
\end{table}

\clearpage
\section{A survey of Improved Tafel Analysis Method}

The IUPAC definition of Tafel analysis introduced by R. Guidelli, R.G. Compton and others,\cite{IUPACTafel1,IUPACTafel2} defines a cathodic Tafel analysis as:
\begin{equation}
    \alpha = \frac{-RT}{F}\frac{d ln(I)}{dE} 
    \label{TafelCathodic}
\end{equation}
with the corresponding definition of the anodic Tafel analysis as:
\begin{equation}
    \beta = \frac{RT}{F} \frac{d ln(I)}{dE}
    \label{TafelAnodic}
\end{equation}
where $R$ is the Gas constant, $T$ is the temperature, and $F$ is the Faraday constant. $I$ and $E$ are experimentally measured current and potential, respectively.  In practice, fitting part of current-voltage curve from CV using linear regression to extract transfer coefficient is the most common and the conventional approach for Tafel analysis.\cite{TafelPractice} Thus, conventional Tafel analysis is essentially fitting part of CV where cathodic/anodic current dominates and $dln(I)$ and $E$ are assumed to be linearly correlated.\cite{ChenTafelAnalysis1,ChenTafelAnalysis2} In essence, any approach fitting current - potential data to extract kinetic information including transfer coefficient, is a form of Tafel analysis. Since Differentiable Electrochemistry correlates experimental current-potential curves to its kinetic information, it is a form of Tafel analysis.

Tafel analysis is an important approach to elucidate electrocatalytic mechanism and evaluate electrokinetics. Conventional Tafel analysis that fits the logarithm of current with overpotential, however, has three major limitations.
\begin{enumerate}
\item The conventional Tafel analysis is an approximation to Butler-Volmer equation, applicable only to fully cathodic reaction. Conventional Tafel analysis only applicable when cathodic or anodic reaction dominates. 
\item The conventional Tafel analysis neglects the effect of mass transport, while electrochemistry is an interplay of interfacial reaction and mass transport. 
\item The conventional Tafel analysis does not specify where to extract the current--potential curve, leaving the analysis procedure more arbitrary and the results more subjective.  
\end{enumerate}
With increasing awareness to these limitations, there are numerous efforts, as tabulated in Table S \ref{tab:NewTafelAnalysis}, that aims to (partially) address the three flaws. From Table S \ref{tab:NewTafelAnalysis}, it is shown that only Differentiable Electrochemistry framework addressed all three flaws.

\clearpage
\begin{table}
    \centering
    \begin{tabular}{|p{0.3\linewidth} | p{0.45\linewidth} | p{0.15\linewidth}| }
    \toprule
         Approaches &  References & Flaws Addressed\\ \midrule
         Gradient-free optimization with covariance matrix adaptation &Corpus, K. R. M.; Bui, J. C.; Limaye, A. M.; Pant, L. M.; Manthiram, K.; Weber, A. Z.; Bell, A. T. \textit{Joule} \textbf{2023}, 7 (6), 1289-1307 & 2,3 \\ \midrule
         Tafel Slope Plot Method &van der Heijden, O.; Park, S.; Vos, R. E.; Eggebeen, J. J. J.; Koper, M. T. M.  \textit{ACS Energy Lett}. \textbf{2024}, 9 (4), 1871-1879.  & 3 \\ \midrule
         Differentiating the Current-Potential Curves &a) M. Corva, N. Blanc, C. J. Bondue, K. Tschulik, \textit{ACS Catal.}\textbf{ 2022}, 12, 13805-13812; b) P. Khadke, T. Tichter, T. Boettcher, F. Muench, W. Ensinger, C. Roth, \textit{Sci. Rep.} \textbf{2021}, 11, 8974.&3\\ \midrule
         Mass-Transport Corrected Fluxes& C. Batchelor‐McAuley, D. Li, R. G. Compton,\textit{ ChemElectroChem} \textbf{2020}, 7, 3844-3851 & 2 \\ \midrule
         Identifying “Mass-Transport-Free” Region &D. Li, C. Lin, C. Batchelor-McAuley, L. Chen, R. G. Compton, \textit{J. Electroanal. Chem}. \textbf{2018}, 826, 117-124 & 2 \\ \midrule
         Linear Transformation of the BV equation & Z. Lukács, T. Kristóf,\textit{ Electrochem. Commun.} \textbf{2023}, 154, 107556.& 1\\ \midrule
         Taylor Expansion of the BV equation &P. Agbo, N. Danilovic, \textit{J. Phys. Chem. C} \textbf{2019}, 123, 30252-30264.&1\\  \midrule
         Differentiable Electrochemistry with Gradient-based Optimization & This Work & 1,2,3\\ \midrule
    \bottomrule
    \end{tabular}
    \caption{A list of recently proposed new Tafel analysis method and the flaws of conventional Tafel analysis they are trying to resolve.}
    \label{tab:NewTafelAnalysis}
\end{table}

\clearpage
\section{Proof-of-Concept Case Studies}
This work consists of three proof-of-concept electrochemical cases:
\begin{enumerate}
    \item voltammetry in weakly supported media at a macroelectrode with migration-diffusion mass transport; 
    \item voltammetry of a dissociative CE reaction at a (hemi-)spherical electrode with convergent diffusion mass transport; 
    \item hydrodynamic voltammetry with convection-diffusion mass transport; 
\end{enumerate}
In the first case, the Poisson--Nernst--Planck (PNP) equation coupled with Butler--Volmer (BV) or Marcus--Hush--Chidsey (MHC) formalism, are solved differentially to recover interfacial kinetics such as electrochemical rate constants ($k_0$), transfer coefficient ($\alpha$) or reorganization energy ($\lambda$). Notably, the integral in the MHC formalism is solved numerically using Gauss-Hermite quadrature without simplifying the formalism .\cite{RN13} The second case is grounded in the experiment of acetic acid dissociation and proton reduction to extract the kinetics and thermodynamics parameters essential for understanding water splitting reaction.\cite{RN14} The third case formulates differentiable machine learning to extract kinematic viscosity ($\nu$) from hydrodynamic voltammetry at a Rotating Disk Electrode (RDE), where $\nu$ is one of the most important characteristics for fuel and lithium battery electrolyte.\cite{RN15,RN16} In addition, its noise robustness is evaluated for practical energy application. 

\subsection{Electrokinetics parameters from voltammetry in weakly supported media}
As a starting point, we briefly outline the underlying partial differential equations to be solved using implicit finite difference method within a differentiable programming framework. Using automatic differentiation, the gradients of errors between experimental and simulation observables with respect to the properties characterizing the theoretical behavior are computed and then used to iteratively update them through gradient-based optimization. The theories of voltammetry in weakly supported media, chronoamperometry of acetic acid reduction, and hydrodynamic voltammetry, are discussed next. 

We consider the electrochemically reversible, one-electron reduction of species $A^{Z_A}$ to $B^{Z_B}$ with electro-inactive cation and anion, $M^{z_M}$ and $N^{z_N}$, added as supporting electrolyte. The electrochemical reaction is $A^{z_A}+e^-\rightleftharpoons B^{z_B}$ where $z_B=z_A-1$ and $z_M=-z_N$. The reduction of $A^{z_A}$ to  in weakly supported media correlates both the concentration gradient and the electric potential gradient. Consequently, the linear mass transport of a charged species $j=(A,B,M\ or\ N)$ in solution is subject to diffusion and migration, and the flux $j_j$ is described by the Nernst-Planck equation
\begin{equation}
    \begin{array}{cc}
         j_j = -\left[D_j\left(\frac{\partial^2c_j}{\partial x^2}\right) + D_j\frac{z_jF}{RT}c_j\left(\frac{\partial \phi}{\partial x}\right)\right]& \\
         \frac{\partial^2 \phi}{\partial x^2} = -\frac{\rho}{\varepsilon_s \varepsilon_0},\ \rho = F\sum_j z_jc_j &
    \end{array}
\end{equation}
where $c_j$,$z_j$ and $D_j$ are concentration, electric charge and diffusion coefficient of species $j$, and $x$ is the distance to the planar electrode. $\phi$ is the electric potential.  $F$,$R$, and $T$ are Faraday Constant, Gas Constant, and temperature, respectively.
The electric potential $\phi$ is described by Poisson equation relating potential with local electric charge density, $\rho$, estimated by local ion concentration,  where $\varepsilon_s$ and $\varepsilon_0$ are the relative permittivity of solvent medium and the permittivity of free space. 

The  electric potential at the electrode surface is described by zero-field approximation, as the size of the electrode is assumed to be significantly larger than the thickness of the electrochemical double layer.\cite{RN17}  Two electrode kinetics theories are employed, starting with the BV kinetics\cite{RN18,RN19}
\begin{equation}
\left\{
    \begin{array}{cc}
         k_{red}^{BV}=k_0\exp\left(-\alpha\left(\theta-\phi_0\right)\right)&  \\
         k_{ox}^{BV}=k_0\exp\left(\left(1-\alpha\right)\left(\theta-\phi_0\right)\right)& 
    \end{array}
\right.
\label{BV}
\end{equation}
where $k_0$ and $\alpha$ are the standard electrochemical rate constant and transfer coefficient.  $\theta$ and $\phi_0$ are the dimensionless applied overpotential and the loss of driving force in weakly supported media and defined as
\begin{equation}
    \begin{array}{cc}
         \phi_0=\frac{F}{RT}\Delta\phi=\frac{F}{RT}\left( \phi_{PET}-\phi_{bulk}\right)& 
    \end{array}
\end{equation}
where $E$ and $E^0_f$ are the applied potential and formal potential, respectively, and $\Delta\phi$ is the potential difference between plane of electron transfer ($\phi_{PET}$) and the bulk solution ($\phi_{bulk}$), quantifying the loss of driving force due to weakly supported media.

The MHC kinetics is\cite{RN20,RN21}
\begin{equation}
\left\{
    \begin{array}{cc}
         k_{red}^{MHC}=k_0\frac{\bI\left(\left(\theta-\phi_0\right),\Lambda\right)}{\bI\left(0,\Lambda\right)}&  \\
         k_{ox}^{MHC}=k_0\frac{\bI\left(\left(\theta-\phi_0\right),\Lambda\right)}{\bI(0,\Lambda)}& 
    \end{array}
\right.    
\end{equation}
where $\Lambda$ is the dimensionless reorganization energy defined as $\Lambda=\frac{F}{RT}\lambda$. The integral is defined as
\begin{equation}
\begin{array}{cc}
         \bI\left(\theta,\Lambda\right)=\int^\infty_{-\infty}\frac{\exp\left(-\frac{\Delta G^{\ddagger}\left(\epsilon\right)}{RT}\right)}{1+\exp(\mp\epsilon)}d\epsilon&  \\
     \frac{\Delta G^\ddagger(\epsilon )}{RT}=\frac{\Lambda}{4}\left(1\pm\frac{\theta-\phi_0+\epsilon}{\Lambda}\right)^2& 
\end{array}
\end{equation}
where the upper sign stands for reduction and the lower sign for oxidation and $\Delta G^\ddagger$ is the activation energy of the redox process. In simulation, the integral is evaluated using Gauss Hermite quadrature as shown in the Dimensionless Equations section of Supporting Information.  

The loss function calculates the difference between ground truth (experimental voltammogram) and the simulated voltammogram 
\begin{equation}
    \mathcal{L}(k_0,\alpha\ or \lambda)=\sum_{t=0}^{t=2t_{switch}}\left(I_t-\hat{I}_t \left(k_0,
    \alpha\ or\  \lambda \right)\right)^2
    \label{LossFunction}
\end{equation}
where $\hat{I}$ is the overall simulation processes solving for transient current as a function of $k_0$,$\alpha$ or $\lambda$ . Since $I_t$ is treated as a constant, the loss function is fully differentiable to calculate the gradient of the loss with respect to the parameters with automatic differentiation. The parameters are then updated iteratively using stochastic gradient descent (SGD) with momentum. After $\sim$150 iterations, parameters that best describe the ground truth data are located.

\begin{figure}
    \centering
    \includegraphics[width=0.8\textwidth]{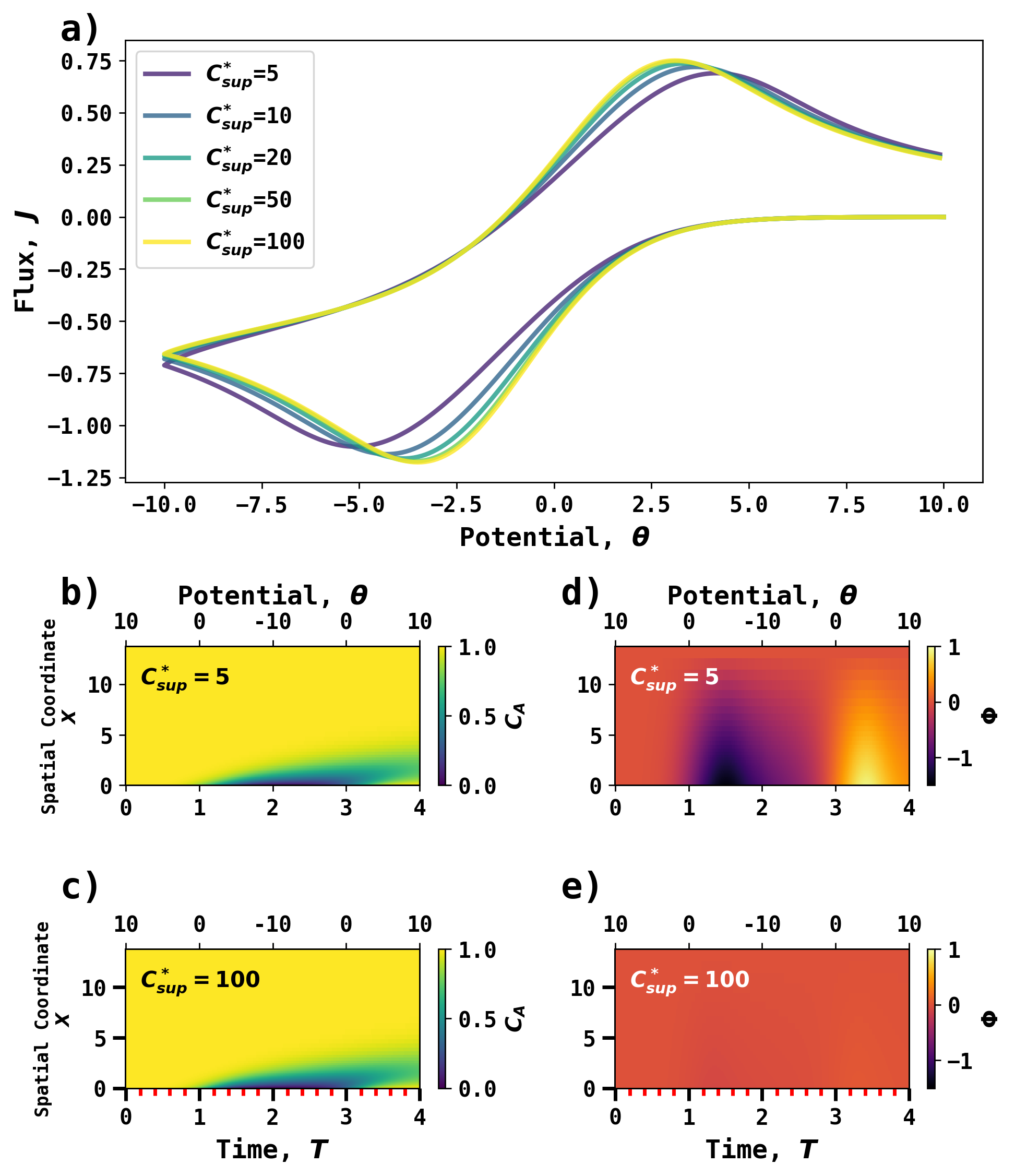}
    \caption{Cyclic voltammetry in the weakly supported media illustrated with dimensionless parameters. (a) Voltammograms at support ratios from 5 to 100.   (b, c) Concentration profile at support ratios of 5 and 100. (d, e) Solution electric potential at support ratios of 5 and 100. }
    \label{fig:WeaklySupportedMediaCV}
\end{figure}

To extract kinetics parameters in voltammetry in weakly supported media, the sensitivity of voltammograms to the relative concentration of support electrolyte is first investigated. Cyclic voltammograms at different support ratios (the relative concentration of supporting electrolyte relative to electroactive species) for one-electron reduction of species $A$ with $z_A=0$ and $z_B=-1$ are simulated and shown in Figure S \ref{fig:WeaklySupportedMediaCV}a when $K_0=1$,$\alpha=0.5$ and $\sigma=10$, where $K_0$ and $\sigma$ are the dimensionless electrochemical rate constants and scan rate, respectively. The charge of supporting electrolytes are $z_M=1$ and $z_N=-1$ respectively. As shown in Figure S \ref{fig:WeaklySupportedMediaCV}a, lowering the support ratio decreases both the cathodic and anodic peak currents and the peaks are shifted to higher overpotentials. In other words, with decreasing support ratio, there is a potential drop in solution that reduced the electrochemical driving force experienced by species A to attract/repel it from the electrode. The distorted voltammograms are the consequences of the discrepancy between applied potential and the real potential difference at the electron transfer plane. Figure S \ref{fig:WeaklySupportedMediaCV}b-e compares the concentration profiles and the electric potentials at support ratios of 5 and 100. Although there is no visible difference in concentrations of A as shown in Figure S \ref{fig:WeaklySupportedMediaCV}b,c, the difference in electric potentials in solution explains the distorted voltammograms. As shown in Figure S \ref{fig:WeaklySupportedMediaCV}d, there is a significant solution phase electric potential at a support ratio of 5 as compared with Figure S \ref{fig:WeaklySupportedMediaCV}e with a support ratio of 100. The negative charge accumulated near the electrode during the cathodic scan causes a more negative potential at the plane of electron transfer and a smaller driving force in reduction. Similarly, the positive charge accumulated during the anodic scan decreases the driving force and shifts the anodic peak to higher overpotentials. Given the importance and difficulty of parameter estimation in weakly supported media, the kinetic parameters are extracted at $C_{sup}^*=5$ from simulated ground truth. 

\begin{figure}
    \centering
    \includegraphics[width=1.0\textwidth]{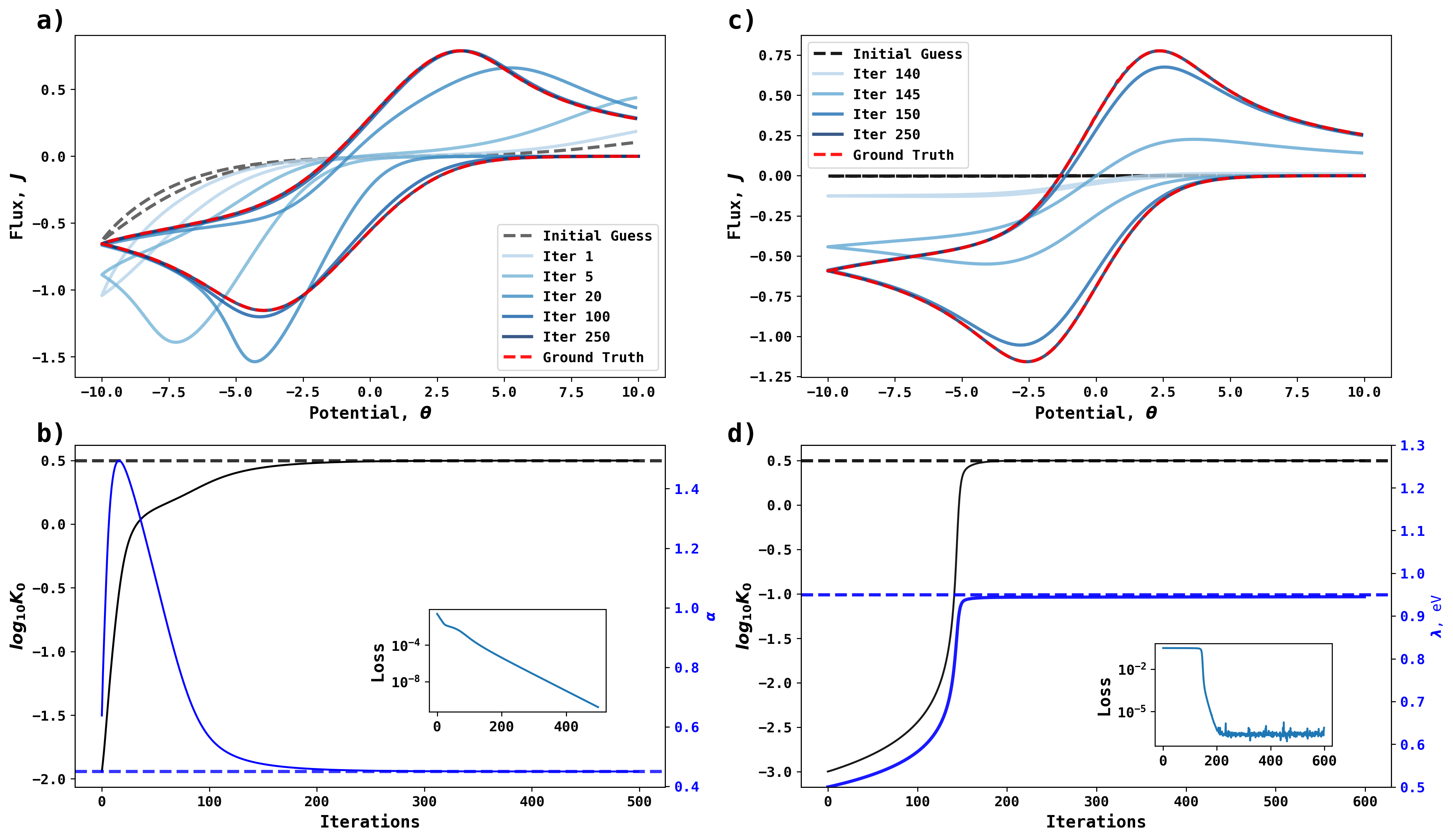}
    \caption{Parameter estimation in weakly supported voltammetry with BV (a, b) or MHC (c, d) kinetics at $C_{sup}^*=5$. (a, c) The ground truth voltammogram (red dashed), the initial guess (black-dashed) and the estimated voltammograms using differentiable simulation at selected iterations with (a) BV or (c) MHC kinetics.  (b) Left y-axis: The optimization trajectory of $K_0$ in log scale (black solid) and its ground truth (black dashed). Right y-axis: The optimization trajectory of transfer coefficient, $\alpha$ (blue solid) and its ground truth (blue dashed). The losses were shown in the inset. (d) Left y-axis: The optimization trajectory of $K_0$ in log scale (black solid) and its ground truth (black dashed). Right y-axis: The optimization trajectory of reorganization energy, $\lambda$ (blue solid) and its ground truth (blue dashed). The losses were shown in the inset. }
    \label{fig:WeaklySupportedMediaDiffEC}
\end{figure}

In the first example, a weakly supported voltammetry with BV kinetics ($K_0=10^{0.5}$,$\alpha=0.45$) with $\sigma=10$ and $C_{sup}^*=5$ is simulated as the ground truth (red trace Figure S \ref{fig:WeaklySupportedMediaDiffEC}a). The target is to estimate $K_0$ and $\alpha$ given only the ground truth voltammogram. The simulated voltammogram with an initial guess of $K_0=10^{-2}$  and $\alpha=0.5$ is shown as the dashed-black trace in Figure S \ref{fig:WeaklySupportedMediaDiffEC}a. Note that the initial guess of $K_0$ is intentionally chosen to be 2.5 orders of magnitude below the ground truth, escalating the optimization difficulty and more closely resembling real experimental scenarios.  With a learning rate of 0.5 and a momentum of 0.5, 250 iterations of differentiable simulations are performed and selected voltammograms during iterations are shown in Figure S \ref{fig:WeaklySupportedMediaDiffEC}a, where good overlap with the ground truth is achieved after 200 iterations. Even though the initial guess significantly underestimates the kinetics and the cathodic transfer coefficient, subsequent optimizations are correctly guided by differentiable simulation gradients to increase $K_0$ and optimize $\alpha$.  The optimization trajectory and the losses are shown in Figure S \ref{fig:WeaklySupportedMediaDiffEC}b and its inset. The non-monotonic optimization trajectory of $\alpha$ is due to the low sensitivity of $\alpha$ at very small $K_0$, showing strong evidence that differentiable simulations are numerically robust and ideal for parameters varying in orders of magnitude. After 250 iterations, the estimated values are $K_0=10^{0.450}$,$\alpha=0.450$.  The errors in $K_0$  and $\alpha$ are less than 0.01\% and negligible, and the loss decreases by 10 orders of magnitude to as small as 1$0^{-11}$, suggesting the high accuracy and efficiency of differential simulations. 

In the second example, a weakly supported voltammetry with MHC kinetics ($K_0=10^{0.5}$ and $\lambda=0.95\ \mathrm{eV}$) is simulated as the ground truth (red dashed trace, Figure S \ref{fig:WeaklySupportedMediaDiffEC}c ). Like the BV study, the target is to estimate $K_0$ and $\lambda$ using only the target voltammogram. Intentionally starting from an arbitrarily low initial guess of $K_0=10^{-3}$ and $\lambda=0.5\ \mathrm{eV}$ where no (visible) electrochemical reaction occurs (black dashed trace,Figure S \ref{fig:WeaklySupportedMediaDiffEC}c), differentiable simulations still manages to calculate the gradients and optimize in the correct direction, as evidenced by selected voltammograms during optimization (blue traces,Figure  S \ref{fig:WeaklySupportedMediaDiffEC}c). The optimization trajectories are also shown in Figure S \ref{fig:WeaklySupportedMediaDiffEC}d. Once the parameters pass a threshold where electrochemical reaction happens in the potential window, optimization accelerates from the ~120th iteration and the losses plateaus at the 200th iteration at $10^{-6}$. The identified parameters at the end of optimization are $K_0=10^{0.500}$ and $\lambda=0.946\ \mathrm{eV}$ (0.5\% error). With an ill-conditioned initial guess, highly nonlinear PNP mass transport equation, and the more complicated MHC kinetics requiring integration, differentiable simulation still manages to trace every data point on the voltammogram back to its parameters, highlighting the capabilities of differentiable simulations and potentials in more complicated systems. 

\subsection{Kinetic and thermodynamic parameters from acetic acid dissociation steady state currents}

In the second electrochemical study, differentiable simulations identify kinetics and thermodynamics from hydrogen evolution reaction in acetic acid. The steady state currents are obtained from previous chronoamperometry experiments on a microelectrode as $-30.5\pm0.4$, $-55.2\pm0.3$, $-102\pm1$ and $-238\pm4\ \mathrm{nA}$ for $10.0$, $20.0$, $40.0$ and $100.0\ \mathrm{mM}$ of bulk acetic acid in $0.1 \mathrm{M}$ $\ch{KNO3}$ recorded at $t=10s$.  The four currents collectively are the targets of differential simulation. While conventional data-driven machine learning relies on parameter sweep of the entire (multi-dimensional) parameter space to build a surrogate model between currents and parameters, differentiable simulation starts with an initial guess of parameters to compute the error of the simulated currents from experimental currents, along with the gradients of the error respective to the parameters. The parameters are then updated with gradient descent to minimize the error until convergence. 

\begin{figure}
    \centering
    \includegraphics[width=0.8\textwidth]{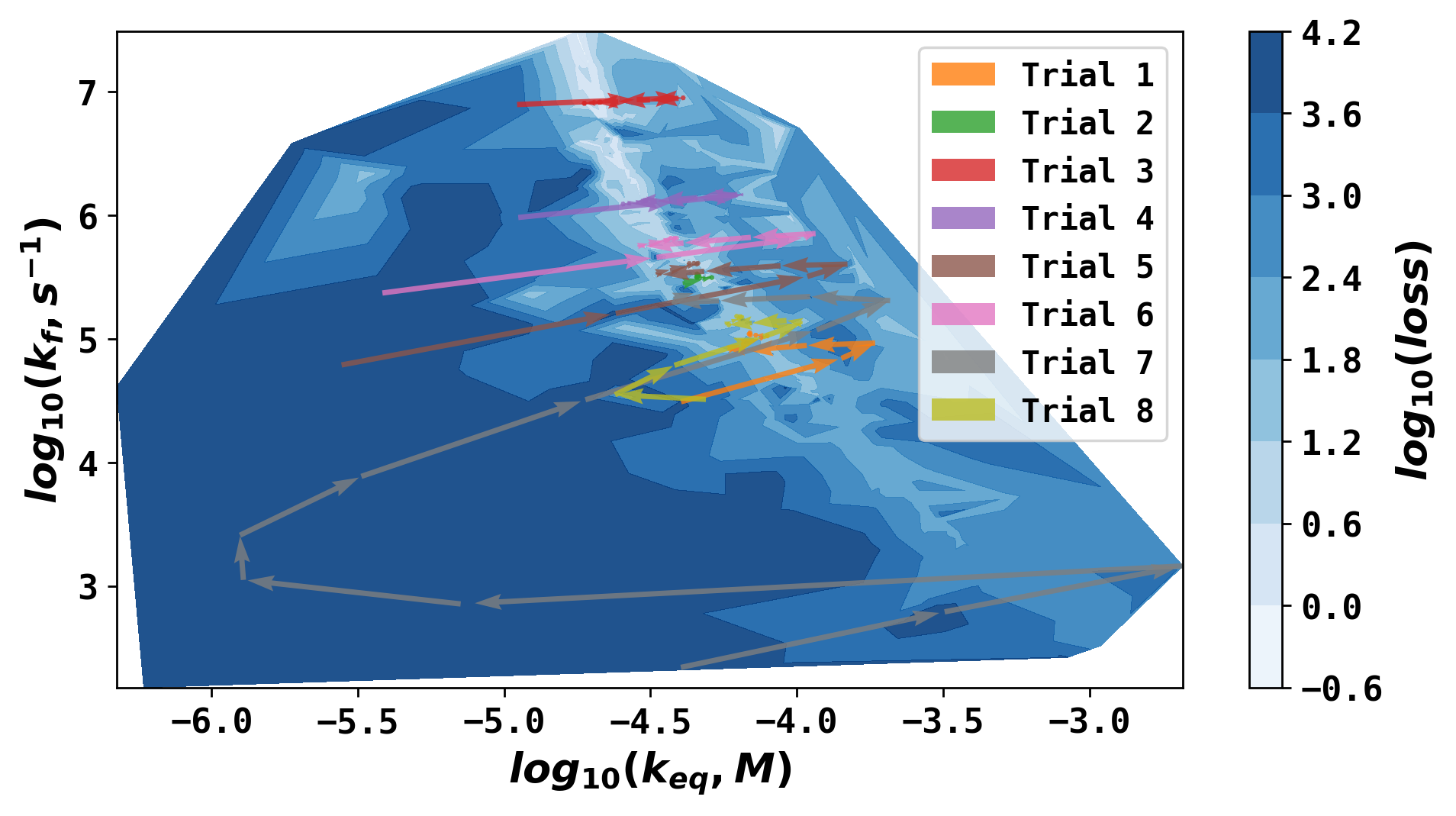}
    \caption{The loss landscape of the experimental acetic acid steady-state currents from the simulated currents as a function of $k_{eq}$ and $k_f$ values along with 8 differentiable simulation optimization trajectories to minimize the loss }
    \label{fig:DiffDissociative_CE}
\end{figure}

50 rounds of independent differentiable simulations and optimizations are performed, started from an initial guess of $k_{eq}$ between $10^{-7}$ to $10^{-3}$  $\mathrm{M}$ and $k_f$ between $10^2$ to $10^8 s^{-1}$ for 30 iterations with an learning rate of $10^{-4}$.  Figure S \ref{fig:DiffDissociative_CE} shows the loss landscape of the steady-state currents collected from all 50 rounds of simulations ranging almost 5 orders of magnitude. The loss landscape is rugged and highly nonlinear, characterized by numerous local minima and saddle points, yet differentiable simulation still efficiently locates the global minima within less than 10 iterations, as evidenced by 8 optimization trajectories shown in Figure S \ref{fig:DiffDissociative_CE}. The optimized results do not fall into the same locations, but distribute along a very deep "valley". The inherent noise in experimental results, the imperfection of mathematical model, and possibly the non-uniqueness of solution, maybe the contributing factors to the less ideal behavior.  The ensemble predictions of $k_{eq}$ and $k_f$ are $10^{-4.37\pm0.25}\ \mathrm{M}$ and $10^{5.74\pm0.79}\ \mathrm{s^{-1}}$. The predicted values agree well with literature values in a slightly different electrolyte ($0.1\ \mathrm{M}\ \ch{KCl}$) of $k_{eq}=10^{-4.54}\ \mathrm{M}$  and $k_f=10^{5.95}\ \mathrm{s^{-1}}$.\cite{RN23,RN24} In this example, differentiable simulations are significantly more efficient over conventional data-driven machine learning: by analytically computing the gradients of simulations respective to parameters of interest ($k_{eq}$ and $k_f$), the parameters are optimized within 30 iterations. Considering that each iteration requires four simulations to account for four different concentrations, only 120 simulations are required, while previous data-driven machine learning required 20,000 simulations as training data as shown in Figure S \ref{fig:DiffDissociative_CE}b, c.\cite{RN22} Moreover, surrogate models are unnecessary for differentiable simulations, cutting the cost of training a surrogate model.

\clearpage
\subsection{Kinematic Viscosity from Hydrodynamic Voltammetry}
In this section, the kinematic viscosity was extracted from simulated voltammograms. The simulated voltammograms are noisy with 0\% to 5\% added Gaussian noise to increase the difficulty of measurement and examine the noise robustness of differentiable electrochemistry simulations. 

\begin{figure}
    \centering
    \includegraphics[width=0.5\linewidth]{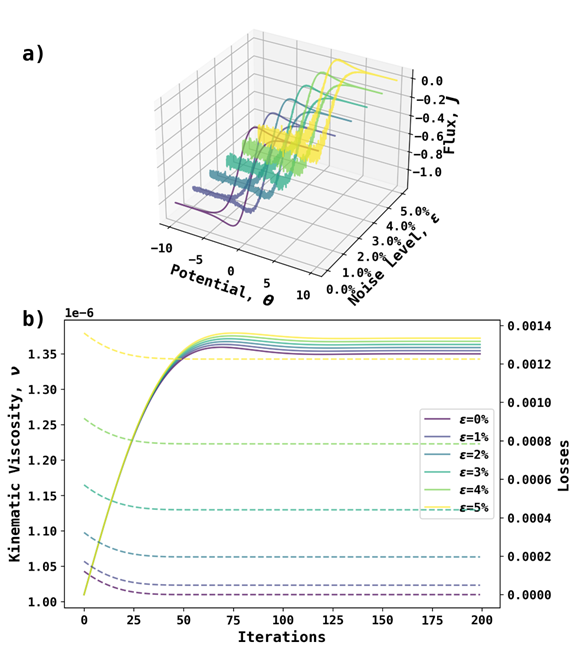}
    \caption{Differentiable simulation of hydrodynamic voltammetry. (a) The target voltammograms at Gaussian noise level from 0 to 5\%.  (b) Starting from an initial guess of $\nu=10^{-6}\ m^2s^{-1}$, the estimated $\nu$ as a function of differentiable simulation iterations (solid lines) and the corresponding losses (dashed lines) at different noise levels.}
    \label{fig:HydrodynamicNoise}
\end{figure}

The target voltammograms were generated assuming a kinematic viscosity of $1.35\times10^{-6}\ \mathrm{m^2s^{-1}}$, diffusion coefficient of $10^{-9} \mathrm{m^2s^{-1}}$, rotational frequency of 1200 rpm, and a scan rate of 0.1 V/s with varying levels of Gaussian noises from 0\% to 5\%. The resulting voltammograms in dimensionless form were reported in Figure S \ref{fig:HydrodynamicNoise}a. Differentiable simulations were applied to find the kinematic viscosity that best describes the hydrodynamic voltammogram, by calculating the loss between the target and simulated voltammograms and minimizing the difference with the gradient of the loss respective to the parameter.  Starting with an initial guess of $10^{-6}\  \mathrm{m^2s^{-1}}$ and a learning rate of $10^{-8}$, the learning results are shown in Figure S \ref{fig:HydrodynamicNoise}b. When the target voltammogram had no noise, the estimated $\nu$ agreed well with ground truth $\nu$ after 40 iterations, and the loss is almost zero. Increasing the target noise to 5\%, the loss stayed at 0.0008 and the estimated $\nu=1.37\times10^{-6} \mathrm{m^2s^{-1}}$ was slightly higher than the ground truth value. However, this is still a satisfactory agreement considering the noisiness of the target voltammogram as shown in the yellow trace of Figure S \ref{fig:HydrodynamicNoise}a. This case shows that hydrodynamic voltammetry with convection-diffusion mass transport can be end-to-end differentiated to optimize any parameters (kinematic viscosity in this case). In addition, differentiable simulations are highly efficient and robust, as only 50 iterations are needed to reach the ground truth and a 5\% noise on the target voltammogram only altered the prediction by 1.5\%. 

\section{Simulation Theories}
\subsection{Marcus-Hush-Chidsey kinetics}\label{MHCTheory}
The Marcus-Hush-Chidsey (MHC) kinetics is defined as: 
\begin{equation}
    \begin{array}{cc}
         K_{red}^{MHC} = K_0\frac{\boldsymbol{I}\left(\theta,\Lambda\right)}{\boldsymbol{I}\left(0,\Lambda\right)}&  \\
         K_{ox}^{MHC} = K_0\frac{\boldsymbol{I}\left(\theta,\Lambda\right)}{\boldsymbol{I}\left(0,\Lambda\right)}& 
    \end{array}
\end{equation}
and $\boldsymbol{I}$ is an integral over the energy levels: 
\begin{equation}
    \boldsymbol{I}\left(\theta,\Lambda\right) = \int^\infty_{-\infty} \frac{\exp{\left(-\frac{G^{\ddagger}\left(\theta, \Lambda, \epsilon\right)}{RT}\right)}}{1+\exp{\left(\mp\epsilon\right)}}d\epsilon
\end{equation}
where $\epsilon$ is a dimensionless variable. The upper sign in $\mp$ represents reduction and the lower sign represents oxidation. The activation energy of the reduction/oxidation processes is calculated from: 
\begin{equation}
    \frac{\Delta G^{\ddagger}}{RT} = \frac{\Lambda}{4}\left(1\pm \frac{\theta+\epsilon}{\Lambda} \right)^2
\end{equation}
The integration $\boldsymbol{I}$ is solved using Gaussian-Hermite quadrature. For the reduction process, the integration is:
\begin{equation}
    \boldsymbol{I}_{red}=\int^\infty_{-\infty}\frac{\exp{\left(-\frac{\Lambda}{4}\left(1+\frac{\theta+\epsilon}{\Lambda}\right)^2\right)}}{1+\exp{\left(-\epsilon\right)}}d\epsilon
\end{equation}
Let $y=\frac{\sqrt{\Lambda}}{2}\left(1+\frac{\theta+\epsilon}{\Lambda}\right)$, the integration becomes:
\begin{equation}
    \boldsymbol{I}_{red}=\int^{\infty}_{-\infty}\exp{\left(-y^2\right)}\frac{2\sqrt{\Lambda}}{1+\exp{\left(-\Lambda\left(y\frac{2}{\sqrt{\Lambda}}-1\right)-\theta\right)}} dy
\end{equation}
For the oxidation process, the integration is:
\begin{equation}
    \boldsymbol{I}_{ox}=\int^\infty_{-\infty}\frac{\exp{\left(-\frac{\Lambda}{4}\left(1-\frac{\theta+\epsilon}{\Lambda}\right)^2\right)}}{1+\exp{\left(\epsilon\right)}}d\epsilon
\end{equation}
Let $y=\frac{\sqrt{\Lambda}}{2}\left(1-\frac{\theta+\epsilon}{\Lambda}\right)$, the integration becomes:
\begin{equation}
    \boldsymbol{I}_{ox}=\int^\infty_{-\infty}\exp{\left(-y^2\right)}\frac{-2\sqrt{\Lambda}}{1+\exp{\left(-\Lambda\left(y\frac{2}{\sqrt{\Lambda}}-1\right)-\theta\right)}}dy
\end{equation}
The transformed integration function is then solved to the 50\textsuperscript{th} degree using the sample locations and weights generated from \textit{numpy.polynomial.hermite.hermgauss} function.
\subsection{Voltammetry in weakly supported media}
The dimensionless parameters for simulations are defined in Table S \ref{tab:NPPDimensionlessTable}:

\begin{table}
    \centering
    \begin{tabular}{|c|c|}
    \toprule
        Parameter & Dimensionless Definition\\  \midrule
        Spatial coordinate & $X=\frac{x}{r_e}$ \\ \midrule
        Potential & $\theta = \frac{F(E-E^0_f )}{RT}$\\ \midrule
        Standard Electrochemical Rate Constant & $K_0=\frac{k_0r_e}{D_A}$\\ \midrule
        Reorganization Energy & $\Lambda = \lambda \frac{F}{RT}$\\ \midrule
        Diffusion Coefficient &  $d_j = \frac{D_j}{D_A}$\\ \midrule
        Solution Potential &  $\Phi = \phi \frac{F}{RT}$\\ \midrule
    \end{tabular}
    \caption{Definitions of dimensionless parameters. $F$, $R$ and $T$ are Faraday constant, Gas Constant, and temperature respectively.}
    \label{tab:NPPDimensionlessTable}
\end{table}

The dimensionless Nernst-Planck-Poisson equation is:
\begin{equation}
\begin{array}{cc}
    \frac{\partial C_j}{\partial T} = d_j \left(\frac{\partial^2C_j}{\partial X^2}\right) + d_jZ_j\left(\frac{\partial C_j}{\partial X}\frac{\partial \Phi}{\partial X} + C_j \frac{\partial^2\Phi}{\partial X^2}\right) &\\
    \frac{\partial^2\Phi}{\partial X^2} = -R_e^2\sum_jz_jC_j&\\
\end{array}
\end{equation}
where $R_e$ is a dimensionless parameter represents the relative scale of the electrode compared to the Debye length:
\begin{equation}
    R_e = r_e \sqrt{\frac{F^2C^*_{ref}}{RT\varepsilon_s\varepsilon_0}}
\end{equation}
where $\varepsilon_0$ is the relative permittivity of the solvent (80 for water at room temperature), $\varepsilon_0$ is the permittivity of free space ($8.85\times 10^{-12} F m^{-1}$).

The dimensionless boundary conditions at the electrode surface, considering the zero-field approximation $\left(\left(\frac{\partial \phi}{\partial x}\right)_{x=0}=0\right)$, are:
\begin{equation}
\begin{array}{cc}
     T=0, X\geq0&  \\
     T>0, X=X_{max}& 
\end{array}
\begin{cases}
C_A=1 \\
C_B=0 \\
\Phi(T=0)=0 \\ 
\Phi(X=X_{max}) + X_{max}\left(\frac{\partial\Phi}{\partial X}\right) = 0 \\ 
C_M = C^*_{sup},\ C_N = C^*_{sup} + z_A \ \text{if} \ z_A>0 \\ 
C_M = C^*_{sup} - z_{A}, \ C_N = C^*_{sup} \ \text{if} \ z_A<0 \\
\end{cases}
\end{equation}

\begin{equation}
    T>0, X=0 
    \begin{cases}
        d_A\left(\frac{\partial C_A}{\partial X}\right)_{X=0} = K_{red}C_A(X=0) - K_{ox}C_B(X=0) \\ 
        d_B\left(\frac{\partial C_B}{\partial X}\right)_{X=0} = -d_A\left(\frac{\partial C_A}{\partial X}\right)_{X=0} \\ 
        \left( \frac{\partial \Phi}{\partial X}\right)_{X=0} = 0 \\ 
        \left( \frac{\partial C_M}{\partial X}\right)_{X=0} = 0 \\ 
        \left( \frac{\partial C_N}{\partial X}\right)_{X=0} = 0 \\ 
    \end{cases}
\end{equation}
where $C^*_{sup} = \frac{c^*_{sup}}{c^*_{A}}$ is the support ratio of the electrolyte solution. The dimensionless current is given by:
\begin{equation}
    J = \frac{Ir_e}{FAD_Ac^*_A} = -\left(\frac{\partial C_A}{\partial X}\right)_{X=0}
\end{equation}
$K_{red}$ and $K_{ox}$ are the rates of reduction and oxidation at the electrode surface, and are described with either Butler-Volmer (BV) or Marcus-Hush-Chidsey (MHC) theories. Note that the effective potential experienced by the analyte is affected by solution phase potential at the electron transfer distance. The BV kinetics are:
\begin{equation}
    \begin{array}{cc}
         K_{red}^{BV} = K_0\exp{\left(-\alpha\left(\theta-\Phi_0\right)\right)}&  \\
         K_{ox}^{BV} = K_0\exp{\left(\left(1-\alpha\right)\left(\theta-\Phi_0\right)\right)}& \\ 
    \end{array}
\end{equation}

The MHC kinetics, taking the effective surface potential$(\theta-\Phi_0)$ into consideration, is:
\begin{equation}
\begin{array}{cc}
     K_{red}^{MHC} = K_0 \frac{\boldsymbol{I}\left(\theta-\Phi_0, \Lambda\right)}{\boldsymbol{I}\left(0, \Lambda\right)} &  \\
     K_{ox}^{MHC} = K_0 \frac{\boldsymbol{I}\left(\theta-\Phi_0, \Lambda\right)}{\boldsymbol{I}\left(0, \Lambda\right)}& \\
\end{array}
\end{equation}

\subsection{Voltammetry of $\ch{Fe^{3+}}/\ch{Fe^{2+}}$ Redox Couple}
The electrochemical reduction of $\ch{Fe^{3+}}/\ch{Fe^{2+}}$ redox couple on Platinum electrode is:
\begin{equation}
    \ch{Fe^{3+} + e- <=> Fe^{2+}} 
\end{equation}
The electrochemical kinetics at the electrode surface is described with Butler-Volmer equation:
\begin{equation}
\begin{cases}
\begin{aligned}
    D_{\ch{Fe^{3+}}}\left(\frac{\partial c_{\ch{Fe^{3+}}} }{\partial x}\right)_{x=0} &= k_0\exp{\left(-\frac{\alpha F\left(E-E_{ref}\right)}{RT}\right)}c_{\ch{Fe^{3+}},x=0} \\
    &- k_0\exp{\left(\frac{\beta F \left(E-E_{ref}\right)}{RT}\right)}c_{\ch{Fe^{2+}},x=0} \\ 
\end{aligned}\\ 
D_{\ch{Fe^{2+}}}\left(\frac{\partial c_{\ch{Fe^{2+}}} }{\partial x}\right)_{x=0} = - D_{\ch{Fe^{3+}}}\left(\frac{\partial c_{\ch{Fe^{3+}}} }{\partial x}\right)_{x=0}
\end{cases}
\end{equation}
where $E$ and $E^0_f$ are the applied potential and the reference potential set at 0.4336 V. $k_0$, $\alpha$, and $\beta$ are the standard electrochemical rate constant, cathodic, and anodic transfer coefficients, which are parameters to be discovered during Differentiable Electrochemistry simulation and optimization. The mass transport is assumed to be diffusion dominated:
\begin{equation}
    \begin{cases}
        \frac{\partial c_{\ch{Fe^{3+}}}}{\partial t}  = D_{\ch{Fe^{3+}}} \frac{\partial^2c_{\ch{Fe^{3+}}}}{\partial x^2}\\ 
        \frac{\partial c_{\ch{Fe^{2+}}}}{\partial t}  = D_{\ch{Fe^{2+}}} \frac{\partial^2c_{\ch{Fe^{2+}}}}{\partial x^2}\\ 
    \end{cases}
\end{equation}

Differentiable Electrochemistry simulations are performed with dimensionless parameters. The dimensionless electrochemical boundary conditions at electrode surface $(X=0)$ are:
\begin{equation}
    \begin{cases}
        d_{\ch{Fe^{3+}}}\frac{\partial C_{\ch{Fe^{3+}}}}{\partial X} = K_0\exp{\left(-\alpha\theta\right)}C_{\ch{Fe^{3+}}} - K_0\exp{\left(\beta\theta\right)}C_{\ch{Fe^{2+}}}\\
        d_{\ch{Fe^{2+}}}\frac{\partial C_{\ch{Fe^{2+}}}}{\partial X} = -d_{\ch{Fe^{3+}}}\frac{\partial C_{\ch{Fe^{3+}}}}{\partial X} \\ 
    \end{cases}
\end{equation}
The dimensionless mass transport equation is:
\begin{equation}
    \begin{cases}
        d_{\ch{Fe^{3+}}} = \frac{\partial^2 C_{\ch{Fe^{3+}}}}{\partial X^2}\\ 
        d_{\ch{Fe^{2+}}} = \frac{\partial^2 C_{\ch{Fe^{2+}}}}{\partial X^2}\\ 
    \end{cases}
\end{equation}

The definitions of dimensionless parameters are shown in Table S \ref{tab:FeCoupleDimensionlessTable}, which are slightly different from other simulation theories.

\begin{table}
    \centering
    \begin{tabular}{|c|c|}
    \toprule
        Concentration of species $j$ & $C_j = \frac{c_j}{c^*_{\ch{Fe^{3+}}}}$ \\ \midrule
        Diffusion coefficient of species $j$ & $d_j = \frac{D_j}{D_{ref}}$\\ \midrule
        Spatial Coordinate & $X=\frac{x_i}{x_{ref}}$ \\ \midrule
        Time & $T=\frac{D_{ref}t}{x_{ref}^2}$ \\ \midrule
        Potential & $\theta = \frac{F}{RT}(E-E_{ref})$ \\ \midrule
        Scan Rate & $\sigma = \frac{x_{ref}^2}{D_{ref}}\frac{F}{RT}\nu$\\ \midrule
        Flux & $J = \frac{x_{ref}}{c^*_{\ch{Fe^{3+}}}D_{ref}}j$ \\ \midrule
        Standard electrochemical rate constant & $K_0 = \frac{k_0x_{ref}}{D_{ref}}$ \\ \midrule
    \end{tabular}
    \caption{The definition of dimensionless parameters for redox reaction of $\ch{Fe^{3+}}/\ch{Fe^{2+}}$ couple.  $D_{ref}=10^{-9} m^2s^{-1}$, $C^*_{\ch{Fe^{3+}}}=4.85\ mM$ and $x_{ref}=0.85\ mm$.}
    \label{tab:FeCoupleDimensionlessTable}
\end{table}

\subsection{Voltammetry of $\ch{Ru(NH3)_6^{3+}}$/$\ch{Ru(NH3)_6^{2+}}$ Redox Couple}
The $\ch{Ru(NH3)_6^{3+}}$/$\ch{Ru(NH3)_6^{2+}}$ redox couple (commonly known as the RuHex) is a classic electrochemistry reaction and an ideal playground for Differentiable Electrochemistry simulation. The electrochemical reaction is:
\begin{equation}
    \ch{Ru(NH3)_6^{3+} + e- <=> Ru(NH3)_6^{2+} }
\end{equation}

Electrochemical reaction of the RuHex couple is reversible and can be characterized as:
\begin{equation}
    \frac{c_{\ch{Ru(NH3)_6^{3+}}}}{\left(c_{\ch{Ru(NH3)_6^{3+}}}^*-c_{\ch{Ru(NH3)_6^{2+}}}\right)} = \exp{\left(E-E^0_f\right)}
\end{equation}
where $c_{\ch{Ru(NH3)_6^{3+}}}^*=0.96\ \mathrm{mM}$ is the bulk concentration of $\ch{Ru(NH3)_6^{3+}}$ and $E^0_f$ is the formal potential of the Redox couple. The experiment was performed in $0.1\ \mathrm{M}$ KCl electrolyte at 298 K on a Glassy Carbon Electrode (GCE) with a radius of $1.5\ \mathrm{mm}$. The mass transport equation is characterized by 1-D diffusion equation. 
\begin{equation}
    \begin{cases}
        \frac{\partial c_{\ch{Ru(NH3)_6^{3+}}}} {\partial t} = D_{\ch{Ru(NH3)_6^{3+}}} \frac{\partial^2c_{\ch{Ru(NH3)_6^{3+}}}}{\partial x^2}\\
        \frac{\partial c_{\ch{Ru(NH3)_6^{2+}}}} {\partial t} = D_{\ch{Ru(NH3)_6^{2+}}} \frac{\partial^2c_{\ch{Ru(NH3)_6^{2+}}}}{\partial x^2}\\
    \end{cases}
\end{equation}

The definition of dimensionless parameters is similar to that of $\ch{Fe^{3+}}/\ch{Fe^{2+}}$ redox couple and not reiterated here.

\subsection{Chronoamperometry of acetic acid reduction}
The general scheme of proton reduction in acetic acid solution is:
\begin{equation}
    \begin{array}{cc}
         \ch{CH3COOH <=> [k_f][k_b] CH3COO-}& \\
         \ch{2 H+ + 2 e- <=> H2}& \\ 
    \end{array}
\end{equation}
The detailed experimental procedure and equations are shown in ref.\cite{RN22}. The problem is briefly described here in terms of dimensionless parameters, concerning the mass transport equation coupled with nonlinear chemical kinetics:
\begin{equation}
    \begin{cases}
        \begin{split}
        \frac{\partial C_{\ch{CH3COOH}}}{\partial T} &= d_{\ch{CH3COOH}}\frac{\partial^2 C_{CH3COOH}}{\partial R^2} + \frac{2}{R}\frac{\partial C_{\ch{CH3COOH}}}{\partial R} \\
        &- K_fC_{\ch{CH3COOH}} + K_bC_{\ch{H+}}C_{\ch{CH3COO-}}\\
        \end{split}\\
        \begin{split}
            \frac{\partial C_{\ch{H+}}}{\partial T} &= d_{\ch{H+}}\frac{\partial^2 C_{\ch{H+}}}{\partial R^2} + \frac{2}{R}\frac{\partial C_{\ch{H+}}}{\partial R} \\
            &+K_fC_{\ch{CH3COOH}} - K_bC_{\ch{H+}}C_{\ch{CH3COO-}}\\
        \end{split}\\ 
        \frac{\partial C_{H^2}}{\partial T} = d_{\ch{H2}}\frac{\partial^2 C_{\ch{H2}}}{\partial R^2} + \frac{2}{R}\frac{\partial C_{\ch{H2}}}{\partial R} \\ 
        \begin{split}
            \frac{\partial C_{\ch{CH3COO-}}}{\partial T} &= d_{\ch{CH3COO-}}\frac{\partial^2C_{\ch{CH3COO-}}}{\partial R^2} + \frac{2}{R} \frac{\partial C_{\ch{CH3COO-}}}{\partial R} \\
            &+K_fC_{\ch{CH3COOH}} - K_bC_{\ch{H+}}C_{\ch{CH3COO-}} \\ 
        \end{split}\\

    \end{cases}
\end{equation}

\subsection{Hydrodynamic voltammetry}
Recall that the mass transport equation for a rotating disk electrode is:
\begin{equation}
    \begin{array}{cc}
         \frac{\partial c_j}{\partial t} = D\frac{\partial^2c_j}{\partial y^2} - v_y \frac{\partial c_j}{\partial y} &  \\
         v_y\approx -Ly^2& \\ 
         L = 0.51023\left(2\pi f\right)^{\frac{3}{2}} \nu^{-\frac{1}{2}}
    \end{array}
\end{equation}
where $v_y$ is the solution speed in $y$ direction perpendicular to the electrode surface, and $L$ is the hydrodynamic number as a function of rotating frequency $f, \ Hz$ and $\nu,\ m^2s^{-1}$, kinematic viscosity.  
To solve this mass transport equation, the coordinate $y$ is transformed to $W$ as:
\begin{equation}
    W = \left(\frac{L}{D}\right)^{\frac{1}{3}} y 
\end{equation}
So that the convection-diffusion mass transport equation is: 
\begin{equation}
    \frac{\partial C_A}{\partial T} = \frac{\partial^2 C_A}{\partial W^2} +W^2\frac{\partial C_A}{\partial W}
\end{equation}
To solve the problem, the Hale transformation is necessary\cite{britz2005digital,GAVAGHAN1997147}:
\begin{equation}
    U = \frac{\int^W_0 \exp\left(-\frac{1}{3}W^3\right)dW}{\int^{\infty}_0 \exp\left(-\frac{1}{3}W^3\right)dW}
\end{equation}
such that the value $U=0$ corresponds to electrode surface and $U=1$ corresponds to a distance infinitely far from electrode.  During simulation, the outer boundary of simulation is at $U_{sim}=0.9999$ , a distance sufficiently far away from electrode surface and unperturbed by surface reaction there.

The boundary conditions after Hale transformation are: 
\begin{equation}
    \begin{array}{cc}
         C_A=1 &  \\
         C_A=1& \\ 
         C_A=f(\theta)&\\
    \end{array}
    \begin{array}{cc}
         T=0,\ U_{sim}\geq U\geq 0&  \\
         T>0, U=U_{sim}& \\
         T>0, U=0 &\\
    \end{array}
\end{equation}

Moreover, with Hale transformation, the differential equation simplifies to:
\begin{equation}
    \frac{\partial C}{\partial T} = \frac{\left(\exp{-\frac{2}{3}W^3}\right)}{\left[\int^{\infty}_0 -\frac{1}{3}W^3dW\right]^2}
\end{equation}
where $\left[\int^{\infty}_0 -\frac{1}{3}W^3dW\right]^2 = \left[\frac{\Gamma{(\frac{1}{3})}}{3^{\frac{2}{3}}}\right]^2=1.65894$, where  $\Gamma$ is the error function. 

\clearpage
\begin{table}
    \centering
    \begin{tabular}{|c|c|}
    \toprule
         Parameter&Dimensionless Form \\ \midrule
         Concentration&$C_j=\frac{c_j}{c^*_A}$ \\ \midrule
         Spatial coordinate in $y$ direction for 1-D simulation & $W=\left(\frac{L}{D}\right)^{\frac{1}{3}}y$ \\ \midrule
         Fluid velocity in W direction under the Levich approximation& $V_W=-W^2$ \\ \midrule
         Time & $T=\left(L^2D\right)^{\frac{1}{3}}t$\\ \midrule
         Potential& $\theta = \left(\frac{F}{RT}\right)\left(E-E^0_f\right)$ \\ \midrule
         Scan rate ($\boldsymbol{\nu}$)& $\sigma = \frac{F\boldsymbol{\nu}}{RT}\left(L^2D_A\right)^{-\frac{1}{3}}$\\ \midrule
         Flux & $J_{RDE} = \frac{I_{RDE}\sqrt{1.65894}}{F\left(\pi r_e^2\right)c_A^*\left(D^2_AL\right)^{\frac{1}{3}}}$ \\ \midrule
         Electrochemical rate constant & $K_0 = \sqrt{1.65894}k_0\left(D^2L\right)^{-\frac{1}{3}} $ \\ \midrule
    \end{tabular}
    \caption{Definition of dimensionless parameters. $F$, $R$, and $T$ are the Faraday Constant, the Gas Constant and temperature, respectively. $E$ and $E^0_f$ are the potential and the formal potential respectively.}
    \label{tab:HydrodynamicDimensionlessTable}
\end{table}

\subsection{Voltammetry with Adsorbed Species} \label{AdsorptionTheory}
\begin{figure}
    \centering
    \includegraphics[width=0.5\linewidth]{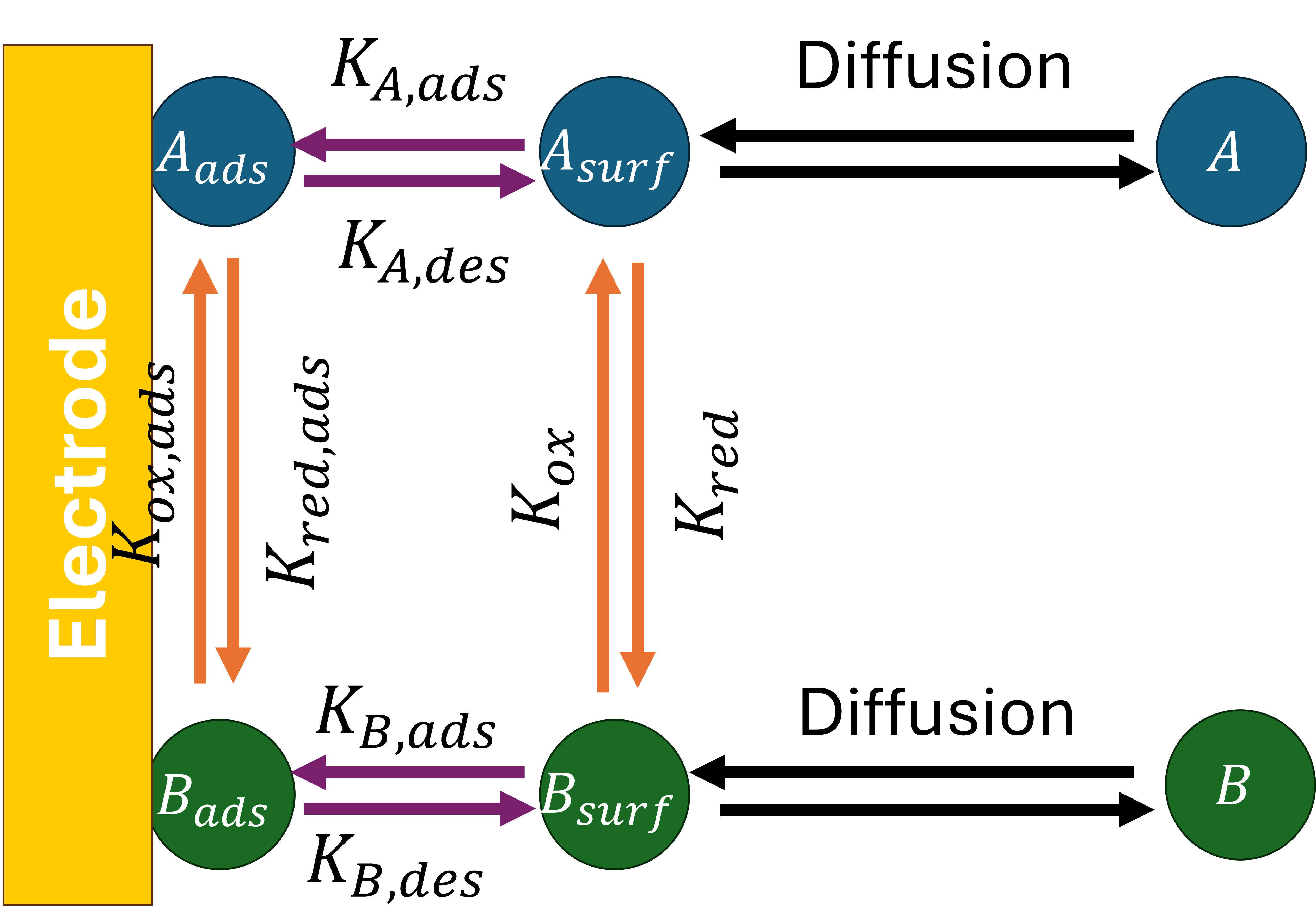}
    \caption{The mechanism of voltammetry with adsorbed/desorbed species. Electrochemical reactions is labeled with orange arrows; conversion between adsorbed and surface species is labeled with purple arrows; mass transport is labeled as black arrows. }
    \label{fig:AdsorptionMechanism}
\end{figure}

If both electroactive species ($A$, $B$) adsorb on the electrode surface (without partial charge transfer), and electron transfer takes places in both solution phase and surface-bound, we are solving the electrochemical system with adsorption. 
The electrochemical reaction with adsorption mechanism is shown in Figure S \ref{fig:AdsorptionMechanism}, where $K_{red/ox}$ and $K_{red/ox,\ ads}$ are the dimensionless electrochemical rate constants for solution-phase and surface-bound species. 

To account for surface adsorption of electroactive species $A$ and $B$, the surface coverage of two species, $\Gamma_A$ (mol m\textsuperscript{-2}) and $\Gamma_B$ (mol m\textsuperscript{-2}) must be resolved, along with the concentration profiles. 

To account for adsorption/desorption dynamics, Langmuir adsorption isotherm is applied. The rates of adsorption and desorption, of species $j\in[A,B]$ that compete for the same surface sites are given by: 
\begin{equation}
    \begin{aligned}
        v_{j,ads} &= k_{j,ads}c_{x=0,j}\left[\Gamma_{j,\text{max}}-\left(\Gamma_A+\Gamma_B\right)\right] \\
        v_{j,des} &= k_{j,des}\Gamma_j
    \end{aligned}
\end{equation}
where $k_{j,ads},\ mol^{-1}m^3s^{-1}$ and $k_{j, des}, \ s^{-1}$ are the adsorption and desorption rate constants of species $j$. $\Gamma_{j,\text{max}}$ is the maximum surface coverage, which is assumed to be equal for both species: $\Gamma_{A,\text{max}}=\Gamma_{B,\text{max}}=\Gamma_{\text{max}}$. Coupling Langmuir isotherm with diffusion mass transport, the governing equations are:
\begin{equation}
    \begin{aligned}
        \frac{d\Gamma_A}{dt} &= v_{A,ads} - v_{A,des} - k_{red,ads}\Gamma_A + k_{ox,ads}\Gamma_B   \\
        \frac{d\Gamma_B}{dt} &= v_{B,ads} - v_{B,des} + k_{red,ads}\Gamma_A - k_{ox,ads}\Gamma_B   \\
        D_A\left(\frac{\partial c_A}{\partial x}\right)_{x=0} &= k_{red}c_{A,x=0} - k_{ox}c_{B,x=0} + k_{A,ads}c_{A,x=0}\left[\Gamma_{max} - \left(\Gamma_A +\Gamma_B \right)\right] - k_{A,des}\Gamma_A   \\ 
        D_B\left(\frac{\partial c_B}{\partial_x}\right)_{x=0} &= -k_{red}c_{A,x=0} + k_{ox}c_{B,x=0} + k_{B,ads}c_{B,x=0}\left[\Gamma_{max} - \left(\Gamma_A + \Gamma_B \right)\right] - k_{B,des}\Gamma_B  \\ 
    \end{aligned}
    \label{eq:DimensionalAdsorption}
\end{equation}
where $k_{red/ox,ads}$ is the electrochemical rate constant and $v_{j, ads/des}$ is the rate of adsorption/desorption of the corresponding species $j$. The formal potentials of the solution and adsorbed species ($E^0_f, E^0_{f,ads}$) are usually different. Note that the latter are related through the adsorption equilibrium constants:
\begin{equation}
    E^0_{f,ads} = E^0_f + \frac{RT}{F}\ln{\left(\frac{k_{A,ads}/k_{A,des}}{k_{B,ads}/k_{B,des}}\right)}
\end{equation}

During simulation, Equation \ref{eq:DimensionalAdsorption} is normalized, and become:
\begin{equation}
    \begin{aligned}
         \frac{d\xi_A}{\Delta T} &= -K_{red,ads}\xi_A + K_{ox,ads}\xi_B + K_{A,ads}p_sC_{A,X=0}\left[1- \left(\xi_A + \xi_B\right)\right] -K_{A,des}p_s\xi_A   \\
         \frac{d\xi_B}{\Delta T} &= K_{red,ads}\xi_A - K_{ox,ads}\xi_B + K_{B,ads}p_sC_{B,X=0}\left[1-\left(\Gamma_A+\Gamma_B\right)\right] -K_{B,des}p_s\xi_B   \\
         d_A\left( \frac{C_A}{\Delta X}\right)_{X=0}  &= K_{red}C_{A,X=0} - K_{ox}C_{B,X=0} + K_{A,ads}C_{A,X=0}\left[1- \left(\xi_A + \xi_B\right)\right] - K_{A,des}\xi_A  \\ 
         d_B\left( \frac{C_B}{\Delta X}\right)_{X=0} &= -K_{red}C_{A,X=0} + K_{ox}C_{B,X=0} + K_{B,ads}C_{B,X=0}\left[1- \left(\xi_A + \xi_B\right)\right] - K_{B,des}\xi_B  \\
    \end{aligned}
\end{equation}
where $p_s$ is the dimensionless saturation number and $\xi_j$ is the relative surface coverage. The dimensionless parameters are defined in Table S \ref{tab:AdsorptionDimensionlessTable}. The electrochemical rates constants for solution and adsorbed species, using Butler-Volmer kinetics, are:
\begin{equation}
\begin{aligned}
     K_{red} &= K_0 \exp{(-\alpha\theta)}   \\
     K_{ox} &= K_0\exp(\beta\theta) \\ 
     K_{red,ads} &= K_{0,ads}\exp{\left(-\alpha_{ads}\left(\theta-\theta_{0,ads}\right)\right)}  \\
     K_{ox,ads} &= K_{0,ads}\exp{\left(-\alpha_{ads}\left(\theta-\theta_{0,ads}\right)\right)}  \\
\end{aligned}
\end{equation}

Lastly, the initial conditions of simulation, assuming pre--equilibrium under Langmuir isotherm, is:
\begin{equation}
    T=0\begin{cases}
        C_A = C_A^* \\
        C_B = C_B^* \\ 
        \xi_A = \frac{K_{A,ads}/K_{A,des}C^*_A}{1+K_{A,ads}/K_{A,des}C_A^*+K_{B,ads}/K_{B,des}C_B^*} \\ 
        \xi_B = \frac{K_{B,ads}/K_{B,des}C_B^*}{1+K_{A,ads}/K_{A,des}C_A^*+K_{B,ads}/K_{B,des}C_B^*} \\ 
    \end{cases}
\end{equation}

After solving the partial differential equation, the dimensionless flux is:
\begin{equation}
    J = -\left[\left(\frac{\partial C_A}{\partial C_B}\right)_{X=0} - \frac{1}{\beta}\left(\frac{d\xi_A}{dT}\right)\right]
\end{equation}

\begin{table}
    \centering
    \begin{tabular}{|c|c|}
        \toprule
        Parameter & Definition\\ \midrule
        Relative surface coverage & $\xi = \frac{\Gamma_j}{\Gamma_{max}}$ \\  \midrule
        Adsorption rate constant & $K_{j,ads}=\frac{k_{j,ads}\Gamma_{max}r_e}{D_A}$\\ \midrule
        Desorption rate constant & $K_{j,des}=\frac{k_{j,des}\Gamma_{max}r_e}{c^*_AD_A}$ \\ \midrule
        Saturation parameter & $p_s=\frac{c^*r_e}{\Gamma_{max}}$ \\ \midrule
        Electrochemical rate constant, solution species & $K_{red/ox} =\frac{k_{red/ox}r_e}{D_A}$ \\  \midrule
        Electrochemical rate constant, adsorbed species & $K_{red/ox,ads}=\frac{k_{red/ox,ads}}{r_e^2}$ \\ \midrule
    \end{tabular}
    \caption{Definition of dimensionless parameters for electrochemical reaction coupled with adsorption simulations. }
    \label{tab:AdsorptionDimensionlessTable}
\end{table}

\subsubsection{Model Validation}
\begin{figure}
    \centering
    \includegraphics[width=0.75\linewidth]{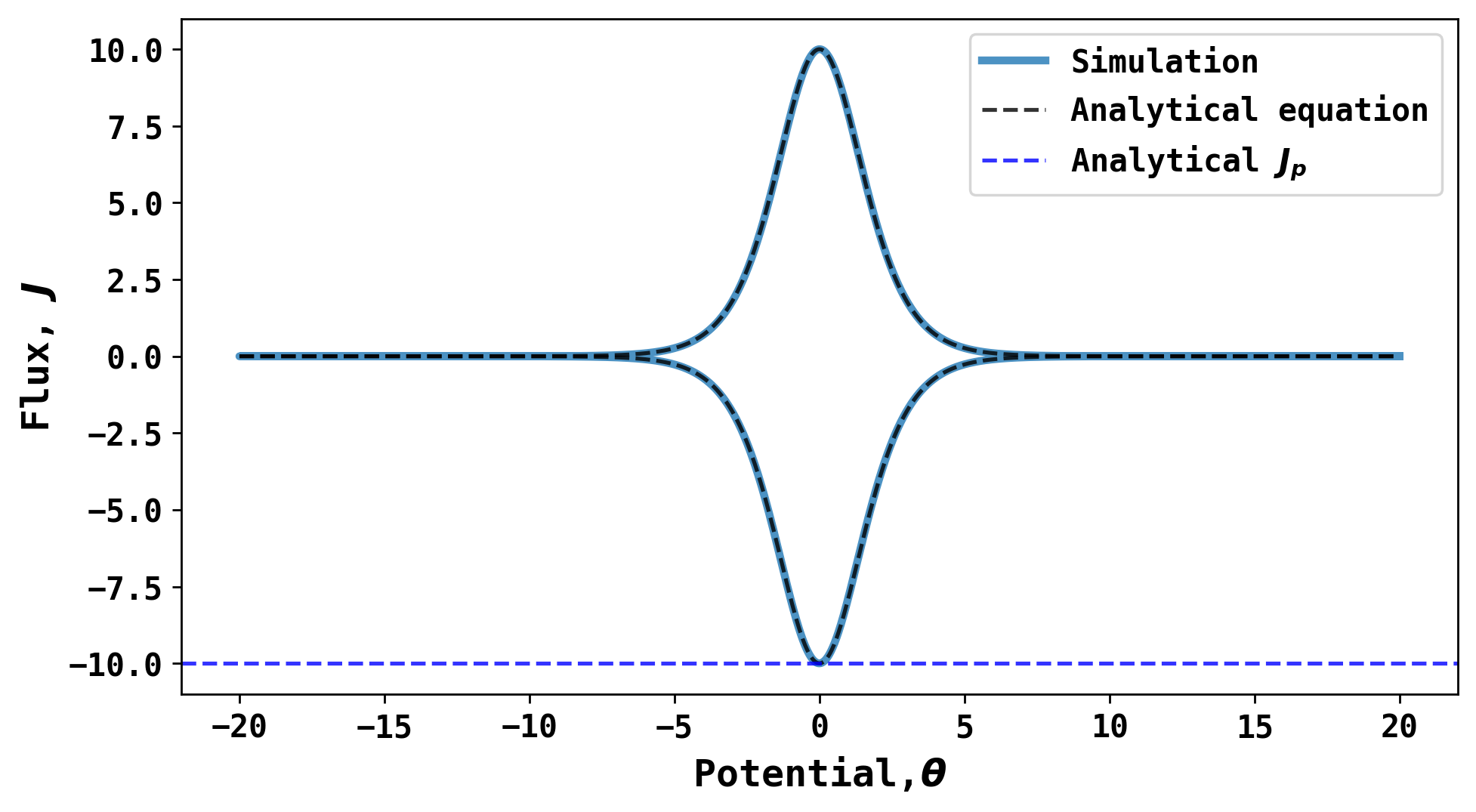}
    \caption{A sample voltammogram (blue curve) considering almost exclusively adsorptive behavior. The relevant simulation parameters are: $\sigma=40$, $K_0=0\text{ (no electrochemical reaction for solution phase species)}$, $K_{0,ads} \text{ (fully reversible electrochemical reaction for adsorbed species)}$, $K_{A,des}=K_{B,des}=10^{-3} \text{ (no desorption)}$, $K_{A,ads}=10^3$, $K_{B,ads}=1$. The black-dashed line is the analytical expression of voltammograms and the blue-dashed line is the analytical peak flux.}
    \label{fig:FullyAdsorptiveFigure}
\end{figure}

The analytical equation for a adsorptive voltammogram considering only a mono-layer of electroactive species $A$ undergoing a one-electron, fully reversible electron process, is:
\begin{equation}
    J = \mp \frac{\exp{\left(-\theta\right)}}{\left(1+\exp{\left(-\theta\right)}\right)^2}
\end{equation}
where the upper sign refers to the cathodic scan and the lower sign refers to the anodic scan. The analytical expression for the peak flux is:
$J_p=\mp \frac{\sigma}{\beta}$
where peak flux scales linearly with scan rates. A voltammetry simulation dominated by electrochemical reaction of an adsorbed monolayer is shown in Figure S \ref{fig:FullyAdsorptiveFigure}, which agrees well with fully adsorptive analytical equation and analytical peak flux. 
 
\clearpage
\section{Discretization Schemes}
\subsection{Discretizing NPP equations with BV or MHC kinetics}
Recall that the dimensionless migration diffusion equation is:
\begin{equation}
    \frac{\partial C_j } {\partial T}=d_j \frac{\partial^2 C_j}{\partial X^2}+d_j z_j  \left(\frac{\partial C_j}{\partial X}  \frac{\partial\Phi}{\partial X}+C_j  \frac{\partial ^2} {\Phi/\partial X^2}\right)
\end{equation}
With an expanding spatial grid and uniform time grid, the discretized mass transport equation for species $j$ at timestep $k$ is:
\begin{equation}
\begin{split}
    \frac{C_{i,j}^k-C_{i,j}^{k-1}}{\Delta T} &= d_j\left(\frac{\frac{C_{i+1,j}^{k}-C_{i,j}^{k}}{\Delta X_+}-\frac{C_{i,j}^k-C_{i-1,j}^k}{\Delta X_{+}}{}}{\frac{1}{2}\left[ \Delta X_+ + \Delta X_{-} \right]}\right) \\&+ d_jz_j\left(\frac{C_{i+1,j}^{k}-C_{i-1,j}^{k}}{\Delta X_+ + \Delta X_{-}} \frac{\Phi_{i+1}^{k}-\Phi_{i-1}^{k}}{\Delta X_+ + \Delta X_{-}} \right) \\&+ d_jz_jC_{i,j}^k\left(\frac{\frac{\Phi_{i+1}^k-\Phi_i^k}{\Delta X{+}} - \frac{\Phi_i^k-\Phi_{i-1}^k}{\Delta X_{-}}}{\frac{1}{2}\left[ \Delta X_+ + \Delta X_{-} \right]} \right)
\end{split}
\end{equation}
The discretized NPP equation is rewritten as follows for the implicit scheme:
\begin{equation}
    \begin{split}
        C_{i,j}^{k-1} &= \left( \alpha_i C_{i-1,j}^k+\beta_i C_{i,j}^k+\gamma_i C_{i+1,j}^k  \right) \\
        &+ z_j\delta_j\left( C_{i+1,j}^k \Phi_{i+1}^k-C_{i+1}^k \Phi_{i-1}^k-C_{i-1,j}^k \Phi_{i+1}^k+C_{i-1,j}^k \Phi_{i-1}^k  \right)\\
        &+ z_jC_{i,j}^k\left( \alpha_i \Phi_{i-1}^k+\beta_i \Phi_i^k+\gamma_i \Phi_{i+1}^k \right)\\
        \alpha_i &= -d_j\left(\frac{2\Delta T}{\Delta X_{-}^2 +\Delta x_{+}\Delta x_{-} }\right) \\ 
        \beta_i  &= d_j \left(\frac{2 \Delta T}{\Delta X_{+}^2 + \Delta X_{+}\Delta X_{-}} +\frac{2 \Delta T}{\Delta X_{-}^2 + \Delta X_{+}\Delta X_{-}} \right) \\ 
        \gamma_i &= -d_j\left(\frac{2\Delta T}{\Delta X_{+}^2 +\Delta x_{+}\Delta x_{-} }\right)\\ 
        \delta_i &= -d_j\left(\frac{\Delta T}{\left(\Delta X_{+} + \Delta X_{-}\right)^2}  \right) \\ 
    \end{split}
\end{equation}

\begin{figure}
    \centering
    \includegraphics[width=0.5\linewidth]{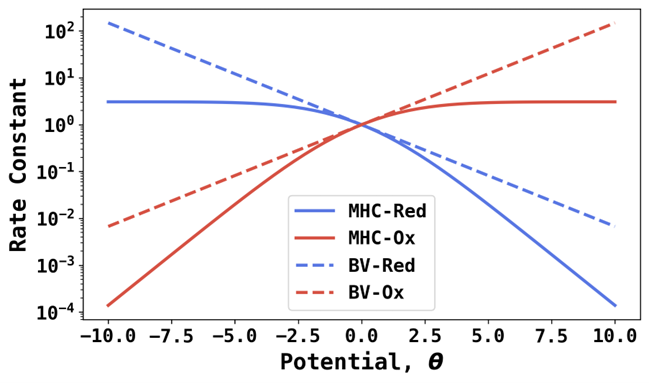}
    \caption{Comparing the electrochemical rate constants of MHC (solid) and BV (dashed) as a function of overpotential. The dimensionless rate constants are $K_0=1.0$ for both MHC and BV kinetics. $\alpha=\beta=0.5$ for BV kinetics and $\Lambda=1$ for MHC kinetics. }
    \label{fig:MHCvsBV_SI}
\end{figure}

Recall that the Poisson equation is:
\begin{equation}
    \frac{\partial^2 \Phi}{\partial X^2} = -Re^2 \sum_jz_jC_j
\end{equation}
The discretized Poisson equation is thus:
\begin{equation}
    \frac{\frac{\Phi_{i+1}^{k} - \Phi_{i}^k}{\Delta X_+} - \frac{\Phi_i^k - \Phi_{i-1}^k}{\Delta X_{-}}}{\frac{1}{2}\left[\Delta X_{+} + \Delta X_{-}  \right]} = -Re^2 \sum_jz_jC_j
\end{equation}
which can be rewritten as:
\begin{equation}
    \begin{array}{cc}
         \alpha_i\Phi_{i-1}^k + \beta_i\Phi^k_{i} + \gamma_i\Phi_{i+1}^{k} + Re^2\sum_jz_jC_{i,j} = 0   \\
         \alpha_i = \frac{2}{\Delta X_{-}^2 + \Delta X_{+}\Delta X_{-}}  \\ 
         \beta_i = -\left(\frac{2}{\Delta X^2 + \Delta X_{+}\Delta X_{-}}\right) \\ 
         \gamma_i = \frac{2}{\Delta X_{+}^2 + \Delta X_{+}\Delta X_{-}} \\ 
        
    \end{array}
\end{equation}
Next, the electrochemical boundary conditions are also discretized. For Butler-Volmer kinetics:
\begin{equation}
\begin{split}
    d_A\left(\frac{\partial C_A}{\partial X}\right)_{X=0} &\approx d_A\left(\frac{C_{i=1,A}^k-C^k_{i=0,j}}{\Delta X}\right) \\
    &=K_0\exp\left(-\alpha\left(\theta^k-\Phi_{i=0}^k\right)\right)C_{i=0,A}^k - K_0\exp\left(\beta\left(\theta^k-\Phi_{i=0}^k\right)\right)C_{i=0,B}^k
\end{split}
\end{equation}
For Marcus-Hush-Chidsey kinetics, 
\begin{equation}
\begin{split}
    d_A\left(\frac{\partial C_A}{\partial X}\right)_{X=0} &\approx d_A\left(\frac{C_{i=1,A}^k-C^k_{i=0,j}}{\Delta X}\right) \\
    &= K_{red}^{MHC}\left(\theta^k,\Phi_{i=0}^k,\Lambda\right)C_{i=0,A}^k - K_{ox}^{MHC}\left(\theta^k,\Phi_{i=0}^k,\Lambda\right) \\
    K_{red}^{MHC}&= K_0 \frac{I\left(\theta^k-\Phi_{i=0}^k,\Lambda \right)}{I\left(0 - \Phi_{i=0}^k, \Lambda\right)} \\  
    K_{ox}^{MHC} &= K_0\frac{\left(\theta - \Phi_{i=0}^k,\Lambda\right)}{\left(0-\Phi_{i=0}^k,\Lambda\right)}
\end{split}
\end{equation}
A supplementary figure illustrating the behavior of BV and MHC kinetics is shown in Figure S \ref{fig:MHCvsBV_SI} . Notably, BV rate constants grow exponentially with overpotential and MHC kinetics plateaued at high overpotential. 

\subsection{Discretizing the convection diffusion equation}
\begin{figure}
    \centering
    \includegraphics[width=0.5\linewidth]{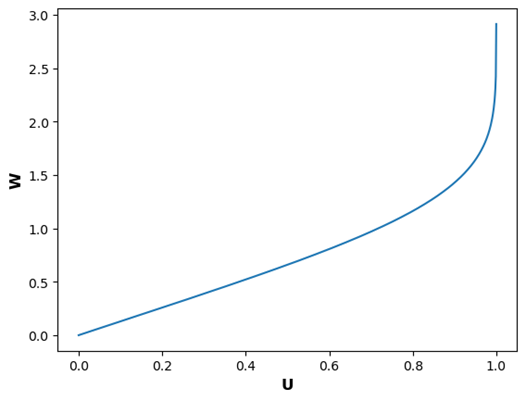}
    \caption{The numerical mapping between $U$ and $W$.}
    \label{fig:UandW}
\end{figure}

The convection-diffusion problem is solved with backward implicit scheme such that for species $A$ at the spatial index $i$ and time index $k$,\cite{VoltammetrySimulation} the solution is given by:
\begin{equation}
    \frac{C_i^k - C_i^{k-1}}{\Delta T} = \frac{\exp\left(-\frac{2}{3}W_i^3\right)}{1.65894}\left(\frac{C_{i-1}^k - 2C_i^k + C_{i+1}^k}{\Delta U^2}\right)
\end{equation}
To numerically solve this equation, we first need to establish the correlation between $U$ and $W$ to calculate the factor $\exp\left(-\frac{2}{3}W_i^3\right)$ for each $U_i$ update. The mapping between $U$ and $W$ is solved using scipy.odeint package and shown in Figure S \ref{fig:UandW}. 

The discretized equation above can be rewritten as:
\begin{equation}
\begin{split}
    \alpha_{j,i} &= -\frac{\exp{\left(-\frac{2}{3}W_i^3\right)}}{1.65894}\frac{\Delta T}{\Delta U^2} \\ 
    \beta_{j,i} &= 2\frac{\exp{\left(-\frac{2}{3}W_i^3\right)}}{1.65894}\frac{\Delta T}{\Delta U^2} +1 \\ 
    \gamma_{j,i} &= -\frac{\exp{\left(-\frac{2}{3}W_i^3\right)}}{1.65894}\frac{\Delta T}{\Delta U^2} \\ 
    \delta_{j,i}^k &= C_{j,i}^{k-1}
\end{split}
\end{equation}

Once the concentration profile has been obtained, the current response is calculated from:
\begin{equation}
    J_{RDE} = -\left(\frac{\partial C}{\partial U}\right)_{U=0} =  \frac{I\sqrt{1.65894}}{FAc^*\left(D^2L\right)^{\frac{1}{3}}}
\end{equation}

\section{Differentiable Electrochemistry for Voltammetry of Adsorbed Species}
Parameter estimation for voltammetry of adsorbed species, is considered a tremendous challenge due to the large parameter space and the associated curse of dimensionality.\cite{BondChemElectroChem} 
This section illustrates the use of Differentiable Electrochemistry for parameter estimation for voltammetry of adsorbed species. The model for simulating voltammetry of adsorbed species is described in section \ref{AdsorptionTheory}. According to the mode, a total of \textbf{10} parameters will be estimated simultaneously from a set of voltammograms. The list of parameters and their symbols are shown in Table S \ref{tab:AdsorptionDiffEC_List_of_Parameters}. This challenge aims to evaluate the capability of Differentiable Electrochemistry for advanced challenges in electrochemistry and electrochemical energy research, where multi-parameter estimation is demanded.

\begin{table}
    \centering
    \begin{tabular}{|p{0.75\linewidth}|p{0.25\linewidth}|}
    \toprule
         Parameter & Symbol \\ \midrule
         Solution phase electrochemical rate constant in logarithmic scale& $\log_{10}{K_0}$ \\ \midrule
         Solution phase cathodic reaction transfer coefficient& $\alpha$ \\ \midrule
         Solution phase anodic reaction transfer coefficient & $\beta$ \\ \midrule
         Adsorbed species electrochemical rate constant in logarithmic scale& $\log_{10}{K_{0,ads}}$ \\ \midrule
         Adsorbed species cathodic transfer coefficient & $\alpha_{ads}$ \\ \midrule
         Adsorbed species anodic transfer coefficient& $\beta_{ads}$\\ \midrule
         Adsorption rate constant of $A$ in logarithmic scale& $\log_{10}{K_{A,ads}}$ \\ \midrule
         Desorption rate constant of $A$ in logarithmic scale&  $\log_{10}{K_{A,des}}$ \\ \midrule
         Adsorption rate constant of $B$ in logarithmic scale& $\log_{10}{K_{B,ads}}$\\ \midrule
         Desorption rate constant of $B$ in logarithmic scale& $\log_{10}{K_{B,des}}$\\ \midrule
    \end{tabular}
    \caption{The list of parameters that for parameter estimation with Differentiable Electrochemistry.}
    \label{tab:AdsorptionDiffEC_List_of_Parameters}
\end{table}

\begin{figure}
    \centering
    \includegraphics[width=1.0\linewidth]{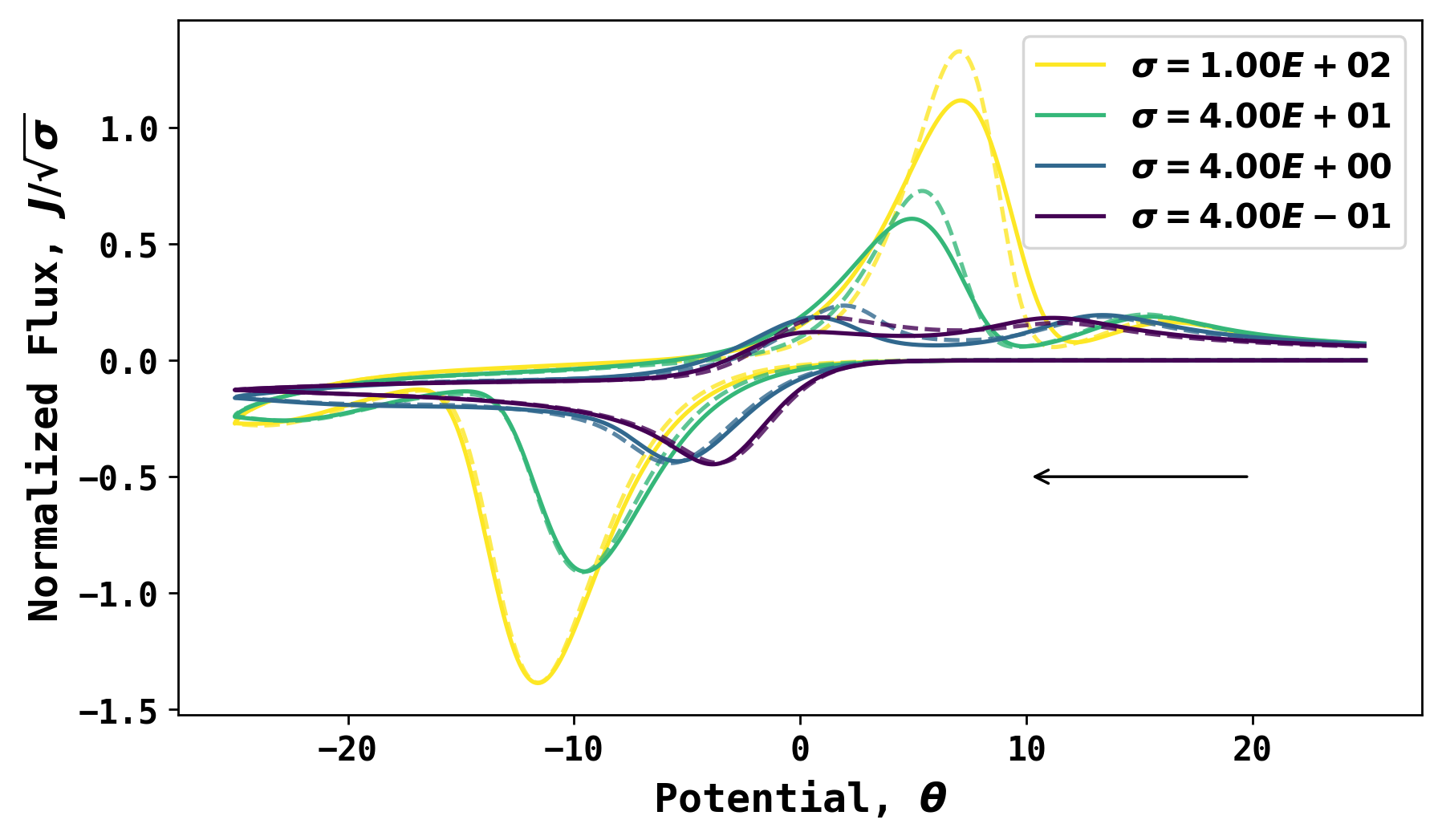}
    \caption{Ground truth (dashed) and fitted (solid) voltammograms for parameter estimation in voltammetry of adsorbed species. The black arrow indicates the direction of scan. }
    \label{fig:DiffECAdsorptionResult}
\end{figure}

To start with, the ground truth voltammograms are generated and shown  Figure S \ref{fig:DiffECAdsorptionResult} as the dashed voltammogram curves. Instead of fitting a single voltammogram (a common practice), there are four voltammograms at different dimensionless scan rates to be fitted simultaneously using Differentiable Electrochemistry to relieve the possibility of non-unique solution. The dimensionless scan rates are: 100, 40, 4 and 0.4. As shown in Figure S \ref{fig:DiffECAdsorptionResult}, the presence of a weak postwave suggest that the reactant adsorbs stronger than the product (i.e. $K_{A,ads}>K_{B,ads}$). The large scale difference of the voltammetric fluxes at different scan rates hints the difficulty of parameter estimation.

\begin{figure}
    \centering
    \includegraphics[width=1.0\linewidth]{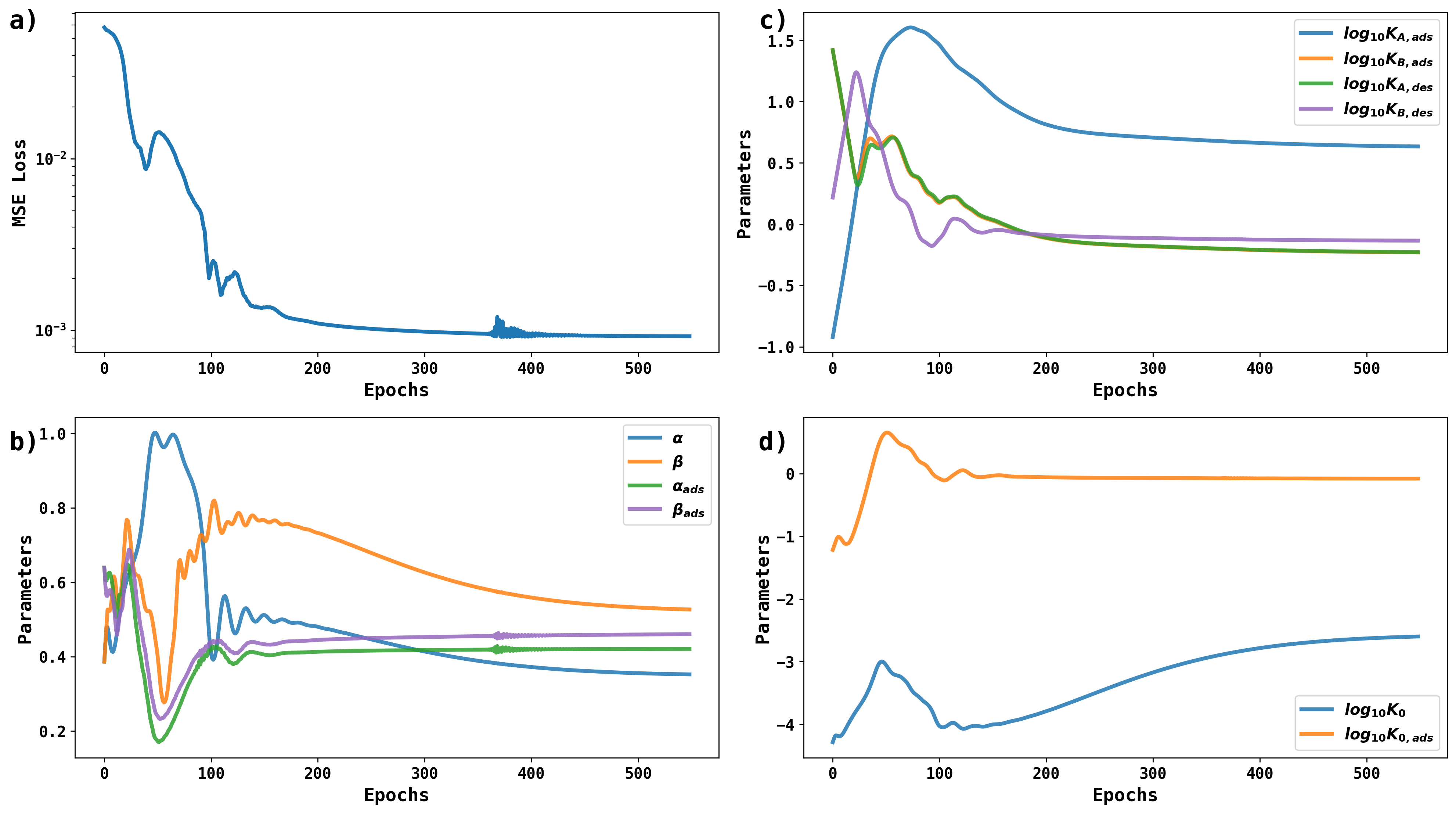}
    \caption{The optimization trajectories for Differentiable Electrochemistry enabled gradient-based discovery of \textbf{10} parameters. (a) The mean squared error (MSE) loss during optimization. (b) The transfer coefficients trajectories during optimization. (c) The adsorption/desorption rate constants trajectories during optimization. (d) The electrochemical rate constants trajectories during optimization.  }
    \label{fig:DiffECAdsorptionHistory}
\end{figure}

Using Differentiable Electrochemistry, \textbf{10} parameters (see Table S \ref{tab:AdsorptionDiffEC_List_of_Parameters}) are simultaneously discovered by simulating voltammograms with guesses, calculating the mean squared error (MSE), calculating the gradients of MSE with respect to parameters using Automatic Differentiation, and lastly gradient-based optimization. A total of 550 optimization steps are performed with a learning rate of 0.05 and an Adam optimizer. The training history about the losses and parameters are shown in Figure S \ref{fig:DiffECAdsorptionHistory}. According to Figure S \ref{fig:DiffECAdsorptionHistory}a, the MSE loss drops by 2 orders of magnitude to $\sim 10^{-3}$ at the end of Differentiable Electrochemistry optimization. The histories transfer coefficients, adsorption/desorption rate constants, and electrochemical rate constants, are shown in Figure S \ref{fig:DiffECAdsorptionHistory}b--d. As shown in Figure S \ref{fig:DiffECAdsorptionHistory}c, $K_{A,ads}$ is indeed larger than $K_{B,ads}$, qualitatively supported by the presence of postwave.  According to Figure S \ref{fig:DiffECAdsorptionHistory}, the loss and most parameters stabilize after 400 epochs, and the resulting voltammograms given the estimated parameters are shown in Figure S \ref{fig:DiffECAdsorptionResult} as the solid lines. The voltammograms fit well in the forward scans, and the shapes of reverse scans are roughly captured. The MSE loss is $\sim9\times 10^{-4}$ between ground truth and simulations. The good performance of Differentiable Electrochemistry for estimation of \textbf{10} parameters by fully differentiating voltammetry with adsorbed species, a highly nonlinear time-stepping problem, shows the initial potential for more complicated challenges in parameter discoveries. This study is not to suggest against the use of gradient-free optimization, but to suggest a synergy between them. For example, gradient-free optimization can be used to explore the loss landscape first, and gradient-based optimization, enabled by Differentiable Electrochemistry, can more easily locate the global optimum. 

In conclusion, by showing Differentiable Electrochemistry enabled gradient-based optimization for a \textbf{10}--parameter discovery, Differentiable Electrochemistry highlight the novelty of making black-box simulator differentiable, and justifies as a new regime in electrochemical simulation.

\section{Differentiable Electrochemistry for $\ch{Fe^{3+}}/\ch{Fe^{2+}}$ Redox Couple }

This section introduces the practice of differentiable electrochemistry simulations for parameter estimation in $\ch{Fe^{3+}}/\ch{Fe^{2+}}$ redox couple.  The five voltammograms at five different scan rates will be differentiated with respect to parameters of interest for gradient-based optimization. In this study, four parameters are discovered: electrochemical rate constant ($k_0$), cathodic transfer coefficient ($\alpha$), anodic transfer coefficient ($\beta$), and average diffusion coefficient $D_{avg}$. The list of simulation parameters and parameters for discovery are shown in Table S \ref{tab:DiffECFe_List_of_Parameters}.

\begin{table}
    \centering
    \begin{tabular}{|c|c|}
    \toprule
         Radius of electrode, $r_e$& 0.85 mm\\ \midrule
         Bulk concentration of $\ch{Fe^{3+}}$, $c^*_{\ch{Fe^{3+}}}$& 4.85 mM \\ \midrule
         Start potential vs. SCE, $E_i$ & 0.8 V\\ \midrule
         Reverse potential vs. SCE, $E_v$ & 0.1 V\\ \midrule
         Reference potential vs. SCE, $E_{ref}$ & 0.4336 V  \\ \midrule
         Scan rate, $\nu$ & [0.01, 0.02, 0.05, 0.1, 0.2] V/s \\ \midrule
         Electrochemical rate constant, $k_0$& Estimated by Differentiable Electrochemistry \\ \midrule
         Cathodic transfer coefficient, $\alpha$& Estimated by Differentiable Electrochemistry \\ \midrule
         Anodic transfer coefficient, $\beta$& Estimated by Differentiable Electrochemistry \\ \midrule
         Average diffusion coefficient, $D_{avg}$& Estimated by Differentiable Electrochemistry \\ \midrule
    \end{tabular}
    \caption{A list of simulation parameters and parameters determined by Differentiable Electrochemistry for $\ch{Fe^{3+}}/\ch{Fe^{2+}}$ redox chemistry. }
    \label{tab:DiffECFe_List_of_Parameters}
\end{table}

\clearpage
\section{Differentiable Electrochemistry with Noisy, Sparse, or Partial Data}
It is of paramount importance to evaluate the performance of Differentiable Electrochemistry when experimental data is noisy, sparse, or incomplete. Such evaluations not only explores the applicability of Differentiable Electrochemistry in realistic research environment, but also builds confidence for researchers to apply such methodology for their domain problems. In this section, the robustness of Differentiable Electrochemistry is evaluated with the voltammograms of $\ch{Fe^{3+}}/\ch{Fe^{2+}}$ redox couple when the voltammograms are noisy, sparse, or incomplete. 

\subsection{Noisy Data}
\begin{figure}
    \centering
    \includegraphics[width=0.75\linewidth]{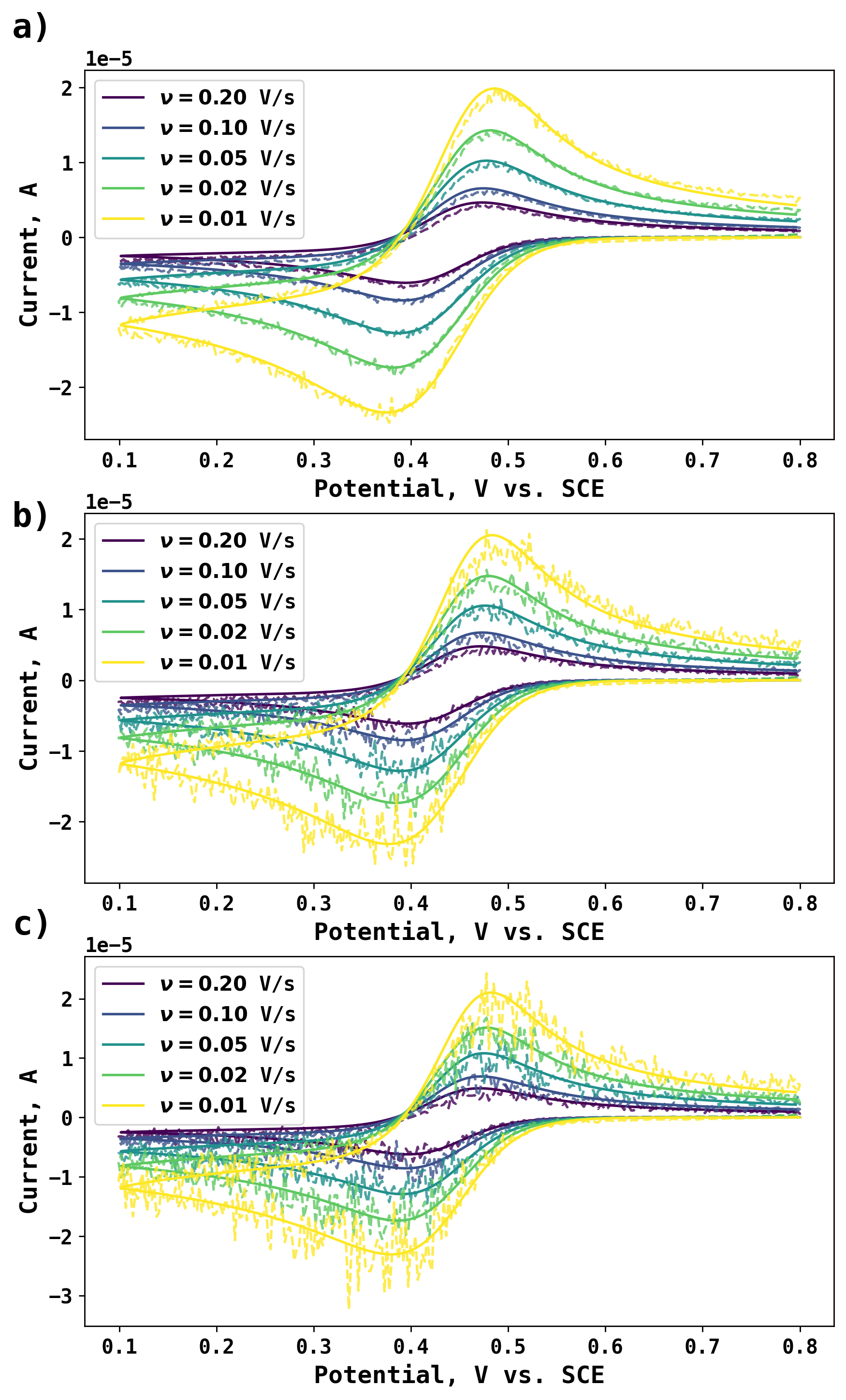}
    \caption{Differentiable Electrochemistry fitting (solid lines) to noisy experimental voltammograms (dashed lines) of $\ch{Fe^{3+}}/\ch{Fe^{2+}}$ redox couple at different scan rates. The multiplicative noise level at (a) $\eta=3\%$,  (b)$\eta=10\%$ , (c)  $\eta=16\%$. }
    \label{fig:DiffECNoiseResult}
\end{figure}

To evaluate the noise robustness of Differentiable Electrochemistry, different levels of multiplicative Gaussian noises are added onto the voltammogram as:
\begin{equation}
    I_{noisy}\left(t\right) = I\left(t\right)\times(1+\eta z\left(t\right)
\end{equation}
where $\eta$ is the relative level of noise ranged from $3\%$ to $16\%$. $z\left(t\right)$ is the standard normal distribution $\mathcal{N}\left(0,1\right)$. The Gaussian noise blended into the voltammograms represents the thermal or shot noise from environment. Testing Differentiable Electrochemistry performance in such scenario helps to evaluate the real-world applicability of Differentiable Electrochemistry for more advanced applications. Figure S \ref{fig:DiffECNoiseResult} shows experimental voltammograms in dashed lines  with increasing level of noise ($3\%$ to $16\%$) from a--c. 

\begin{figure}
    \centering
    \includegraphics[width=1\linewidth]{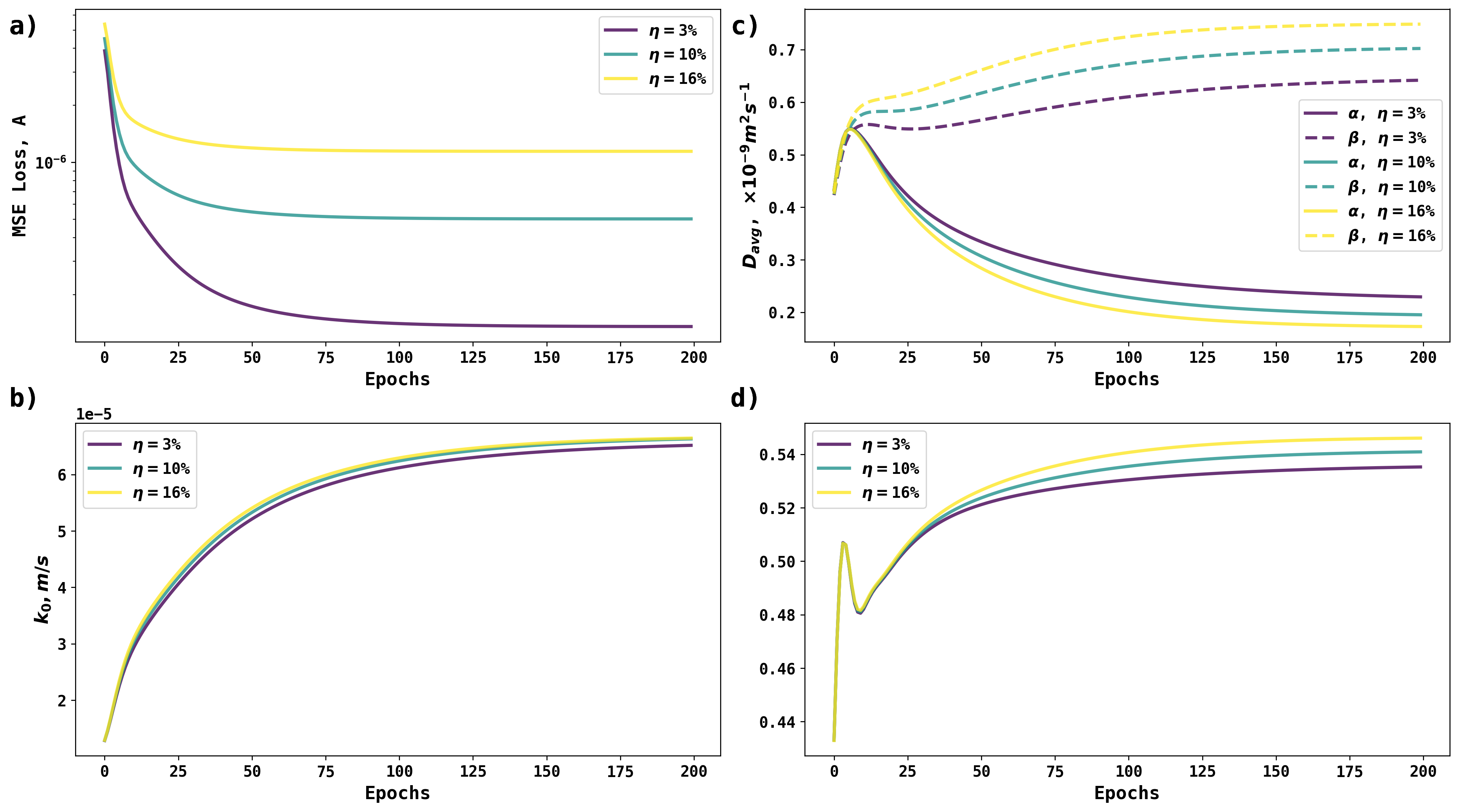}
    \caption{The optimization trajectories of Differentiable Electrochemistry of parameter estimation of $\ch{Fe^{3+}}/\ch{Fe^{2+}}$ couple at different noise levels. (a) Mean squared error of current, (b) electrochemical rate constant, (c) transfer coefficients, and (d) diffusion coefficient trajectories are shown.}
    \label{fig:DiffECFeRedoxEnsembleNoise}
\end{figure}

The optimization trajectories as a function of multiplicative Gaussian noise ($\eta$) is shown in Figure S \ref{fig:DiffECFeRedoxEnsembleNoise}. Figure S \ref{fig:DiffECFeRedoxEnsembleNoise}a shows the MSE loss of current prediction as a function of $\eta$. As expected, increasing $\eta$ increases the aleatoric uncertainty in the data and leads to higher MSE. Figure S \ref{fig:DiffECFeRedoxEnsembleNoise}b examines the prediction of $k_0$ as a function of $\eta$, and shows that prediction of electrochemical rate constant is almost unaffected by $\eta$. The predicted $k_0$ when $\eta=16\%$ is $k_0={6.64\times10^{-5}\ \mathrm{m/s}}$, close to $k_0=(6.54\pm0.02)\times10^{-5}\ \mathrm{m/s}$ when $\eta=0$. The predicted transfer coefficients, however, are more susceptible to noise levels, as shown in Figure S \ref{fig:DiffECFeRedoxEnsembleNoise}c. With increasing $\eta$, the extracted $\alpha$ decreases while the extracted $\beta$ increases. At $\eta=16\%$, the predicted $\alpha$ and $\beta$ are $0.17$ and $0.75$, respectively. Figure S \ref{fig:DiffECFeRedoxEnsembleNoise}b shows the positive influence of $\eta$ on average diffusion coefficients, partly by increasing the absolute peak magnitude. $D_{avg}=5.5\times10^{-9}\ \mathrm{m^2s^{-1}}$ when $\eta=16\%$, close to $D_{avg}=(5.33\pm0.01)\times10^{-9}\ \mathrm{m^2s^{-1}}$ when $\eta=0$. By evaluating the performance of Differentiable Electrochemistry at different level of noises, we show that Differentiable Electrochemistry predictions of $k_0$ and $D$ are resilient to noise, while the predictions of transfer coefficients are more susceptible to the levels of noises. In general, Differentiable Electrochemistry simulation and optimization offers a valuable solution to parameter estimation in experimental electrochemical responses. 


\clearpage
\subsection{Sparse Data}
In voltammetric studies, currents are usually collected at very small potential steps at the scale of $\sim 1\ \mathrm{mV}$. It is also important to understand how Differentiable Electrochemistry perform when data is very sparse and collected at a wide potential interval. This test is very important when sparse data is collected due to limited sampling frequency. Hence, the sparse data scenario is introduced to evaluate Differentiable Electrochemistry handling of sparse data with large potential steps. 

\begin{figure}
    \centering
    \includegraphics[width=0.7\linewidth]{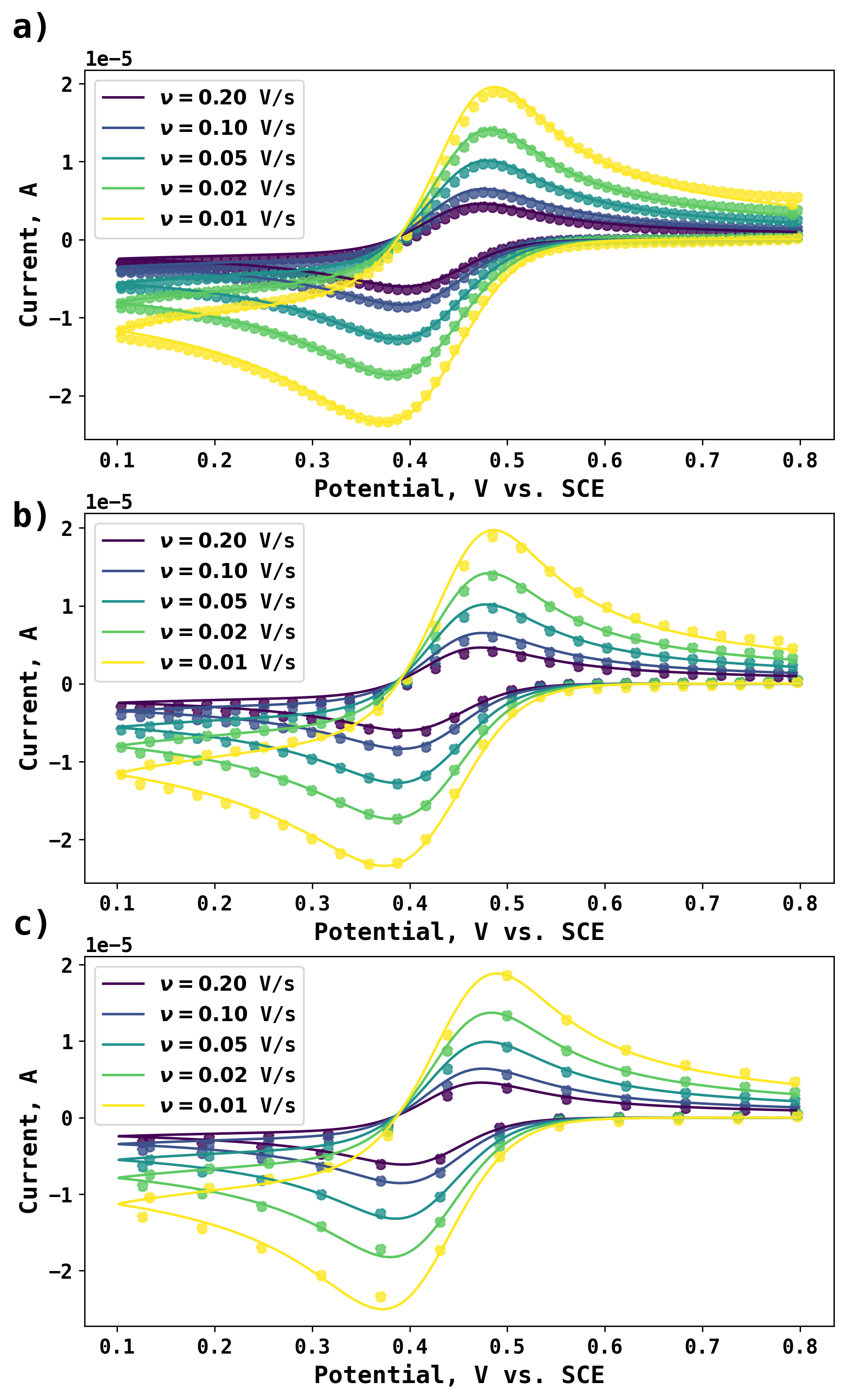}
    \caption{Differentiable Electrochemistry fitting to sparse experimental voltammograms of $\ch{Fe^{3+}}/\ch{Fe^{2+}}$ redox couple at different scan rates. (a) The potential step is $\Delta E = 20\ \mathrm{mV}$. (b) The potential step is $\Delta E = 60\ \mathrm{mV}$. (c) The potential step is $\Delta E = 125\ \mathrm{mV}$. }
    \label{fig:DiffECPartialResult}
\end{figure}

The experimental voltammograms collected at a potential step of $\Delta E= 2.5\ \mathrm{mV}$, and are downsampled to potential steps of $\Delta E = 20, 60, 125 \ \mathrm{mV}$. The Differentiable Electrochemistry fitting to experimental voltammograms at different potential steps is shown in Figure S \ref{fig:DiffECPartialResult} with increasing potential steps from a--c. It is observed that even at $\Delta E = 125\ \mathrm{mV}$, Differentiable Electrochemistry simulator is still able to calculate the gradients of the very sparse experimental data for optimization, and achieving qualitatively satisfactory fitting. 

\begin{figure}
    \centering
    \includegraphics[width=1\linewidth]{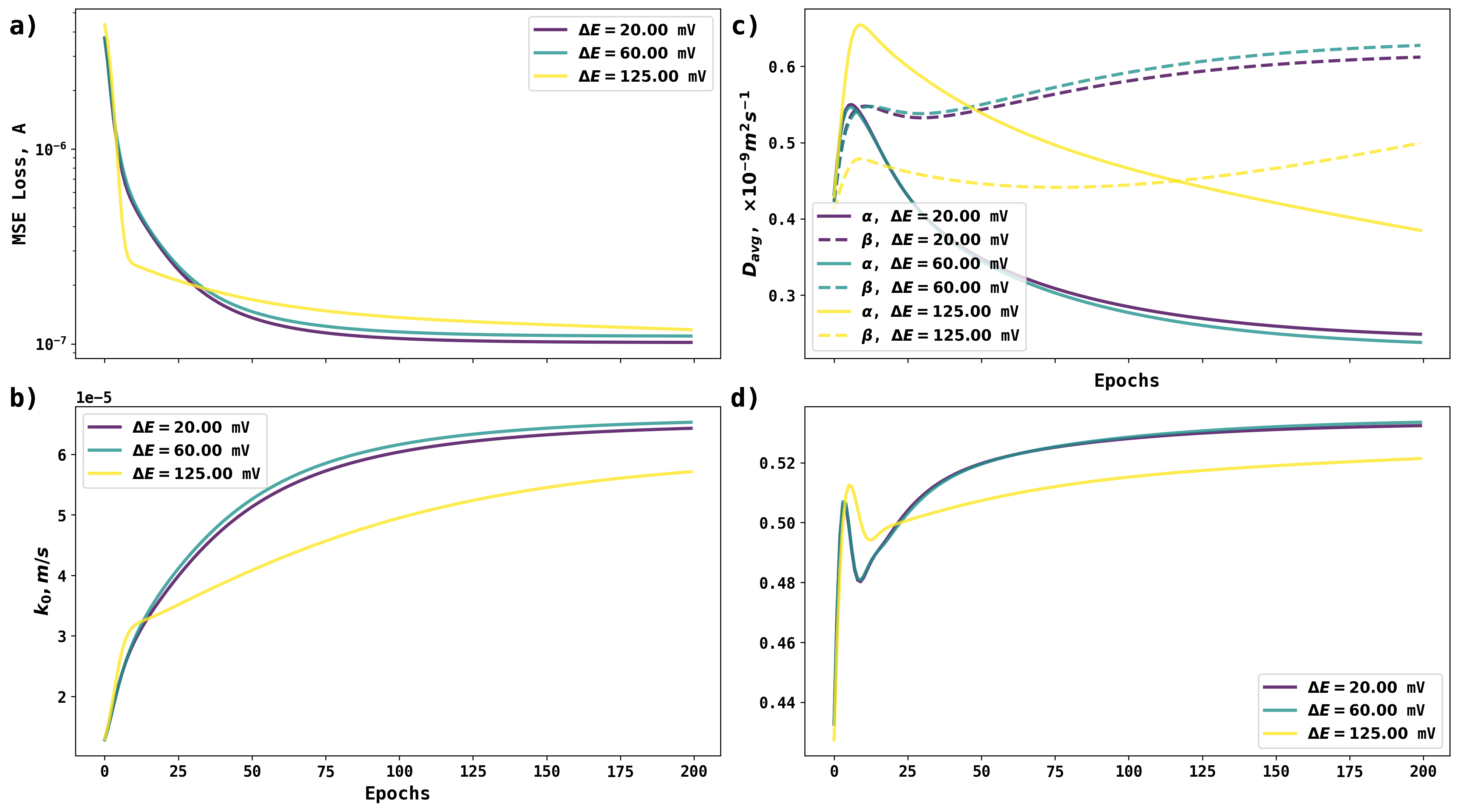}
    \caption{The optimization trajectories of Differentiable Electrochemistry of parameter estimation of $\ch{Fe^{3+}}/\ch{Fe^{2+}}$ couple at different potential step sizes at $\Delta E = 20, 60, 125\ \mathrm{mV}$. (a) Mean squared error of current, (b) electrochemical rate constant, (c) transfer coefficients, and (d) diffusion coefficient trajectories are shown. }
    \label{fig:DiffECFeRedoxEnsemblePartial}
\end{figure}

A more quantitative illustration of Differentiable Electrochemistry parameter estimation with sparse data is shown in Figure S \ref{fig:DiffECFeRedoxEnsemblePartial}. In Figure S \ref{fig:DiffECFeRedoxEnsemblePartial}, the optimization trajectories are almost identical when the potential step is at 20 or 60 mV. The MSE losses increase from $1.0\times10^{-7}\ \mathrm{A}$ when $\Delta E = 20\ \mathrm{mV}$ to $1.2\times10^{-7}\ \mathrm{A}$ when $\Delta E = 125\ \mathrm{mV}$, showing that Differentiable Electrochemistry fitting errors are relatively insensitive to potential steps. The predictions of parameters when $\Delta E = 125\ \mathrm{mV}$ are quite different from parameter predictions when $\Delta E$ is smaller, possibly because of the scarcity of data points near formal potential. 

In conclusion, by applying Differentiable Electrochemistry simulations to sparse experimental voltammograms of $\ch{Fe^{3+}}/\ch{Fe^{2+}}$ redox couple, we show that Differentiable Electrochemistry indeed calculates and traces every data point back to its parameter, offering a robust parameter estimation approach even when data is as sparse as $\Delta E = 60\ \mathrm{mV}$.

\clearpage
\subsection{Partial Data }
This scenario tests Differentiable Electrochemistry when the data is incomplete. This scenario shows that Differentiable Electrochemistry can generate gradients with partial experimental data. This scenario reflects the flexibility of Differentiable Electrochemistry and justifies as a new regime for electrochemistry simulation.  

\begin{figure}
    \centering
    \includegraphics[width=0.8\linewidth]{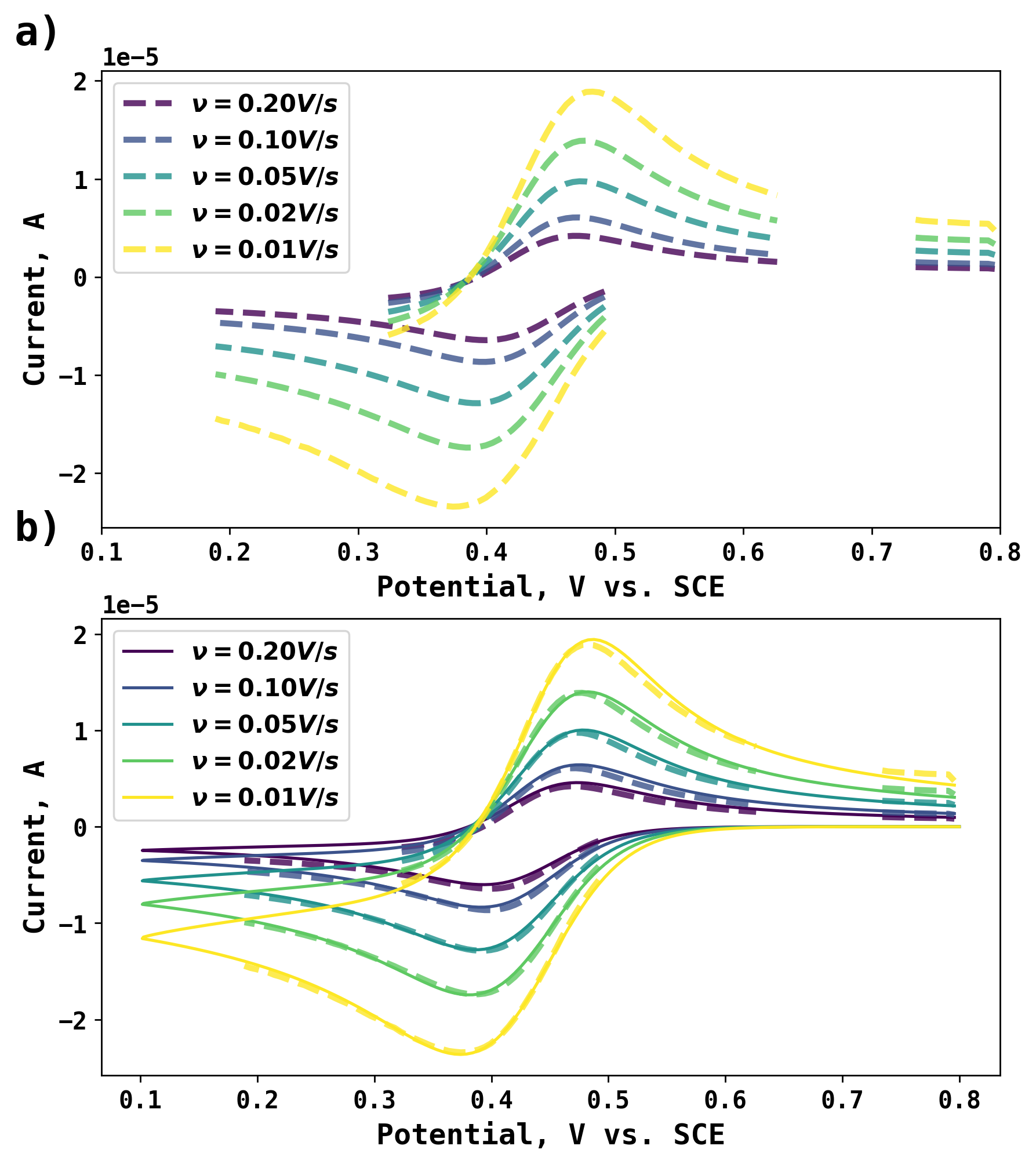}
    \caption{Differentiable Electrochemistry with partial experimental voltammograms. (a) The partial experimental voltammograms and (b) Differentiable Electrochemistry fitting of partial experimental voltammograms. }
    \label{fig:DiffECPartialResult2}
\end{figure}

\begin{figure}
    \centering
    \includegraphics[width=1.0\linewidth]{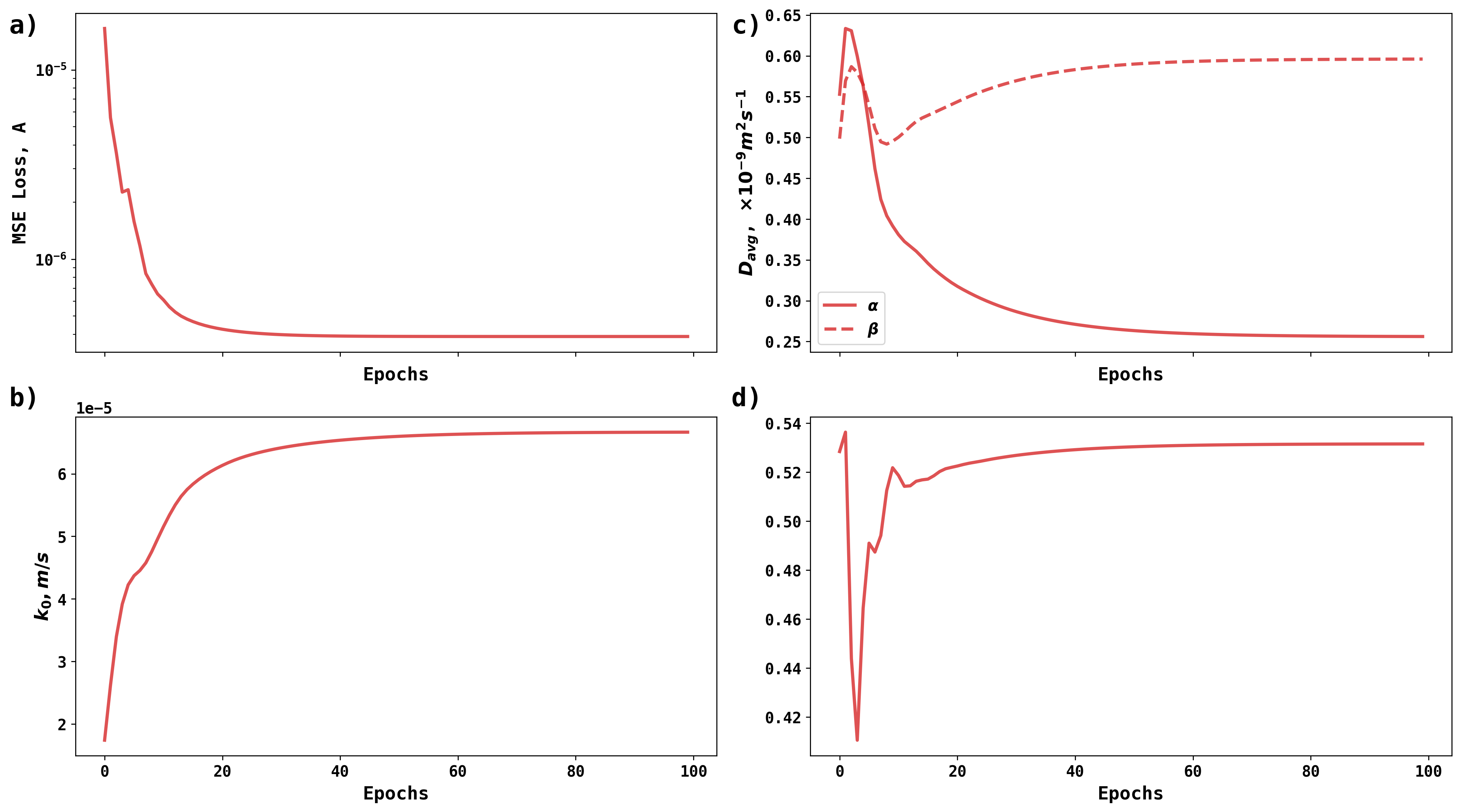}
    \caption{The Differentiable Electrochemistry learning trajectories with partial experimental voltammograms. (a) Mean squared error of current, (b) electrochemical rate constant, (c) transfer coefficients, and (d) diffusion coefficient trajectories are shown.}
    \label{fig:DiffECFeRedoxEnsemblePartial2}
\end{figure}

Parts of voltammograms are intentionally discarded to represent the partial data scenario. It can be visualized in Figure S \ref{fig:DiffECPartialResult2}a, where partial experimental voltammograms are shown as the ground truth for Differentiable Electrochemistry. As shown in Figure S \ref{fig:DiffECPartialResult2}a, parts of the forward and revers scan peaks are included, along with a short wave near the end of reverse scan. Using Differentiable Electrochemistry, the fitted voltammograms are shown in Figure S \ref{fig:DiffECPartialResult2}b, agreeing well with the partial experimental voltammograms at all five scan rates. The learning trajectories with partial experimental voltammograms are shown in Figure S \ref{fig:DiffECFeRedoxEnsemblePartial2}, where the MSE loss is $3\times 10^{-7}\ \mathrm{A}$, slightly larger than the loss with full voltammograms ($10^{-7}\ \mathrm{A}$). The predicted parameters with partial voltammograms are: $k_0=6.66\times10^{-5}\ \mathrm{m/s}$, $\alpha=0.256$, $\beta=0.596$, $D_{avg}=5.31\times10^{-10}\ \mathrm{m^2s^{-1}}$. The predicted results only slightly different from  Differentiable Electrochemistry with the full experimental voltammograms, highlighting the applicability of Differentiable Electrochemistry in various challenging scenarios, including partial experimental data and distinguish itself from conventional methods.

\clearpage
\section{Differentiable Electrochemistry for $\ch{Ru(NH3)_6^{3+}}$/$\ch{Ru(NH3)_6^{2+}}$ Redox Couple}
\begin{figure}
    \centering
    \includegraphics[width=0.8\linewidth]{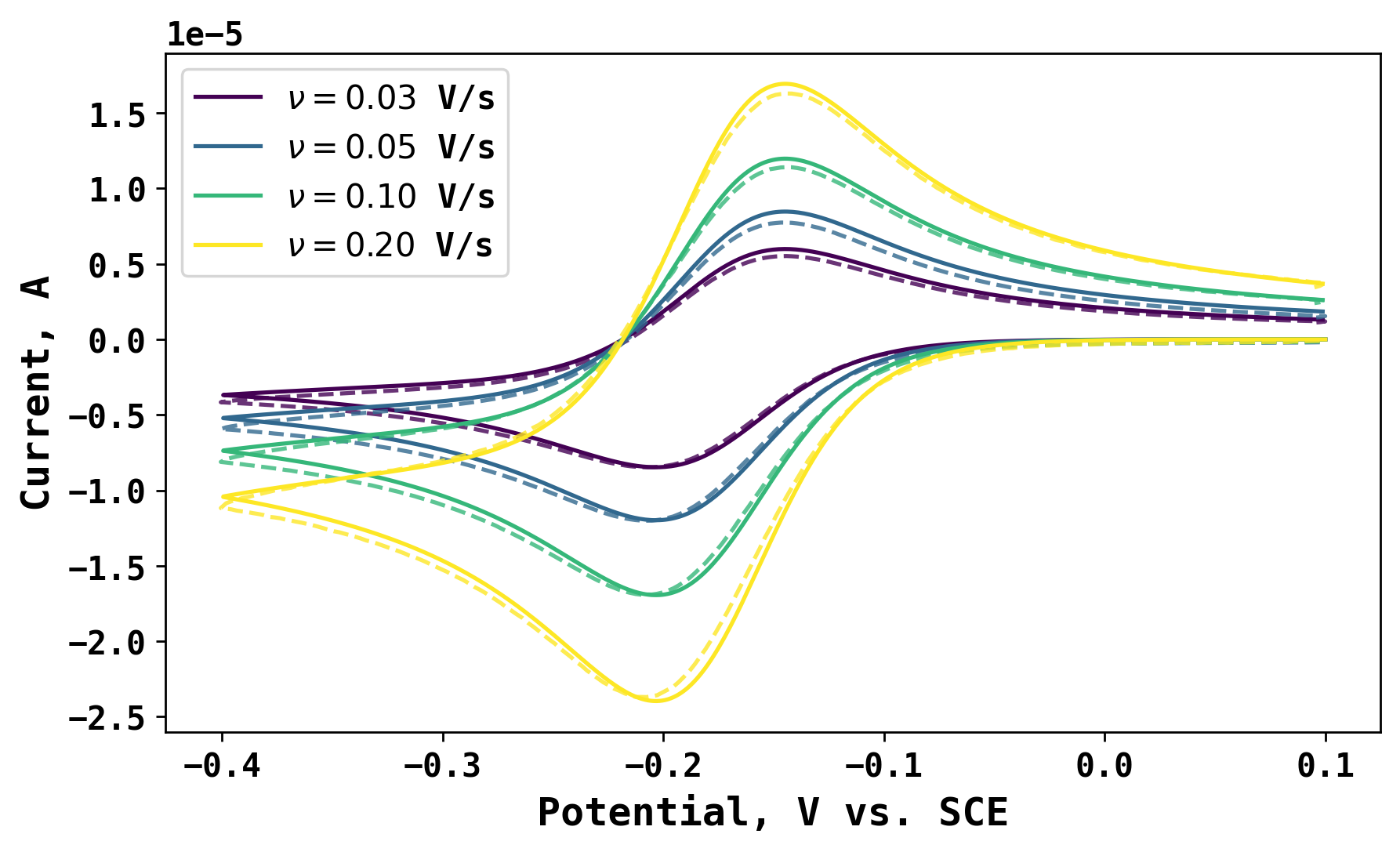}
    \caption{Experimental voltammograms (dashed lines) of $\ch{Ru(NH3)_6^{3+}}$/$\ch{Ru(NH3)_6^{2+}}$ redox coupled are compared with simulated voltammograms using parameters estimated by Differentiable Electrochemistry simulation and optimization (solid lines).  }
    \label{fig:DiffECRuHexRedoxResult}
\end{figure}

This section introduces the practice of differentiable electrochemistry simulations for parameter estimation in the $\ch{Ru(NH3)_6^{3+}}$/$\ch{Ru(NH3)_6^{2+}}$ redox couple. The voltammograms are from ref.\cite{chen2024discovering} and shown in Figure S \ref{fig:DiffECRuHexRedoxResult} as dashed lines, which are the ground truth for Differentiable Electrochemistry. In this study, four voltammograms at four different scan rates are differentiated with respect to their parameters for gradient-based optimization.  Two parameters are discovered: the formal potential $(E^0_f)$ and the average diffusion coefficient $(D_{avg})$ of the two electroactive species.  The list of simulation parameters and parameters for discovery are shown in Table S \ref{tab:RuHexRedoxParameterList}. 

\begin{table}
    \centering
    \begin{tabular}{|c|c|}
    \toprule
        Radius of electrode, $r_e$& $1.5\ \mathrm{mm}$  \\ \midrule
        Bulk concentration of $\ch{Ru(NH3)_6^{3+}}$, $c_{\ch{Ru(NH3)_6^{3+}}}^*$ & $0.96\ \mathrm{mM}$ \\ \midrule 
        Start potential vs. SCE, $E_i$ & $0.1\ \mathrm{V}$ \\ \midrule 
        Reverse potential vs. SCE, $E_v$ & $-0.4\ \mathrm{V}$ \\ \midrule 
        Scan rates & $[0.025,\ 0.05,\ 0.1,\ 0.2]\ \mathrm{V/s}$ \\ \midrule 
        Formal potential, $E^0_f$ & Estimated by Differentiable Electrochemistry \\ \midrule
        Average diffusion coefficient, $D_{avg}$ &  Estimated by Differentiable Electrochemistry  \\ \midrule 
    \end{tabular}
    \caption{A list of simulation parameters and parameters determined by Differentiable Electrochemistry for$\ch{Ru(NH3)_6^{3+}}$/$\ch{Ru(NH3)_6^{2+}}$ redox chemistry. }
    \label{tab:RuHexRedoxParameterList}
\end{table}

\begin{figure}
    \centering
    \includegraphics[width=0.6\linewidth]{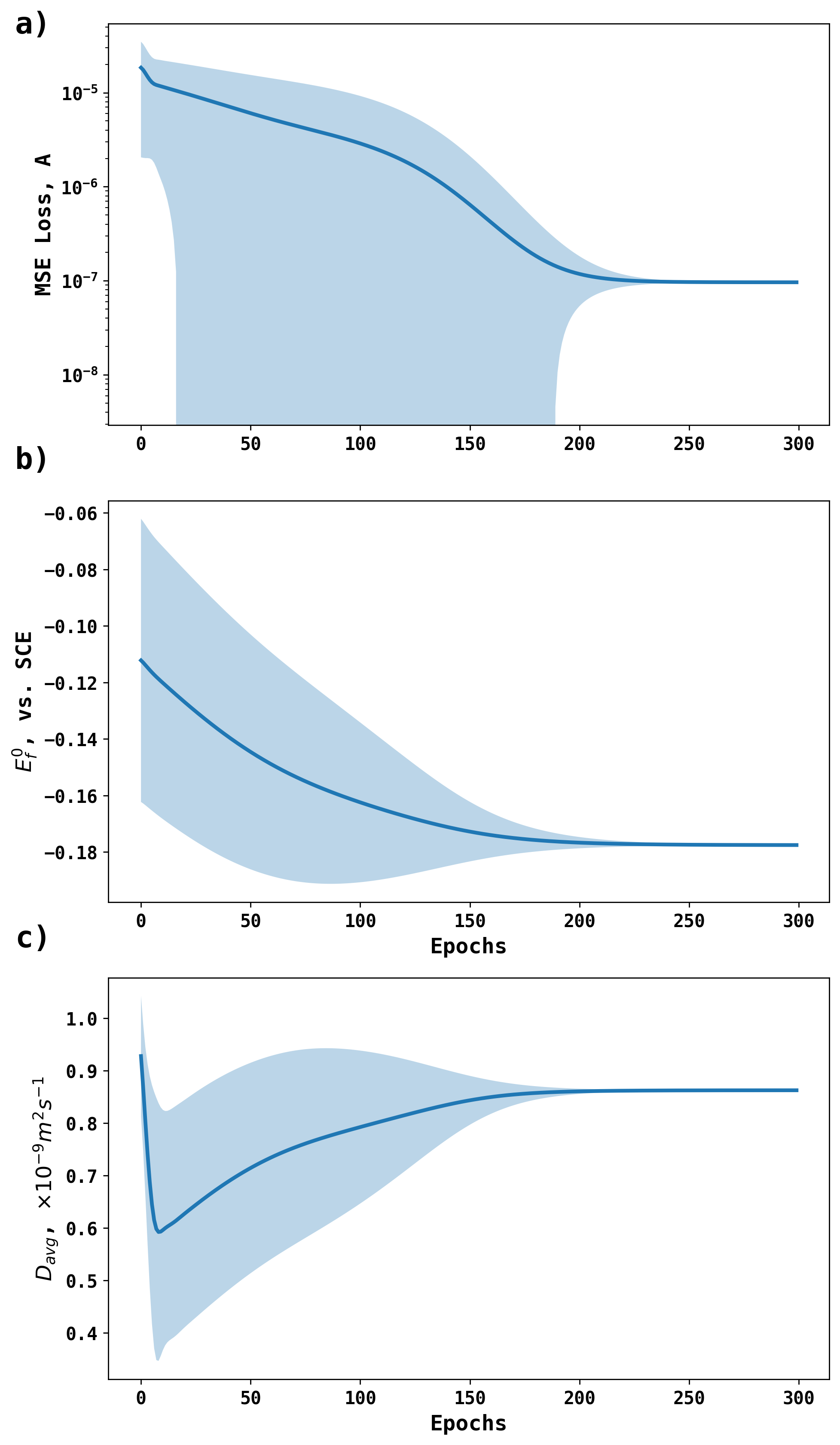}
    \caption{The optimization trajectories of Differentiable Electrochemistry of parameter estimation of $\ch{Ru(NH3)_6^{3+}}$/$\ch{Ru(NH3)_6^{2+}}$ couple. The mean and standard deviation of 30 runs are shown in solid lines and shadowed regions, respectively. (a) Mean squared error of current, (b) formal potentials and (c) diffusion coefficient trajectories are shown.}
    \label{fig:DiffECFeRuHexdoxEnsemble}
\end{figure}

Differentiable Electrochemistry simulation is next introduced to simultaneously reproduce voltammograms at four different scan rates, to calculate gradients of reproduction losses with respect to parameters of interests, and to optimize parameters with gradient-based optimization. The optimization step is repeated for 300 epochs with a stochastic gradient descent (SGD) optimizer and a learning rate of $10^{-3}$. 30 runs of different initial guesses of parameters are performed to evaluate the epistemic uncertainty of Differentiable Electrochemistry. The optimization trajectories are shown in Figure S \ref{fig:DiffECFeRuHexdoxEnsemble}, where the mean trajectories are shown in solid lines and the shadowed area are $\pm1$ standard deviation. As shown in Figure S \ref{fig:DiffECFeRuHexdoxEnsemble}a, the mean squared error of reproducing voltammetric current converged after 200 epochs of optimization and is $~10^{-7}\ \mathrm{A}$ after 300 epochs of optimization. Similarly, the trajectory for formal potential and average diffusion coefficient are shown in Figure S \ref{fig:DiffECFeRuHexdoxEnsemble}b--c. The predicted formal potential is $E^0_f=-0.178\pm{0.001}\ \mathrm{V}$ vs. SCE, and the diffusion coefficient is $(0.863\pm0.001)\ \mathrm{m^2s^-1}$. The simulated voltammograms (solid lines) using the Differentiable Electrochemistry estimated parameters are compared with experimental voltammograms (dashed lines) and show in Figure S \ref{fig:DiffECRuHexRedoxResult}, showing a good fit to experimental results at four different scan rates. As shown in this case, Differentiable Electrochemistry serves a powerful parameter estimation tool for fully reversible electrochemical reactions.

\clearpage
\section{Differentiable Li Kinetics}

In the context of Li metal electrode, the electrochemical reaction is:
\begin{equation}
    \ch{Li+ + e^- <=> Li}
\end{equation}
where the forward reaction is electrodeposition and the reverse is stripping. 
The preference for MHC is based on its applicability at high overpotential, but MHC formalism involves integration calculation and is very challenging to parameterize. To bypass this difficulty, Sripad et al. used an approximate expression for MHC kinetics developed in ref.\cite{RN13} as:
\begin{equation}
    j_{MHC,approx}=j_{0, MHC, approx} \sqrt{\pi \Lambda} \tanh{\left(\frac{\Lambda}{2}\right)}\mathrm{erfc}\left(\frac{\Lambda - \sqrt{1+\sqrt{\Lambda}+\theta^2}}{2\sqrt{\Lambda}}\right)
    \label{eq:approxMHC}
\end{equation}
where $j_{0, MHC,approx}$ is the exchange current density for approximate MHC models. $\Lambda$ and $\theta$ are dimensionless reorganization energy and dimensionless potential as defined in Table S \ref{tab:NPPDimensionlessTable}.

The Marcus-Hush formalism is enforced by implementing a potential dependent transfer coefficient:
\begin{equation}
    \alpha^{MH} = \frac{1}{2} + \frac{\theta}{4\Lambda}
\end{equation}
The current density response with MH formalism is thus:
\begin{equation}
    j_{MH} = j_{0,MH}\exp{\left(\frac{-\alpha_{MH}F}{RT}\eta\right)} - j_{0,MH}\exp{\left(\frac{\left(1.0-\alpha_{MH}\right)F}{RT}\eta\right)}
\end{equation}
where $j_{0,MH}$ is the exchange current density and $\eta$ is the overpotential. 

\begin{figure}
    \centering
    \includegraphics[width=1.0\linewidth]{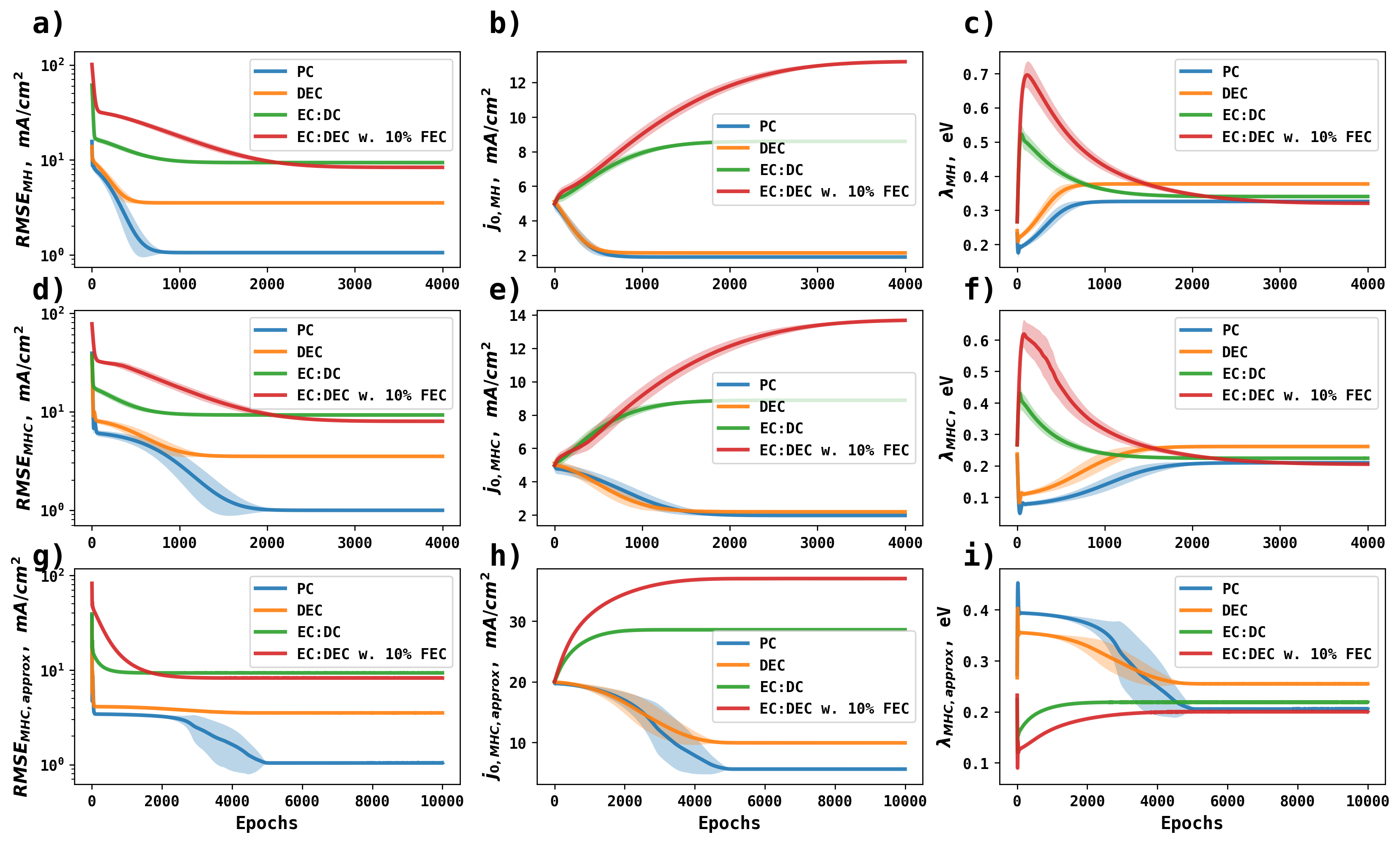}
    \caption{Differentiable Electrochemistry trajectories of explaining Tafel curves with MH (a--c), MHC(d--f), and the approximate MHC formalisms (g-i). The trajectories of root mean squared errors (a,d,g) exchange current densities (b,e,h) and reorganization energies (c,f,i) for four solvents are shown.   }
    \label{fig:DiffECLiEnsemble}
\end{figure}

The closed-form approximation to MHC formalism (Equation \ref{eq:approxMHC}), however, is less accurate at low overpotential and more accurate at high overpotential.\cite{RN13} In this section, Differentiable Electrochemistry simulation is applied to directly parameterize MHC kinetics without approximation or closed-form expression and compared with MH formalism. The MHC formalism in the context of Li metal electrode, is:
\begin{equation}
    j^{MHC} = j_0^{MHC}\frac{\boldsymbol{I}_{red}\left(\theta,\Lambda\right)}{\boldsymbol{I}_{red}\left(0,\Lambda\right)} - j_0^{MHC}\frac{\boldsymbol{I}_{ox}\left(\theta,\Lambda\right)}{\boldsymbol{I}_{ox}\left(0,\Lambda\right)}
\end{equation}
where $\boldsymbol{I}_{red/ox}$ is the integral over the Fermi-Dirac statistics as shown in Section \ref{MHCTheory}. Note that since the voltammetric data is collected in a small potential window, the surface concentration and mass transport were not explicitly considered. The Tafel data are fitted with 30 different initial guesses for 4000 epochs at a learning rate of $10^{-2}$ and Adam optimizer. The optimization trajectories using MH formalism is shown in Figure S \ref{fig:DiffECLiEnsemble}a-c, MHC formalism in Figure S \ref{fig:DiffECLiEnsemble}d-f, and the approximate MHC model in Figure S \ref{fig:DiffECLiEnsemble}h-i.  The root mean squared error, exchange current density, and reorganization energy with different formalisms for different electrolyte are presented. The errors are well converged at the end of Differentiable Electrochemistry optimization. 

The fitted exchange current density and reorganization energy are shown in Table S \ref{tab:DiffECLiExchange} and Table S \ref{tab:DiffECLiReorge}. The root mean squared error (RMSE) fitted using MH, approximate MHC, and MHC formalisms are given in Table S \ref{tab:DiffECLiMSE}, where MHC formalism has lower fitting errors by 0.8\% to 4.3\% compared with approximate MHC formalism.  In conclusion, Differentiable Electrochemistry directly implements and differentiates MHC formalism without the need of approximation, better characterize Li electrodeposition and stripping.

 \begin{table}
     \centering
     \begin{tabular}{|c|c|c|c|c|}
     \toprule
          Solvent&  $j_{0,MH},$ & $j_{0,MHC,approx}$ & $j^{MH}_{0,L.fit}$,& $j_{0,MHC}$  \\ \midrule
          PC     &  $1.90\pm0.01$ & $5.60\pm0.01$   &  2.6  & $1.98\pm0.01$  \\ \midrule
          DEC    &  $2.14\pm0.01$ & $9.94\pm0.01$  &  3.7  & $2.20\pm0.01$ \\ \midrule
          EC:DEC &  $8.59\pm0.01$ & $28.6\pm0.1$  &  10.4  & $8.88\pm0.01$ \\  \midrule
          EC:DEC w. 10\% FEC & $13.2\pm0.1$ & $37.0\pm0.1$ & 16.0 & $13.7\pm0.1$\\  \midrule
     \end{tabular}
     \caption{The exchange current density obtained using Differentiable Electrochemistry with MH formalism, approximate MHC model, $j_{0,MHC,approx}$ $j_{0,MH}$ and MHC formalism, $j_{0,MHC}$. $j^{MH}_{0,L.fit}$ is the value reported by Boyel et al. ref.\cite{boyle2020transient} The unit of exchange current density is $\mathrm{mA/cm^2}$}
     \label{tab:DiffECLiExchange}
 \end{table}

 \begin{table}
     \centering
     \begin{tabular}{|c|c|c|c|c|}
     \toprule
          Solvent&   $\lambda_{MH}$ & $\lambda_{MHC,approx}$ &$\lambda^{MH}_{L.fit}$ & $\lambda_{MHC}$  \\ \midrule
          PC     &   $0.326\pm0.001$ & $0.206\pm0.001$ & $0.34\pm0.01$ & $0.209\pm0.001$ \\ \midrule
          DEC    &   $0.377\pm0.001$ & $0.255\pm0.001$ & $0.34\pm0.01$ & $0.261\pm0.001$\\ \midrule 
          EC:DEC &   $0.340\pm0.001$ & $0.219\pm0.001$ & $0.30\pm0.02$ & $0.224\pm0.001$\\ \midrule 
          EC:DEC w. 10\% FEC &  $0.320\pm0.001$ & $0.200\pm0.001$ & $0.28\pm0.02$ & $0.204\pm0.001$\\ \midrule  
     \end{tabular}
     \caption{The reorganization energy obtained using Differentiable Electrochemistry with MH formalism, $\lambda_{MH}$, approximate MHC model, $\lambda_{MHC,approx}$ and MHC formalism, $\lambda_{MHC}$. $\lambda^{MH}_{L.fit}$ is the value reported by Boyel et al.\cite{boyle2020transient} The unit of reorganization energy is $\mathrm{eV}$.}
     \label{tab:DiffECLiReorge}
 \end{table}

\begin{table}
    \centering
    \begin{tabular}{|c|c|c|c|c|}
    \toprule
        Solvent & $RMSE_{MH}$ & $RMSE_{MHC,approx}$  & $RMSE_{MHC}$   \\ \midrule
         PC&  1.06    & 1.04  & 0.995         \\ \midrule
         DEC&  3.53   & 3.52 & 3.51            \\ \midrule
         EC:DEC& 9.38 & 9.33 & 9.25          \\ \midrule
         EC:DEC w. 10\% FEC& 8.36  & 8.23 &7.98   \\ \midrule
    \end{tabular}
    \caption{The root mean squared error (RMSE) obtained using Differentiable Electrochemistry with MH formalism, $RMSE_{MH}$, approximate MHC model, $RMSE_{MHC,approx}$ and MHC formalism, $RMSE_{MHC}$. The unit of RMSE is $\mathrm{mA/cm^2}$. }
    \label{tab:DiffECLiMSE}
\end{table}

\clearpage
\section{Differentiable Electrochemistry for Hydrogen Evolution Reaction}\label{HERSection}
\subsection{Acetic HER on Pt Electrode}
The Hydrogen Evolution Reaction (HER) in acidic environments is:
\begin{equation}
    \ch{H^+_{aq} + e^- <=> 1/2 H_2}
\end{equation}
The mechanism of HER, in acidic media, is usually described as a combination of the three elementary reactions
\begin{equation}
\begin{aligned}
    \text{Volmer Step}: &\ch{H+ + e- + * <=> H_{ads}}\\
    \text{Tafel Step}: &\ch{H_{ads} + H_{ads} <=> H_2}\\
    \text{Heyrovsky Step}:&\ch{H_{ads} + H+ + e- <=> H_2}
\end{aligned}
\end{equation}

The reaction of HER proceeds via the intermediary of adsorbed hydrogen atoms, which leads to a huge variation in the rate of the reaction dependent on the material of the electrode. Specifically, the mechanism of metal electrodes including Pt, Au, or W follows the Volmer-Heyrovsky mechanism:\cite{UV4} 
\begin{equation}
\begin{aligned}
    \text{Volmer Step}:\ &\ch{H+ + e- + * <=> [fast] H_{ads}}\\
    \text{Heyrovsky Step}:\ &\ch{H_{ads} + H+ + e- -> [slow, rds] H_2(g)}
\end{aligned}
\end{equation}
where the first step is the Volmer reaction, and the second step is the slow, rate-determining Heyrovsky reaction. The expected transfer coefficient can be derived as follows.  First, the rate equation of Volmer step at equilibrium, is:
\begin{equation}
\label{eq:VolmerEq}
    k_{0,V}[\ch{H+}](1-\xi) \exp{\left( -\frac{-\alpha F}{RT}\eta\right)} = k_{0,-V}\xi \exp{\left(\frac{\beta F}{RT}\eta\right)}     
\end{equation}
where $k_{0,V}$ is the electrochemical rate constant of Volmer reaction, and $k_{0,-V}$ corresponds to the reverse reaction. $[\ch{H+}]$ is the concentration of $\ch{H+}$ in solution and assumed to be equal to its activity, and $\xi$ is the relative coverage of $\ch{H_{ads}}$ on electrode surface. $\xi$ is defined as: $\xi=\frac{\Gamma_{\ch{H_{ads}}}}{\Gamma_{max}}$. $\alpha$ and $\beta$ are the cathodic and anodic transfer coefficients and $\alpha+\beta =1$. $\eta$ is the overpotential. By solving Equation \ref{eq:VolmerEq}, the expression for $\xi$ at equilibrium is:
\begin{equation}
    \label{VolmerStepCoverage}
    \xi = \frac{[\ch{H+}]}{[\ch{H+}]+\frac{k_{0,-V}}{k_{0,V}}\exp{\left(\frac{F}{RT}\eta\right)}}
\end{equation}

The rate equation of Heyrovsky reaction, ignoring the reverse reaction, is:
\begin{equation}
    r_{HER} = k_{0,H}\xi [\ch{H+}]\exp{\left(\frac{-\alpha F}{RT}\eta\right)}
\end{equation}
By plugging in the expression of $\xi$ from Equation  \ref{VolmerStepCoverage}, the rate equation is obtained as:
\begin{equation}
    \label{VHRate}
    r_{HER} = k_{0,H} \frac{[\ch{H+}]^2}{[\ch{H+}]+\frac{k_{0,-V}}{k_{0,V}}\exp{\left(\frac{F}{RT}\eta\right)}}\exp{\left(\frac{-\alpha F}{RT}\eta\right)}
\end{equation}
If $[\ch{H+}]$ is small ($[\ch{H+}]\ll\frac{k_{0,-V}}{k_{0,V}}\exp{\left(\frac{F}{RT}\eta\right)}$), the rate equation can be approximated as:
\begin{equation}
    r_{HER}\approx k_{HER}[\ch{H+}]^2\exp\left(-(1+\alpha)\frac{F}{RT}\eta\right) 
\end{equation}
Leading to an ideal apparent transfer coefficient as $1+\alpha$:
\begin{equation}
    \ln{|I_{HER}|} = -\frac{(1+\alpha)F}{RT}\eta + \text{constant}
\end{equation}
Since $\alpha$ is usually taken as 0.5, the apparent transfer coefficient of HER on Pt electrode is expected to be around 
1.5.\cite{prats2021determination} If the Volmer step is slow and the rate determining step, the apparent transfer coefficient of the Volmer-Tafel mechanism is around 2.0.  

\subsection{Simulation of HER in acidic solution}
The objective of extracting transfer coefficient of HER from a linear sweep voltammetry (LSV) on a Pt rotating disk electrode as reported by Koper and colleagues.\cite{AP1} The reported diameter of Pt disk is 5 mm, and the LSV was recorded at a rotation rate of 2500 rpm and a scan rate of 2 mV/s in 1 M $\ch{HClO4}$. 

First, the partial differential equation, boundary conditions and initial conditions for this problem are specified. Since the electrode is at millimeter scale and a fast rotational speed, the governing mass transport equation is the convection-diffusion mass transport equation in 1D:
\begin{equation}
    \begin{array}{cc}
        \frac{\partial c_{\ch{H+}}}{\partial t} = D_{\ch{H+}}\frac{\partial^2c_{\ch{H^+}}}{\partial y^2} - v_y\frac{\partial c_{\ch{H^+}}}{\partial y} &  \\
        v_y=-Ly^2& \\
        L = 0.51023(2\pi f)^{3/2} \nu^{-1/2}&
    \end{array}
\end{equation}
where $C_{\ch{H+}}$ is the concentration of $\ch{H+}$ in solution and $y$ is the distance to the electrode surface. $\nu$ is the kinematic viscosity of the solvent.  Since the pKa of $\ch{HClO4}$ is -7 and a strong acid,\cite{haynes2016crc} the bulk concentration of $\ch{H+}$ is $c^*_{\ch{H+}} = 1M$, the initial and outer boundary condition of this problem is:
\begin{equation}
    \begin{array}{cc}
        c_{\ch{H+}} = 1M & t=0, 0\leq y< y_{sim}  \\
        c_{\ch{H+}} = 1M & 0\leq t \leq t_{sim}, y= y_{sim}
    \end{array}
\end{equation}
The electrode surface $(y=0)$ boundary condition is described as:
\begin{equation}
    j = k_0 \exp\left(-\frac{\alpha F}{RT}\eta\right)C_{\ch{H+},y=0}
\end{equation}
where $k_0$ is standard electrochemical rate constant correlated with exchange current density, $\alpha$ is the cathodic transfer coefficient of HER determined using Differentiable Electrochemistry. The other simulation parameters are tabulated in Table S \ref{tab:HER_LSV_Parameters}

\begin{table}
    \centering
    \begin{tabular}{|c|c|}
    \toprule
         Parameters& Value \\ \midrule
         Scan rate of LSV& 0.002 V/s \\  \midrule
         Rotational frequency, $f$& 2500 rpm \\ \midrule
         Diffusion coefficient of $\ch{H+}$, $D_{\ch{H+}}$ & $9.311\times10^{-9} \ \mathrm{m^2s^{-1}}$ \\ \midrule
         Kinematic viscosity of solvent, $\nu$ & $10^{-6}\ \mathrm{m^2s^{-1}}$ \\ \midrule
         Bulk concentration of $\ch{H+}$, $c^*_{\ch{H+}}$ & 1 M   \\ \midrule
         Formal potential of $\ch{H+}$/$\ch{H2}$ couple& 0 V vs. RHE \\  \midrule
         Cathodic transfer coefficient& Determined by Differentiable Electrochemistry \\ \midrule
         Electrochemical rate constant, $k_0$ & Determined by Differentiable Electrochemistry  \\
    \bottomrule
    \end{tabular}
    \caption{Simulation parameters for LSV of HER on a rotating Pt electrode.}
    \label{tab:HER_LSV_Parameters}
\end{table}

The simulations are performed with dimensionless parameters as defined in Table S \ref{tab:HydrodynamicDimensionlessTable}. The dimensionless mass transport equation is:
\begin{equation}
    \frac{\partial C_{\ch{H+}}}{\partial T} = \frac{\exp\left( -\frac{2}{3}W^3\right)}{[\int_{0}^{\infty}-\frac{1}{3}W^3dW]^2} \frac{\partial^2 C_{\ch{H^+}}}{\partial U^2}
\end{equation}
The dimensionless initial and outer boundary condition is:
\begin{equation}
    \begin{array}{cc}
        C_{\ch{H+}}=1 & T=0, U_{sim}\geq U \geq 0 \\
        C_{\ch{H+}}=1 & T_{sim}>T>0, U=U_{sim} 
    \end{array}
\end{equation}
The dimensionless equation of HER on Pt electrode using Butler-Volmer equation is:
\begin{equation}
    J = K_0 \exp(-\alpha\theta)C_{\ch{H+},U=0}
\end{equation}
where $K_0$ and $\alpha$ are parameters to be learned using Differentiable Electrochemistry. In simulation, $log_{10}{K_0}$ is learned.

\clearpage
\section{Estimating Diffusivity and Transference Number from Operando Fields}
In this section, differentiable finite volume simulation is employed to estimate the salt diffusivity $D$ and the transference number $t_+^0$ based on the concentrated solution theory using operando measurements of salt concentration $c$ and solvent velocity $v_0$ in a polarized LiTFSI/PEO electrolyte.

\subsection{Governing equations}\label{sec:gov-eq}
The spatiotemporal evolution of the salt concentration $c(x,t)$ and solvent velocity $v_0(x,t)$ in the moving electrode reference frame is governed by: \cite{AP2}
\begin{equation}\label{eqn:gov-eq}
    \begin{aligned}
    \frac{\partial c}{\partial t} &= \frac{\partial}{\partial x} \left( 
    D \left( 1 - \frac{\mathrm{d} \ln c_0}{\mathrm{d} \ln c} \right) 
    \frac{\partial c}{\partial x} 
    - t_+^0 \frac{i}{F} - c v_0 \right) \\[6pt]
    \frac{\partial v_0}{\partial x} &= \bar{V} \frac{\partial}{\partial x} \left(
    D \left( 1 - \frac{\mathrm{d} \ln c_0}{\mathrm{d} \ln c} \right) 
    \frac{\partial c}{\partial x} 
    - t_+^0 \frac{i}{F} \right)
    \end{aligned}
\end{equation}
where $D$ is the salt diffusivity, $t_+^0$ is the cation transference number relative to solvent motion, $c_0$ is the solvent concentration, $i$ is the current density, $\bar{V}$ is the salt partial molar volume, and $F$ is Faraday’s constant. The initial and boundary conditions are:
\begin{equation}
    \begin{aligned}
    c &= c_{\mathrm{avg}}, 
    & 0 < x < L, \; t = 0 \\[6pt]
    v_0 &= \bar{V}(1 - t_+^0)\frac{i}{F}, 
    & 0 < x < L, \; t = 0 \\[6pt]
    - D \left( 1 - \frac{\mathrm{d} \ln c_0}{\mathrm{d} \ln c} \right) 
    \frac{\partial c}{\partial x} 
    &= (1 - t_+^0)\frac{i}{F}, 
    & x = 0, L, \; t > 0 \\[6pt]
    v_0 &= 0, 
    & x = 0
    \end{aligned}
\end{equation}

\subsection{Concentration Dependence of Properties and Parametrization of Transference Number}
To solve this set of PDEs, several parameters need to be determined. The dependence of $\left( 1 - \frac{\mathrm{d} \ln c_0}{\mathrm{d} \ln c} \right)$, $\bar{V}$ on $c$ is known from electrolyte density measurements via equation (A3) and Appendix C in \cite{Mistry2022v0effect}. The current density $i(t)$ is obtained from Figure S1(a) in the SI of \cite{AP2}. The diffusivity $D$ and the transference number $t_+^0$ are the only two unknowns, and they are interrelated according to the following equation:
\begin{equation}
\frac{D \left( 1 - \dfrac{\mathrm{d} \ln c_0}{\mathrm{d} \ln c} \right)}{(1 - t_+^0)}
= - \frac{i_{\mathrm{ss}}}{F \, \dfrac{\partial c_{\mathrm{ss}}}{\partial x}}
\end{equation}
where $c_{\mathrm{ss}}(x)$ and $i_{\mathrm{ss}}$ represent the steady state concentration profile and the current density from the experiment measurements, respectively. A quadratic polynomial is used to fit $c_{\mathrm{ss}}(x)$ in Figure S4(c) of \cite{AP2} and the resulting functional relationship between $D$ and $t_+^0$ is obtained from Figure S4(f) of \cite{AP2}. Therefore, the salt diffusivity can be fully determined once the concentration dependence of the transference number is established. The same technique for approximating $t_+^0$ is adopted as equation (7) in \cite{AP2}, where $N_p=2$, i.e.,
\begin{equation}
    t_+^0 = p_0 + p_1 \left( \frac{c - c_{\mathrm{avg}}}{c_{\mathrm{avg}}} \right)
\end{equation}

\subsection{Finite Volume Scheme}
The problem defined in section \ref{sec:gov-eq} is solved using finite volume method with staggered storage. The computational domain of the electrolyte of length $L$ is discretized into $N$ uniform cells, where $[x_{i-1/2},x_{i+1/2}]$ denotes the $i$th cell for $i \in [1,2,\dots,N]$. Concentrations are stored at cell centroids, while velocities are defined at the interfaces between adjacent control volumes. The central difference is used for spatial derivatives, and the first-order Euler method is employed for time marching. Let $\Delta t$ be the time step and $\Delta x = L/N$ be the cell size. First, the concentration of cell $i$ at time $t^{n+1}$ is obtained by:
\begin{equation}
    c_i^{n+1} = c_i^{n} - \frac{\Delta t}{\Delta x} \left( f_{i+1/2}^{n} - f_{i-1/2}^{n} \right)
\end{equation}
where $f_{i+1/2}^{n}$ is the flux at interface $i+1/2$ at time $t^n$, which is numerically approximated by:
\begin{equation}
    f_{i+1/2}^{n} = g_{i+1/2}^{n} + c_{i+1/2}^{n} v_{0,i+1/2}^{n}
\end{equation}
with
\begin{equation}
    \begin{aligned}
        g_{i+1/2}^{n} 
        &= -D \left(c_{i+1/2}^{n}\right)
        \left(1 - \frac{\mathrm{d} \ln c_0}{\mathrm{d} \ln c} \Bigg|_{c_{i+1/2}^{n}}\right)
        \frac{c_{i+1}^n - c_i^n}{\Delta x}
        + t_+^0 \left(c_{i+1/2}^{n}\right)\frac{i(t^n)}{F} \\[6pt]
        c_{i+1/2}^{n} 
        &= \frac{c_i^n + c_{i+1}^n}{2}
    \end{aligned}
\end{equation}
Next, the solvent velocity at interface $i+1/2$ at time $t^{n+1}$ is solved by integrating the velocity equation over cell $i$, assuming that $\bar{V}$ is constant in each cell:
\begin{equation}
    v_{0,i+1/2}^{n+1} = v_{0,i-1/2}^{n+1} - \bar{V} (c_{i}^{n+1}) \left(g_{i+1/2}^{n+1} - g_{i-1/2}^{n+1} \right)
\end{equation}
The concentration and velocity states can be evolved to presribed time by performing these two steps iteratively. All relevant parameters are tabulated in Table S \ref{tab:sim-params}.
\begin{table}[htbp]
\caption{Simulation parameters for Li-Li symmetric cells with LiTFSI/PEO electrolyte.}
\centering
\begin{tabular}{|c|c|}
\toprule
\textbf{Parameters} & \textbf{Value} \\
\midrule
Electrolyte thickness, $L$ & 3 mm \\ \midrule
Initial salt concentration, $c_{\mathrm{avg}}$ & 1.87 M \\ \midrule
Molar mass of Li$^{+}$ cations, $M^{+}$ & 6.94 g mol$^{-1}$ \\ \midrule
Molar mass of TFSI$^{-}$ anions, $M^{-}$ & 280.15 g mol$^{-1}$ \\ \midrule
Molar mass of PEO, $M^{0}$ & 44.05 g mol$^{-1}$ \\ \midrule
Faraday’s constant, $F$ & 96485 C mol$^{-1}$ \\ \midrule
Number of cells, $N$ & 50 \\ \midrule
\end{tabular}
\label{tab:sim-params}
\end{table}

\subsection{Loss Function and Optimization}
The loss function is defined as the root mean squared error between the simulated concentration fields, $c^{\mathrm{sim}}$, and the operando concentration fields, $c^{\mathrm{op}}$, plus a regularization term of parameters:
\begin{equation}
    \mathrm{Loss}(p_0, p_1) 
    = \frac{1}{c_{\mathrm{avg}}}
    \left[
    \frac{1}{N_t N_x}
    \sum_{n=1}^{N_t} \sum_{i=1}^{N_x} 
    \left( c^{\mathrm{sim}}(x_i, t^n; p_0, p_1) - c^{\mathrm{op}}(x_i, t^n) \right)^2
    \right]^{\tfrac{1}{2}}
    + \lambda \left(p_0^2 + p_1^2\right)
\end{equation}
where $N_t$ is the number of time instants, $N_x$ is the number of spatial points, $\lambda=0.005$ is the regularization coefficient.

All the 9 concentration profiles from Figure 3 in \cite{AP2} are used in the loss function, while the number of spatial points is determined by selecting the cells whose coordinates are located in between the range of the coordinates of the operando data. Interpolation is performed to obtain the values of operando fields at cell centroids of the finite volume mesh.

The regularization term is added to avoid aggressive estimation of parameters due to lack of experimental data on both sides of the electrolyte and to accelerate the optimization process. The BFGS method implemented in the ScipyMinimize solver from the Python package JAXopt is employed with a convergence tolerance of $1\times10^{-6}$.

\begin{figure}
    \centering
    \includegraphics[width=\linewidth]{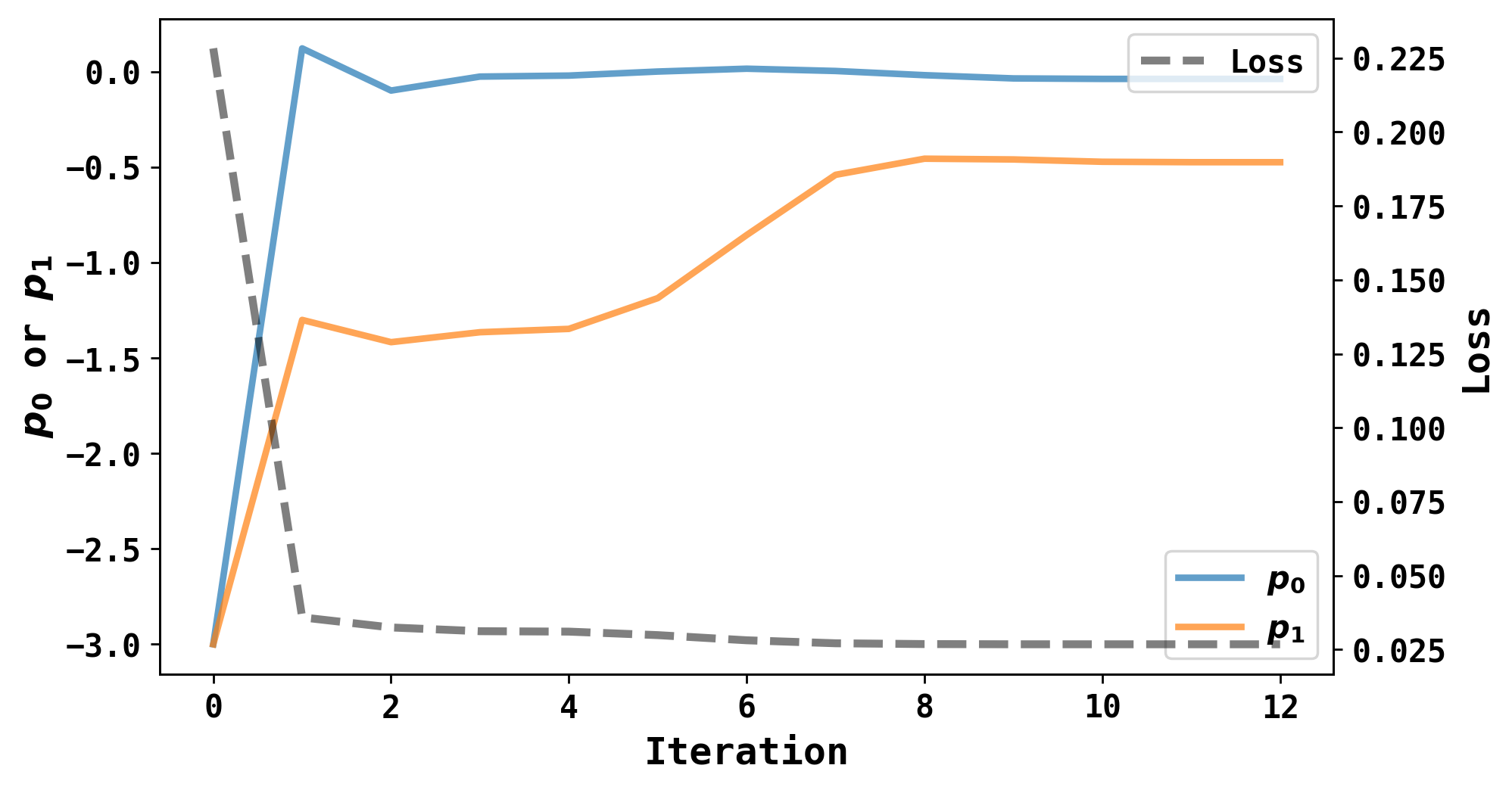}
    \caption{Convergence history of the differentiable simulation with initial guess of $(-3, -3)$.}
    \label{fig:opt-hist}
\end{figure}

\begin{table}[htbp]
\caption{Optimization results with different initial guesses.}
\centering
\begin{tabular}{|c|c|c|c|c|}
\toprule
\textbf{Initial guess} & \textbf{Iteration} & \textbf{Loss} & \textbf{Estimated parameters} & \textbf{Error} \\
\midrule
$(0, 0)$   & 13 & 0.026759 & $(-0.037581,\,-0.474187)$ & $5.59 \times 10^{-8}$ \\
$(-1, 2)$  &  7 & 0.026759 & $(-0.037582,\,-0.474162)$ & $2.80 \times 10^{-7}$ \\
$(-2, 0)$  & 11 & 0.026759 & $(-0.037580,\,-0.474190)$ & $4.20 \times 10^{-7}$ \\
$(-3,-3)$  & 12 & 0.026759 & $(-0.037583,\,-0.474186)$ & $6.41 \times 10^{-7}$ \\
\bottomrule
\end{tabular}
\label{tab:sensitivity-analysis}
\end{table}

The optimization converges within just a few iterations, as shown in Figure s \ref{fig:opt-hist}. To investigate the sensitivity of the final solution about the initial guess of $(p_0,p_1)$, several numerical experiments of different initial values were also done. The results are tabulated in Table \ref{tab:sensitivity-analysis}, where ``Error'' denotes the norm of the final gradient. All numerical experiments give the same result for various initial guesses, which demonstrates the robustness of the optimization. Even though the full trajectory of one differentiable simulation contains 0.66 million time steps, the optimization completes within about 5 minutes using 1 CPU with 16 GB memory, indicating the high efficiency of the implementation.

\subsection{Results}
\begin{figure}
    \centering
    % First figure
    \begin{minipage}[t]{0.48\textwidth}
    \centering
    \begin{overpic}[width=\linewidth]{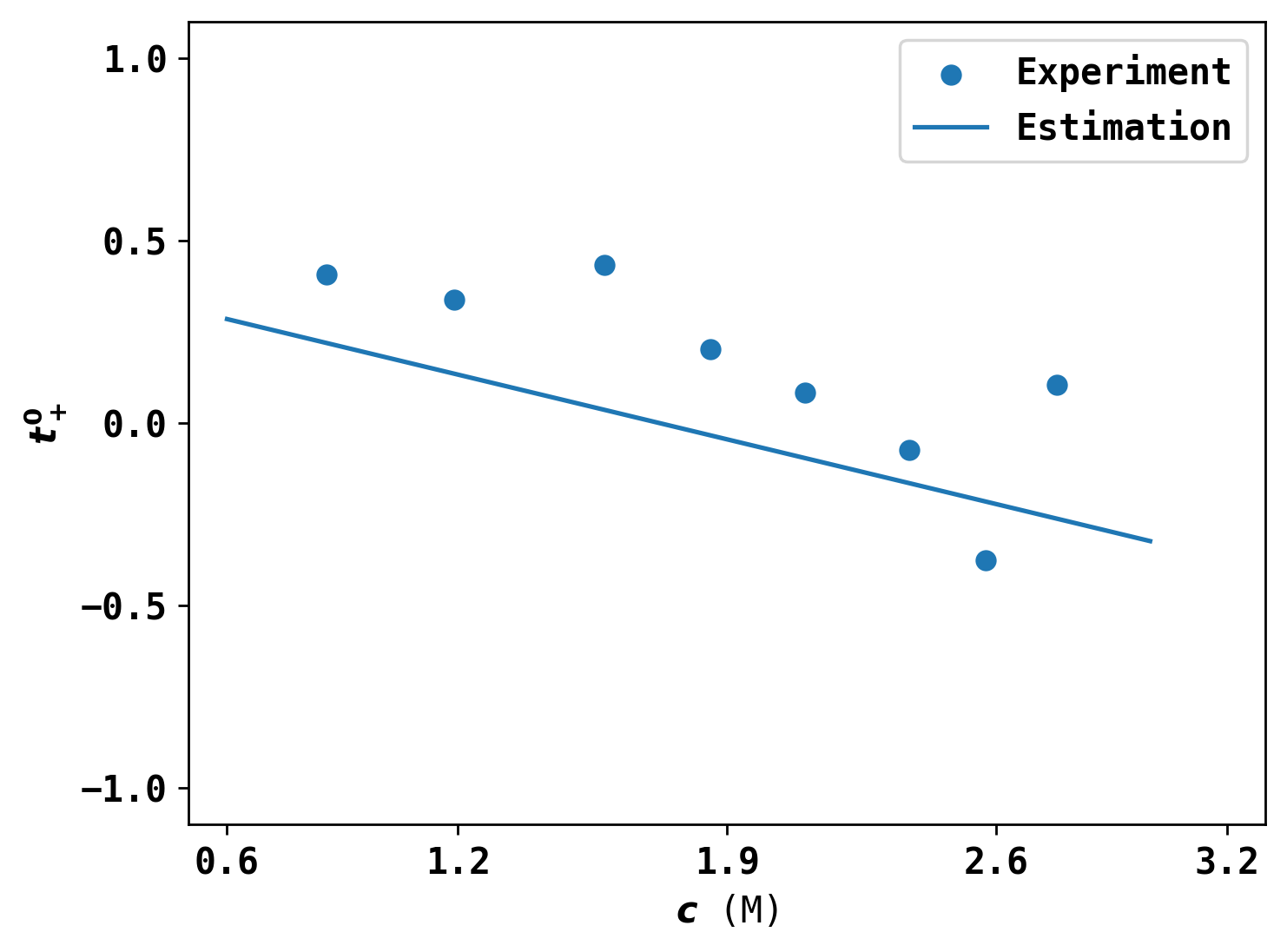}
        \put(-5,70){(a)}
    \end{overpic}
    \end{minipage}
    % Second figure
    \begin{minipage}[t]{0.48\textwidth}
    \centering
    \begin{overpic}[width=0.94\linewidth]{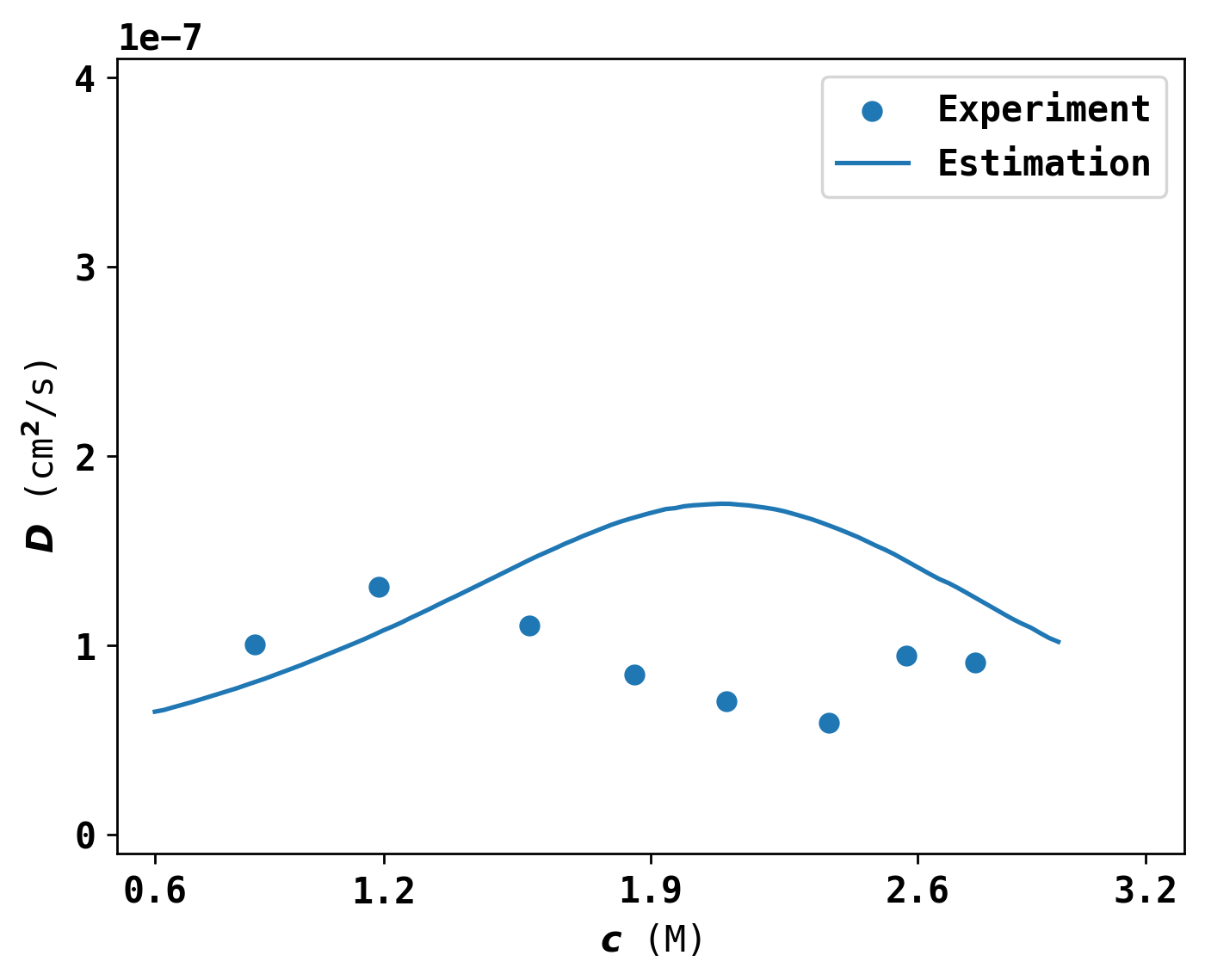}
        \put(-3,74){(b)}
    \end{overpic}
    \end{minipage}
    
    \caption{Estimated (a) transference number and (b) salt diffusivity as functions of concentration. The experiment data is from \cite{Mistry2022v0effect}.}
    \label{fig:tp0-and-D}
\end{figure}

The estimated transference number $t_+^0 (c)$ and salt diffusivity $D(c)$ are shown in Figure S \ref{fig:tp0-and-D}.

\begin{figure}
    \centering
    \includegraphics[width=0.6\linewidth]{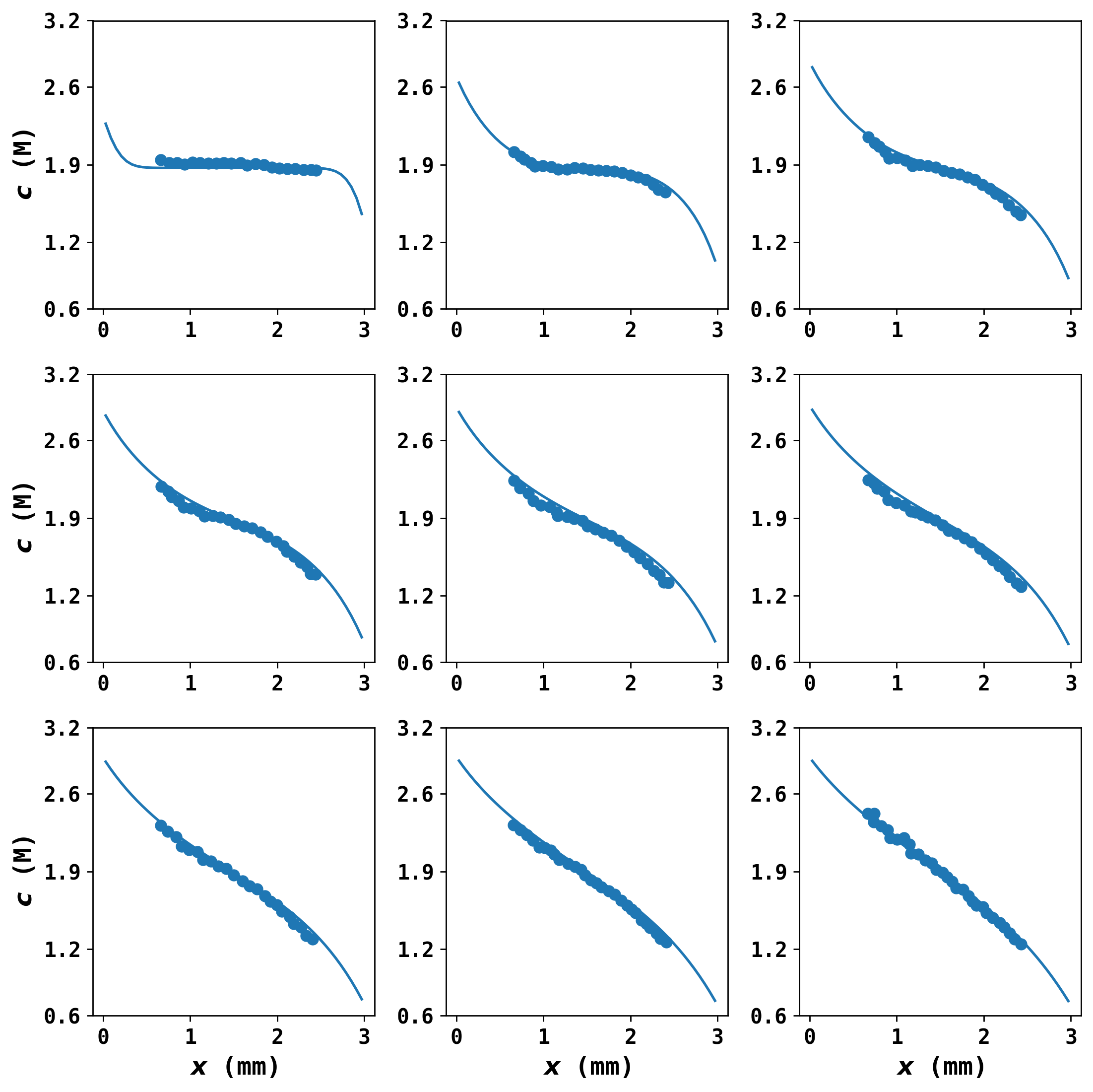}
    \caption{Comparison between the simulated concentration profiles (solid line) and the XAM-measured data (scatter) at 9 selected times. From left to right, top to bottom, the times are $t=20,151,314,412,511,609,707,871,1100$ min, respectively.}
    \label{fig:concentration-comparison}
\end{figure}

To further assess their representativeness, Figure S \ref{fig:concentration-comparison} presents a comparative analysis between the simulated concentration fields using these two estimated properties and operando XAM fields at 9 distinct times. The simulation results match the experiment measurements very well for all time instants, underscoring the reliability of the estimated properties. 

It is noteworthy that the differentiable framework applied so far has not incorporated solvent velocity measurements. To validate the capability of the estimated properties in predicting the solvent motion, the governing equations \ref{eqn:gov-eq} are solved for $v_0 (x,t)$. The relationship between the velocity $v_0^{\prime} (x,t)$ in the stationary laboratory frame and the velocity $v_0 (x,t)$ measured in the moving electrode reference frame is given by:
\begin{equation}
    v_0^{\prime} (x,t) = v_0 (x,t) + v_{\mathrm{interface}}^{\prime}
\end{equation}
Here, $v_{\mathrm{interface}}^{\prime}$ is the velocity of Li/electrolyte interface in the stationary laboratory frame calculated by:
\begin{equation}
    v_{\mathrm{interface}}^{\prime} = -\frac{M_{\mathrm{Li}}i}{\rho_{\mathrm{Li}}F}
\end{equation}
where $M_{\mathrm{Li}}$ is the atomic mass of lithium, and $\rho_{\mathrm{Li}}$ is the mass density of the lithium metal.

\begin{figure}
    \centering
    \includegraphics[width=0.6\linewidth]{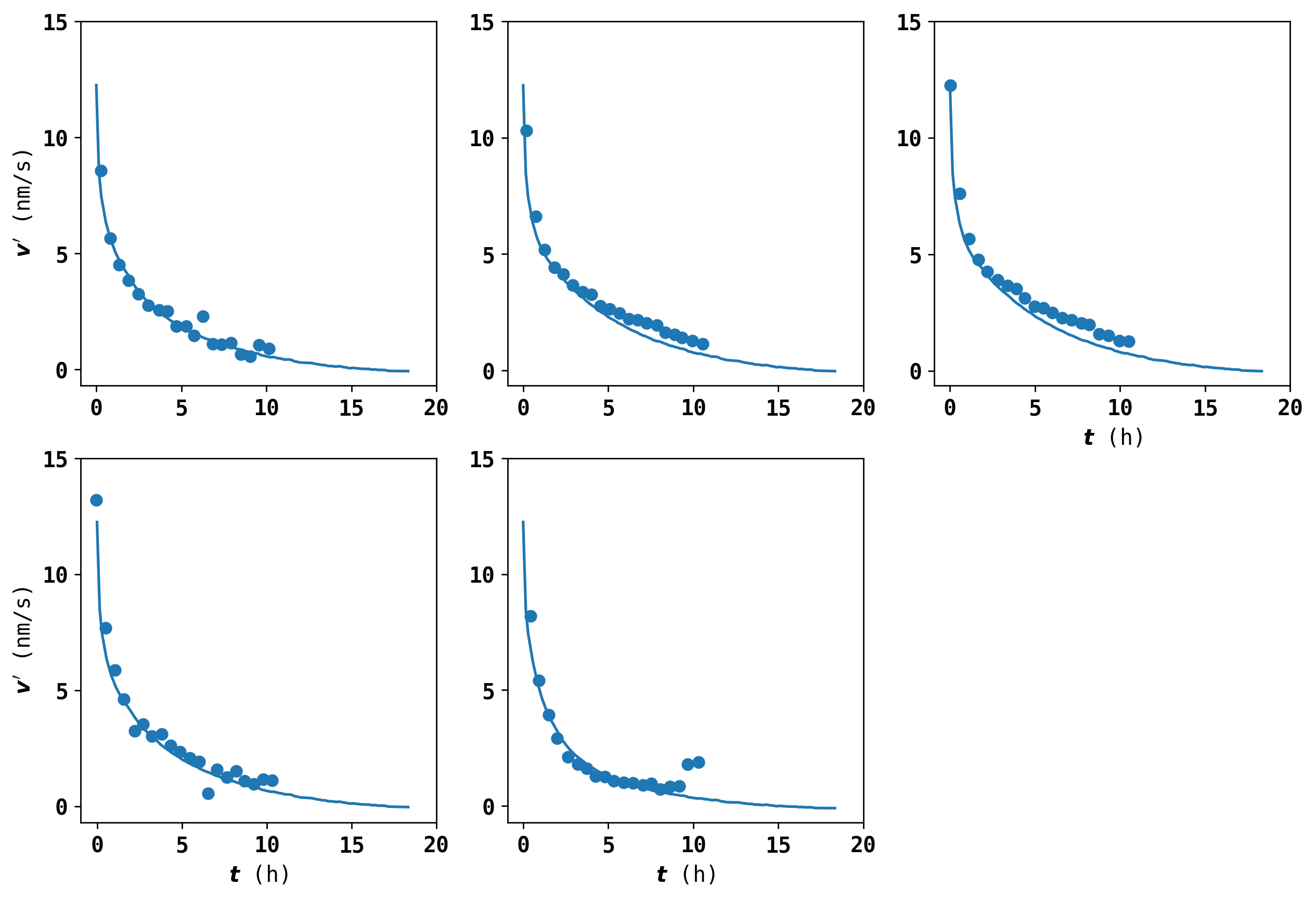}
    \caption{Comparison between the simulated solvent velocity (solid line) and the XPCS measurements (scatter) at 5 different locations along the cell. From left to right, top to bottom, the locations are $x=0.95,1.30,1.65,2.00,2.35$ mm, respectively.}
    \label{fig:velocity-comparison}
\end{figure}

Figure S \ref{fig:velocity-comparison} compares the simulated $v_0^{\prime} (x,t)$ against the XPCS measurements at five different locations along the cell thickness. They match reasonably well.

In a nutshell, the differentiable finite volume solver successfully identifies the concentration dependence of the transference number and the salt diffusivity, which not only provide an accurate description of the concentration fields but also enable reliable predictions of the solvent velocity.

\subsection{Benchmarking Optimization Methods}
To contextualize the efficiency of the differentiable framework, we benchmarked it against four advanced gradient-free optimization methods: particle swarm optimization (PSO), \cite{kennedy1995particle} Bayesian optimization (BO), \cite{frazier2018tutorial} covariance matrix adaptation evolution strategy (CMA-ES), \cite{hansen2016cma} and Nelder-Mead method (NM).\cite{nelder1965simplex} Among these methods, PSO, BO and CMA-ES are stochastic algorithms, while NM and DiffEC are deterministic methods. The corresponding Python packages used for these algorithms are listed in Table S \ref{tab:pkg}. 
\begin{table}[h]
\caption{Algorithms and their corresponding Python packages.}
\centering
\begin{tabular}{|l|l|}
\toprule
Algorithm & Package \\
\midrule
PSO    & \texttt{scikit-opt} \\
BO     & \texttt{scikit-learn} \cite{scikit-learn} \\
CMA-ES & \texttt{pycma} \cite{hansen2019pycma} \\
NM     & \texttt{SciPy} \cite{2020SciPy-NMeth}\\
\bottomrule
\end{tabular}
\label{tab:pkg}
\end{table}

\begin{figure}[htbp]
    \centering
    \begin{overpic}[width=\linewidth]{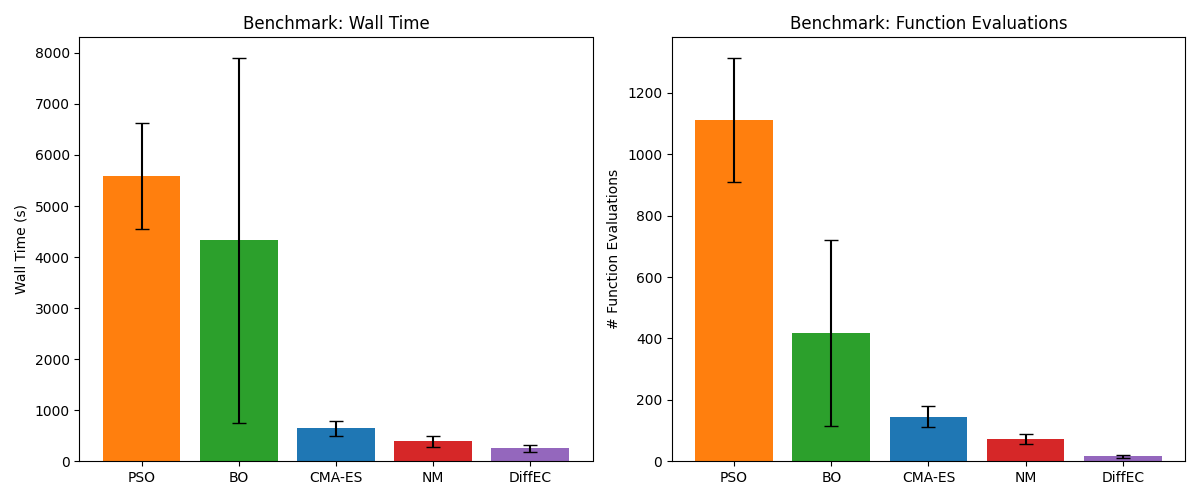}
        \put(0, 40){(a)}
        \put(50, 40){(b)}
    \end{overpic}
    \caption{Benchmarking results of different optimization algorithms applied to parameter estimation of $(p_0, p_1)$: 
    (a) wall time, 
    (b) number of function evaluations. 
    Error bars denote the standard deviation over 5 random initializations for stochastic methods (PSO, BO, and CMA-ES) and 5 initial guesses for deterministic methods (NM and DiffEC). The lower the wall time and number of function evaluations, the better the optimizer performance.}
    \label{fig:benchmark}
\end{figure}

To ensure a fair comparison, all methods employ the same highly optimized solver code implemented in JAX. Two key performance metrics are reported: wall time and the number of function evaluations needed to reach a given target loss 0.02676 on a single CPU, as shown in Figure S \ref{fig:benchmark}. Each methods are initialized with 5 initial guesses for variance estimation. 

The results reveal a stark contrast between gradient-free approaches and the differentiable solver. PSO and BO incur prohibitively large wall times, averaging $5000$--$6000$ seconds, with significant variance. These methods also require hundreds to thousands of function evaluations due to their reliance on repeated forward simulations without gradient information. CMA-ES and NM are the state-of-the-art methods, achieving convergence within several hundred seconds and fewer than 200 evaluations, yet they still demand more resources compared to the differentiable framework. 

By contrast, our DiffEC achieves convergence in only $\sim 251$ seconds on a single CPU, with $16\pm4$ function evaluations. This significant reduction arises from leveraging exact gradients through automatic differentiation of the finite volume solver. The efficiency gain directly translates into reduced computational cost, enabling rapid convergence with minimal evaluations. 

In summary, the differentiable approach provides robust and accurate estimation from operando fields and demonstrates superior efficiency over representative conventional optimization methods in terms of both wall time and function evaluations.

\clearpage
\section{Test and Validation}
This section tests and validates finite difference simulation of voltammetry in weakly supported media, one of the more challenging simulation tasks. 
Simulations of the voltammetry in weakly supported media used expanding spatial grid to save computational resources and to accelerate computational processes. The expanding grid is expressed as:
\begin{equation}
    X(i) = X(i-1) + \Delta X \times \omega^{i-1} \ \text{where} \geq 1
\end{equation}
where $X(i)$ is the dimensionless coordinate at point $i$ and $X(0)=0$, which is the surface of the electrode. The dimensionless spatial step $\Delta X$ and the expanding grid factor $\omega$ collectively determines the fineness of the grid. The time step of simulation is determined by $\Delta \theta$, which is uniform during simulation. Thus, $\Delta X$, $\Delta \theta$ and $\omega$ are systematically investigated to test the convergence of simulations. 

\begin{figure}
    \centering
    \includegraphics[width=0.9\linewidth]{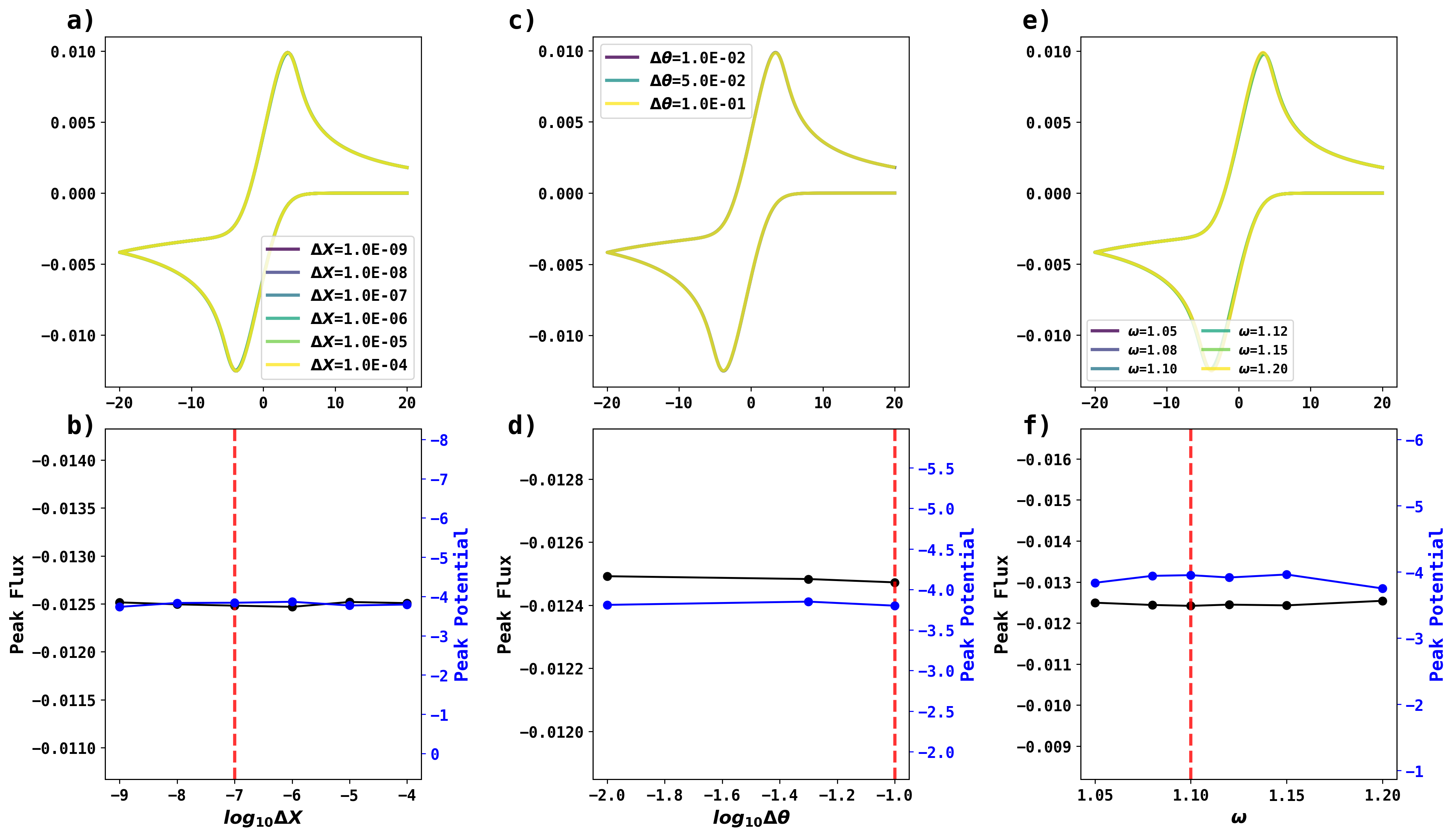}
    \caption{Convergence test when $\sigma=10^{-3}$ and $K_0=10$. (a, b) The voltammograms, peak fluxes and potentials as a function of $\Delta X$. (c, d) The voltammograms, peak fluxes and potentials as a function of $\Delta \theta$. (e, f) The voltammograms, peak fluxes, and peak potentials as a function of $\omega$.}
    \label{fig:Convergence1}
\end{figure}

There are two metrics used to evaluate the convergence of simulation: forward scan peak fluxes and forward scan peak potentials. To test the convergence of simulations, a series of simulations with BV kinetics were performed at a low support ratio of $C_{sup}^*=5$ and a transfer coefficient at $\alpha=0.5$,at dimensionless scan rates of $\sigma=10^{-3}$ and $10^7$, and electrochemical rate constants of $K_0=10$ and $10^5$. The four sets of parameters represent voltammetry with very fast and very slow kinetics and scan rates. The charges of the four species are $z_A=0$, $z_B=-1$, $z_M=1$ and $z_N=-1$. To test convergence, $\Delta X$ was varied between $10^{-4}$ to $10^{-9}$, $\Delta \theta$ was varied between $10^{-1}$ to $10^{-2}$, and $\omega$ was varied between $1.05$ to $1.20$. The results are illustrated at Figure S \ref{fig:Convergence1} to Figure S \ref{fig:Convergence4}, and the red-dashed vertical lines were simulation parameters used in production runs at $\Delta X=10^{-7}$,$\Delta \theta=10^{-1}$ and $\omega=1.10$. The four figures suggested that the spatial and temporal steps are sufficiently small for the differentiable simulation production runs. 

\begin{figure}
    \centering
    \includegraphics[width=0.9\linewidth]{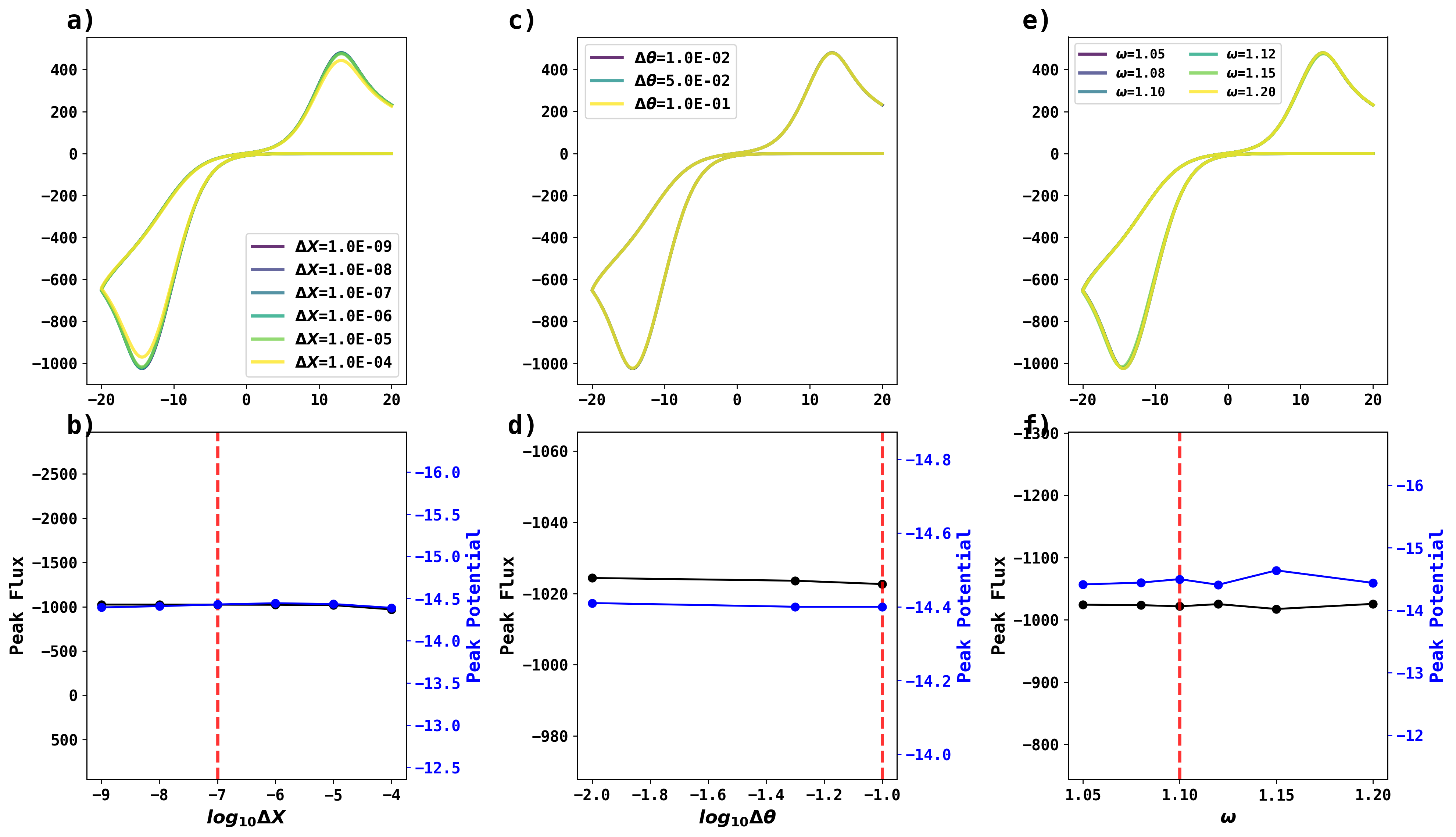}
    \caption{Convergence test when $\sigma=10^{7}$ and $K_0=10$. (a, b) The voltammograms, peak fluxes and potentials as a function of $\Delta X$. (c, d) The voltammograms, peak fluxes and potentials as a function of $\Delta \theta$. (e, f) The voltammograms, peak fluxes, and peak potentials as a function of $\omega$.}
    \label{fig:Convergence2}
\end{figure}

\begin{figure}
    \centering
    \includegraphics[width=0.9\linewidth]{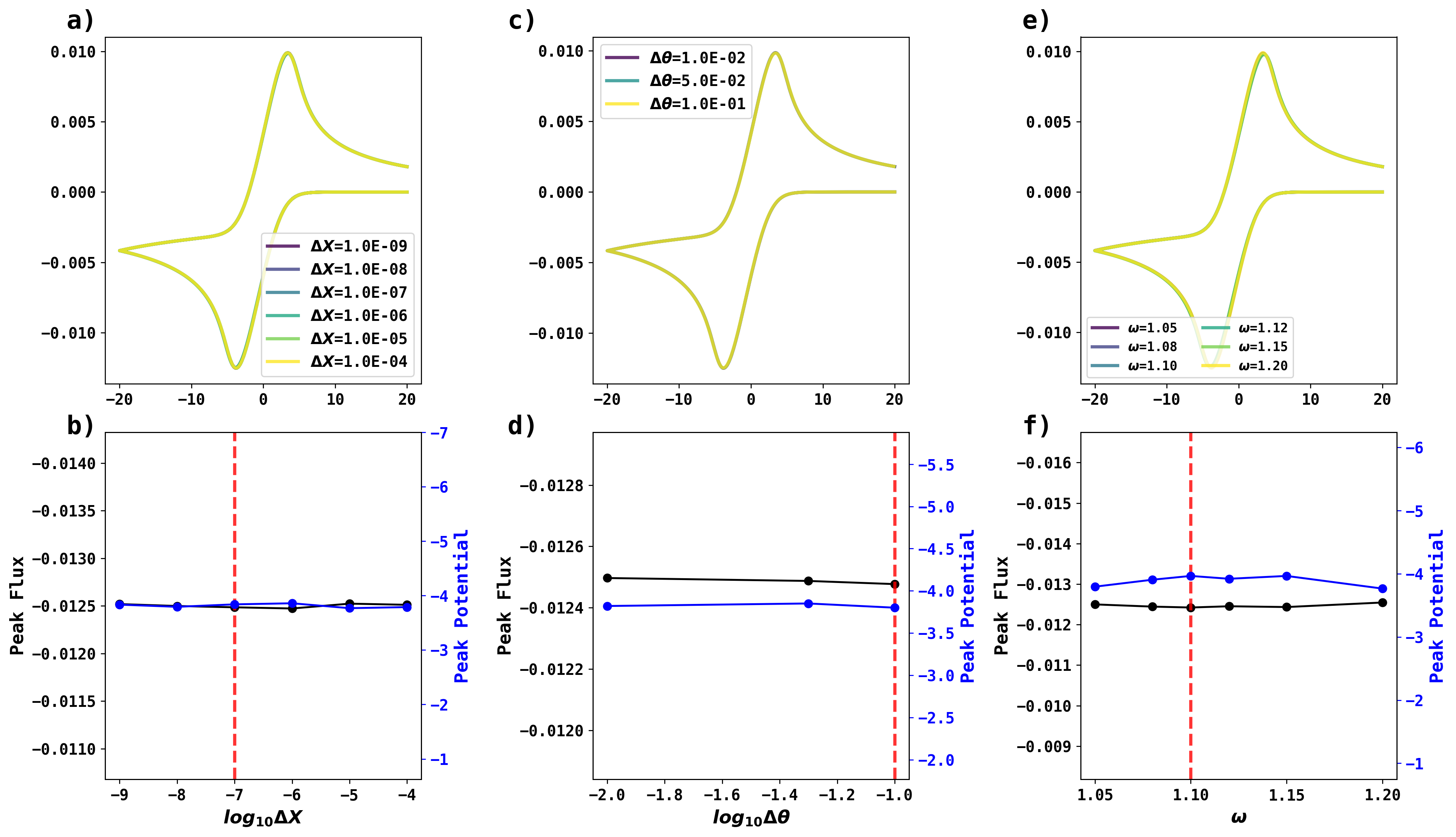}
    \caption{Convergence test when $\sigma=10^{-3}$ and $K_0=10^5$. (a, b) The voltammograms, peak fluxes and potentials as a function of $\Delta X$. (c, d) The voltammograms, peak fluxes and potentials as a function of $\Delta \theta$. (e, f) The voltammograms, peak fluxes, and peak potentials as a function of $\omega$.}
    \label{fig:Convergence3}
\end{figure}

\begin{figure}
    \centering
    \includegraphics[width=0.9\linewidth]{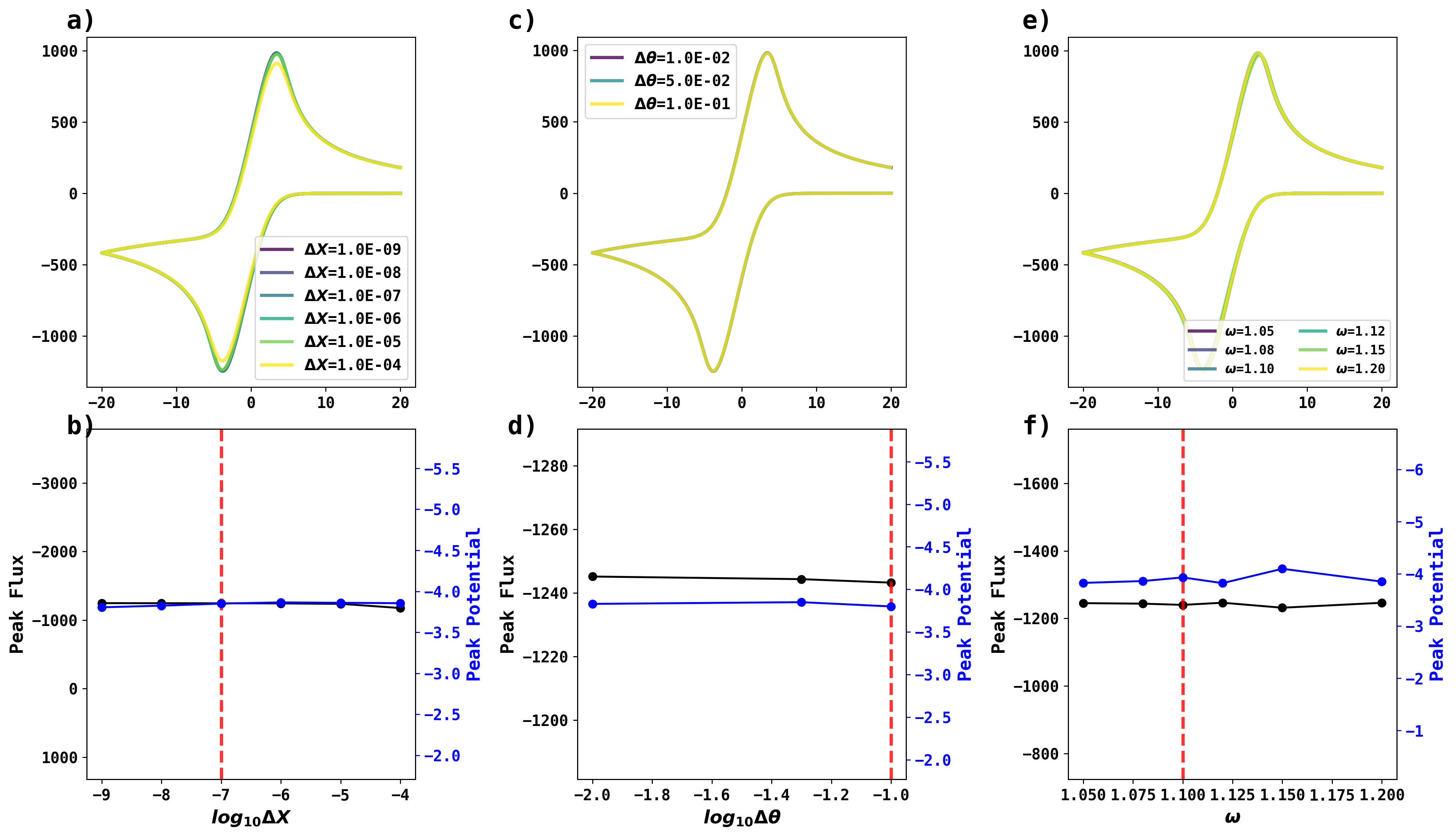}
    \caption{Convergence test when $\sigma=10^{7}$ and $K_0=10^5$. (a, b) The voltammograms, peak fluxes and potentials as a function of $\Delta X$. (c, d) The voltammograms, peak fluxes and potentials as a function of $\Delta \theta$. (e, f) The voltammograms, peak fluxes, and peak potentials as a function of $\omega$.}
    \label{fig:Convergence4}
\end{figure}

\section{Computational Methods}
Differentiable simulations for the weakly supported voltammetry and acetic acid reductions were performed on the Lighthouse cluster at University of Michigan, Ann Arbor with 6 CPU cores and 80 GB of memory per core. The large memory requirement was due to the time marching scheme with small timestep to account for fast electrochemical reactions, and the iterative Newton-Raphson method for nonlinear equations. Despite the large memory requirement, each simulation took only 5 minutes. The linear hydrodynamic voltammetry was simulated using 16 CPU cores and 5 GB of memory per core. The large memory requirement can be resolved with adaptive time steeping method,\cite{shampine1986some} balancing conventional reverse mode automatic differentiation with forward mode automatic differentiation, or with increasing accessibility of semiconductors.\cite{dyer2023gradient} 

\section{Verifiability Checklist}
Verifiability defines the trustworthiness and scope of scientific research in where it is valid. Verifiability in energy research is very important as it enables stakeholders (experimentalists, investors, policy makers, etc.) to trust insights from a model without necessarily reproducing the entire study (although we provided comprehensive information to fully reproduce our study). Verifiability ensures that the theoretical insights are more transparent, interpretable, and meaningful beyond the original authorship. Increasing verifiability maintains scientific integrity and research impact and completing a verifiability checklist is a desired way proposed by Mistry et al. in 2021 to increase verifiability.\cite{Mistry2021Verifiability} A verifiability checklist is thus provided in Table S \ref{tab:Verifiability Test} and the paper meets the all requirements in the variability checklist.

\begin{table}
    \centering
    \begin{tabular}{|p{0.05\linewidth} | p{0.8\linewidth} | p{0.05\linewidth}| }
    \toprule
         1&Have you provided all assumptions, theory, governing equations, initial and boundary conditions, material properties (e.g. open circuit potential) with appropriate precision and literature sources, constant states (e.g. temperature), etc.?   &Y \\ \midrule
         &\textbf{Remarks}: Yes, all assumptions, theory, governing equations, initial and boundary conditions are fully provided in the Supporting Information.  For example, when simulating Hydrogen Evolution Reaction at rotating disk electrode, the assumption of 1D simulation, the mass transport and electrokinetic theory, the convection-diffusion equation, initial and boundary conditions are provided in section \ref{HERSection}, The parameters used in simulation are provided in Table S \ref{tab:HER_LSV_Parameters}.   & \\ \midrule
         2&If your calculation has a probabilistic component (e.g. Monte Carlo, initial configuration of Molecular Dynamics, etc., did you provide statistics (mean, standard deviation, confidence interval, etc.) from multiple ($\geq3$) runs of a representative case?   &Y  \\ \midrule
         &\textbf{Remarks}: All experimental cases, including $\ch{Fe3+}$/$\ch{Fe2+}$, RuHex redox couple, Li electrodeposition and stripping, and Acidic HER, and symmetric Li cell.  & \\ \midrule
         3&If data-driven calculations are performed (e.g.), did you specify dataset origin, the rationale behind choosing it, what information it contains, and the specific portion of it being utilized? Have you described the thought process for choosing a specific modeling paradigm?  & Y \\ \midrule
         &\textbf{Remarks}: Yes, data origin of all data-driven cases are provided in the manuscript and additionally mentioned here: steady state currents of acetic acid reduction,\cite{RN22} LSV of HER on rotating Pt electrode,\cite{AP1} and operando concentrations fields.\cite{AP2}  The thought process of introducing differentiable finite difference simulations is given in the introduction part of the manuscript.  & \\ \midrule
         &  & \\
         4&Have you discussed all sources of potential uncertainty, variability, and errors in the modeling results and their impact on quantitative results and qualitative trends? Have you discussed the sensitivity of modeling (and numerical) inputs such as material properties, time step, domain size, neural network architecture, etc. where they are variable or uncertain?&Y \\ \midrule
         &\textbf{Remarks}: The effects of initial guesses and the noise robustness of the differentiable simulators are included in the manuscript. A convergence test was performed to evaluate the sensitivity of time and space grid size and grid expanding factors and is given in section 6 of the supporting information.& \\ \midrule
         5&Have you sufficiently discussed new or not widely familiar terminology and descriptors for clarity? Did you use these terms in their appropriate context to avoid misinterpretation? Enumerate these terms in the “Remarks”.&Y\\ \midrule
         &\textbf{Remarks}: All terminology used are checked by all authors and are considered widely used terminology in the electrochemical research area. &\\
    \end{tabular}
    \caption{The verifiability checklist of this work. }
    \label{tab:Verifiability Test}
\end{table}

\clearpage
\bibliography{Library}